\documentclass[12pt]{article}
\usepackage{amsmath}
\usepackage{graphicx,psfrag,epsf}
\usepackage{enumerate}
\usepackage{natbib}
\usepackage{url} 
\usepackage{amsthm}

\newcommand{\blind}{1}

\usepackage{hyperref}
\usepackage{dsfont}
\usepackage{bm}
\usepackage{xcolor}

\usepackage{algcompatible}
\usepackage{algorithm}
\usepackage[noend]{algpseudocode}
\usepackage{adjustbox}
\usepackage{graphicx}
\usepackage{amsmath}
\usepackage{amssymb}
\usepackage{multirow}
\usepackage[tight,footnotesize]{subfigure}

\usepackage[switch, modulo]{lineno}
\usepackage{setspace}
\usepackage{algcompatible}
\usepackage{algorithm}
\usepackage[noend]{algpseudocode}

\makeatletter
\def\BState{\State\hskip-\ALG@thistlm}
\makeatother

\usepackage{amsmath}

\DeclareMathOperator*{\argmin}{arg\,min}

\newtheorem{definition}{Definition}
\newtheorem{theorem}{Theorem}
\newtheorem{remark}{Remark}

\begin{document}

\def\spacingset#1{\renewcommand{\baselinestretch}%
{#1}\small\normalsize} \spacingset{1}


\if1\blind
{
  \title{\bf SMART-MC: Characterizing the Dynamics of Multiple Sclerosis Therapy Transitions Using a Covariate-Based Markov Model}
  \author{Beomchang Kim 
  \hspace{.2cm}\\
    {\normalsize \medskip Department of Biostatistics, Virginia Commonwealth University}\\
    Zongqi Xia 
    \hspace{.2cm}\\
    {\normalsize \medskip Department of Neurology, Department of Biomedical Informatics,}\\
    {\normalsize \medskip University of Pittsburgh}\\
    and \\
    Priyam Das\thanks{
    The authors gratefully acknowledge the anonymous reviewers and the Editor for their insightful comments and constructive feedback, which greatly improved the clarity and quality of this manuscript. ZX is supported in part by NINDS R01NS098023 and R01NS124882.}
    \hspace{.2cm}\\
 {\normalsize \medskip Department of Biostatistics, Virginia Commonwealth University}\\
 {\normalsize \medskip Department of Biomedical Informatics, Harvard Medical School}}
  \maketitle
} \fi

\if0\blind
{
  \bigskip
  \bigskip
  \bigskip
  \begin{center}
    {\large\bf SMART-MC: Characterizing the Dynamics of Multiple Sclerosis Therapy Transitions Using a Covariate-Based Markov Model}
\end{center}
  \medskip
} \fi

\bigskip
\begin{abstract}
Treatment switching is a common occurrence in the management of Multiple Sclerosis (MS), where patients transition across various disease-modifying therapies (DMTs) due to heterogeneous treatment responses, differences in disease progression, patient characteristics, and therapy-associated adverse effects. To investigate how patient-level covariates influence the likelihood of treatment transitions among DMTs, we adopt a Markovian framework, \textit{Sparse Matrix Estimation with Covariate-Based Transitions in Markov Chain Modeling} (SMART-MC), in which the transition probabilities are modeled as functions of these covariates. Modeling real-world treatment transitions under this framework presents several challenges, including ensuring parameter identifiability and handling sparse transitions without overfitting. To address identifiability, we constrain each transition-specific covariate coefficient vectors to have a fixed L2 norm. Furthermore, our method automatically estimates transition probabilities for sparsely observed transitions as constants and enforces zero transition probabilities for transitions that are empirically unobserved. This approach mitigates the need for additional model complexity to handle sparsity while maintaining interpretability and efficiency. To optimize the multi-modal likelihood function, we develop a scalable, parallelized global optimization routine, which is validated through benchmark comparisons and supported by key theoretical properties. Our analysis uncovers meaningful patterns in DMT transitions, revealing variations across MS patient subgroups defined by age, race, and other clinical factors.

\end{abstract}

\noindent%
\vfill

\newpage
\spacingset{1.8} 
\section{Introduction}
\label{intro}
\vspace{-0.4cm}
\noindent Multiple sclerosis (MS) is a chronic neurological disorder involving immune-mediated damage to the central nervous system. MS primarily includes relapsing-remitting MS (RRMS), characterized by episodic relapses, and progressive forms such as secondary progressive MS (SPMS) and primary progressive MS (PPMS), which involve worsening disability without remission \citep{Dimitriou2023}. Disease-modifying therapies (DMTs) are central to MS management, aiming to reduce relapses, slow progression, and alleviate symptoms. Treatment strategies evolve as patients transition from relapsing to progressive stages, incorporating neurodegeneration-targeted therapies and guided by clinical factors and patient-specific considerations \citep{Goldschmidt2021}. Recent studies highlight the complexity of modeling MS treatment sequences, particularly regarding therapy transitions. Factors like age at onset, relapse frequency, and progression rate influence decisions on treatment escalation or de-escalation \citep{Macaron2023}. Younger RRMS patients benefit from aggressive therapies to reduce long-term disability, while progressive-stage patients prioritize slowing progression over relapse prevention \citep{Iacobaeus2020}. Patient preferences, side effect tolerance, and quality of life further shape therapeutic choices \citep{Hoffmann2024}.

DMTs have evolved with selection based on disease stage, severity, and individual factors, such as prior treatment response and administration preferences. First-line therapies for RRMS include glatiramer acetate and interferon-beta, while oral options, such as dimethyl fumarate, fingolimod, and teriflunomide, provide convenience \citep{Faissner2019}. B-cell depletion therapies, like rituximab and ocrelizumab, reduce disease activity in RRMS and PPMS but not SPMS \citep{Gelfand2017}. Other options, such as natalizumab for high disease activity and alemtuzumab for refractory cases, address specific patient needs despite adverse events \citep{Simpson2021}. Over time, patients often switch therapies, reflecting the dynamic and individualized nature of MS management.

Despite advancements, understanding factors driving treatment transitions remains complex. Models integrating clinical and patient-specific factors, such as relapse dynamics and biomarker profiles, aim to optimize treatment strategies \citep{Frascoli2022}. While studies have examined DMT sequences and criteria for switching or discontinuing treatments \citep{Gross2019}, comprehensive analyses using longitudinal Electronic Health Records (EHR) are limited. Emerging machine learning approaches leverage EHR data to study MS progression \citep{Branco2022}, but they have rarely focused on the factors driving treatment transitions over the course of treatment. In this study we aim to model MS DMT trajectories within a Markovian framework, investigating the influence of clinical and demographic covariates on treatment transitions.

Markov models have been widely used for temporal sequence modeling in domains such as music prediction \citep{Li2019}, website navigation \citep{Melnykov2016}, and longitudinal data analysis \citep{Haan2017}, as well as for clustering tasks involving click-stream data \citep{Urso2024}, and handwriting classification \citep{Coviello2014}. However, their application to electronic health record (EHR) data remains limited. A recent study by \cite{Das2023} clustered Rheumatoid Arthritis patients based on treatment history by estimating transition probabilities, but did not incorporate patient phenotypes into the clustering. Integrating phenotypic data with treatment sequences could improve interpretability and support individualized treatment transition modeling, particularly in MS.

Parameter estimation in mixture Markov and Hidden Markov models remains challenging due to the large number of constrained parameters and the non-concave likelihood function. Proposed solutions include the Expectation-Maximization (EM) algorithm \citep{Helske2019} and hierarchical EM (HEM; \citealp{Coviello2014}), though these methods often converge to local optima. \cite{Das2023MsiCOR} made the first attempt to incorporate patient-specific covariates into the mixture Markov model, enabling covariate-informed clustering, while simultaneously addressing the non-concave likelihood maximization issue by introducing a Pattern Search (PS; \citealp{Torczon1997}) based global optimization method, which avoids local solutions and improves global maximization. Non-convex benchmark studies demonstrated its superiority over existing global and local optimization algorithms, such as Genetic Algorithm (GA), Sequential Quadratic Programming (SQP) and Interior Point (IP) algorithm \citep{Fraser1957, Nocedal2006}. It also outperformed the EM-based approach in the context of mixture Markov modeling. However, a few caveats remain. First, in their proposed strategy, patient-specific covariates account for clinical and demographic influences on cluster membership. Consequently, transition probabilities are estimated at the cluster level, potentially obscuring the direct influence of diverse phenotypes (e.g., race, sex, age) on treatment transitions. Second, rare or absent transitions, such as from mitoxantrone to glatiramer acetate, are not explicitly constrained to zero, leading to possible non-zero estimates. While sparse regression techniques like LASSO \citep{Tibshirani1996} may address this sparsity, the non-concavity of the likelihood function and the cross-validation requirements arise additional computational challenges.

To elucidate the role of patient-specific covariates in treatment transitions, we propose modeling transition probabilities as functions of covariates instead of clustering patients by phenotype. This allows a more nuanced understanding of phenotype effects on transitions. To mitigate computational challenges of penalized models like LASSO, we estimate transition probabilities as functions of covariates only when empirical transitions exceed the number of phenotypes or a user-defined threshold. For rare transitions, with low empirical counts, probabilities are treated as constants, reducing computational burden while ensuring empirical alignment. Our proposed method, Sparse Matrix Estimation with Covariate-Based Transitions in Markov Chain Modeling (SMART-MC), estimates individualized treatment transition probabilities while addressing sparsity and model identifiability. To address the non-concave likelihood, we develop the Multiple Spherically Constrained Optimization Routine (MSCOR), a parallelizable global optimization algorithm.

The rest of the paper is organized as follows. Section \ref{dataset} outlines the research objectives and describes the dataset. Section \ref{smart_mc} introduces SMART-MC. Section \ref{MSCOR} develops MSCOR and benchmarks its performance. Section \ref{MS_analysis} applies it to estimate the effects of covariates on MS DMT transition probabilities using EHR data. Section \ref{conclusion} concludes with future research directions.
\vspace{-.5cm}

\section{Covariate-driven MS-DMT Transition Dynamics}
\label{dataset}
\vspace{-0.3cm}
Multiple Sclerosis (MS) is a heterogeneous disease where treatment switching is common due to factors such as treatment response, tolerability, side effects, and evolving disease course. 
Over the past decades, numerous DMTs have been developed to manage MS and mitigate the frequency and severity of relapses. However, treatment response varies considerably among patients due to heterogeneity in disease progression, patient characteristics, and adverse effects associated with specific therapies. Consequently, treatment switching is a common clinical occurrence, highlighting the need for a deeper understanding of how patients transition between therapies over time.

Understanding how patient-level clinical and demographic factors influence longitudinal treatment transitions remains a key gap in the MS literature \citep{Weideman2017, Casanova2022}. While prior studies have highlighted predictors of initiating high-efficacy therapies \citep{Ontaneda2017}, few have systematically quantified transition patterns across the full sequence of MS disease-modifying therapies (DMTs) in real-world populations. We structure our investigation around the following research questions:
\begin{itemize}
    \item [(i)] How do clinical factors such as disease duration influence the likelihood of transitioning between first-line injectables (e.g., interferon-beta, glatiramer acetate), oral therapies (e.g., dimethyl fumarate, S1P modulators), and high-efficacy agents (e.g., natalizumab, B-cell depleting therapies)?
    \item [(ii)] How do demographic factors such as age, sex, and race/ethnicity impact treatment sequencing choices?
    \item [(iii)] What are the most frequent transition pathways observed in real-world MS care, and which treatment transitions are most sensitive to patient characteristics?
\end{itemize}
By systematically modeling transition probabilities as a function of patient-level covariates, our framework aims to identify key factors associated with treatment switching and quantify how covariates regulate longitudinal treatment dynamics in MS. In addition, characterizing these patterns may help healthcare providers anticipate drug demand, optimize treatment allocation, and guide insurers and policymakers in developing reimbursement policies that promote evidence-based and cost-effective MS care.

We analyze MS DMT sequence data from the electronic health record (EHR) system of the Massachusetts General and Brigham hospital network (Boston, US), including the Comprehensive Longitudinal Investigation of Multiple Sclerosis at Brigham and Women's Hospital (CLIMB) cohort \citep{Liang2022}. The dataset contains patient-level DMT usage along with clinical and demographic information. To ensure data reliability, we include patients who initiated DMTs on or after January 1, 2006, when electronic prescribing was adopted. To avoid over-counting repeated visits with the same DMT in short intervals, we aggregate observations into three-month periods starting from the DMT initiation date. Within each interval, identical consecutive DMTs are collapsed into a single entry. For example, $A \to A \to A$ becomes $A$, while $A \to A \to B \to B \to A \to C \to C$ reduces to $A \to B \to A \to C$. Consecutive observations of the same DMT are still allowed if they span different intervals.

The final cohort included 822 patients with a mean age of 36.7 (s.d. 10.4) years and a mean disease duration (defined as the time elapsed from the onset of the first neurological symptom to the start of DMT) of 15.4 (s.d. 9.3) months. Of these patients, 74.0\% were female, 90.9\% were White, and 4.9\% were Black. A total of 91.0\% of patients experienced at least one treatment switch during follow-up, undergoing a median of 9.0 transitions with a mean of 10.5 transitions. Twelve distinct DMTs were available in the dataset: alemtuzumab (Ale), cyclophosphamide (Cyc), daclizumab (Dac), dimethyl fumarate (DF), fingolimod (Fin), glatiramer acetate (GA), interferon-beta (IB), mitoxantrone (Mit), natalizumab (Nat), ocrelizumab (Orc), rituximab (Rit), and teriflunomide (Ter). Since Dac has been withdrawn from the market and was rarely prescribed, we excluded it from analysis. Rit and Orc were grouped into a single mechanistic category termed B-cell depletion (BcD); Fin and Ter were grouped as S1P modulators; and Cyc, Mit, and Ale, which are infrequently used and primarily reserved for aggressive MS, were grouped into a category labeled Aggressive/Legacy therapies (AL). Consequently, the analysis considers seven DMT categories, which define the state space of our Markov model.

\begin{figure}[]
	\centering
	\includegraphics[width=.99\textwidth]{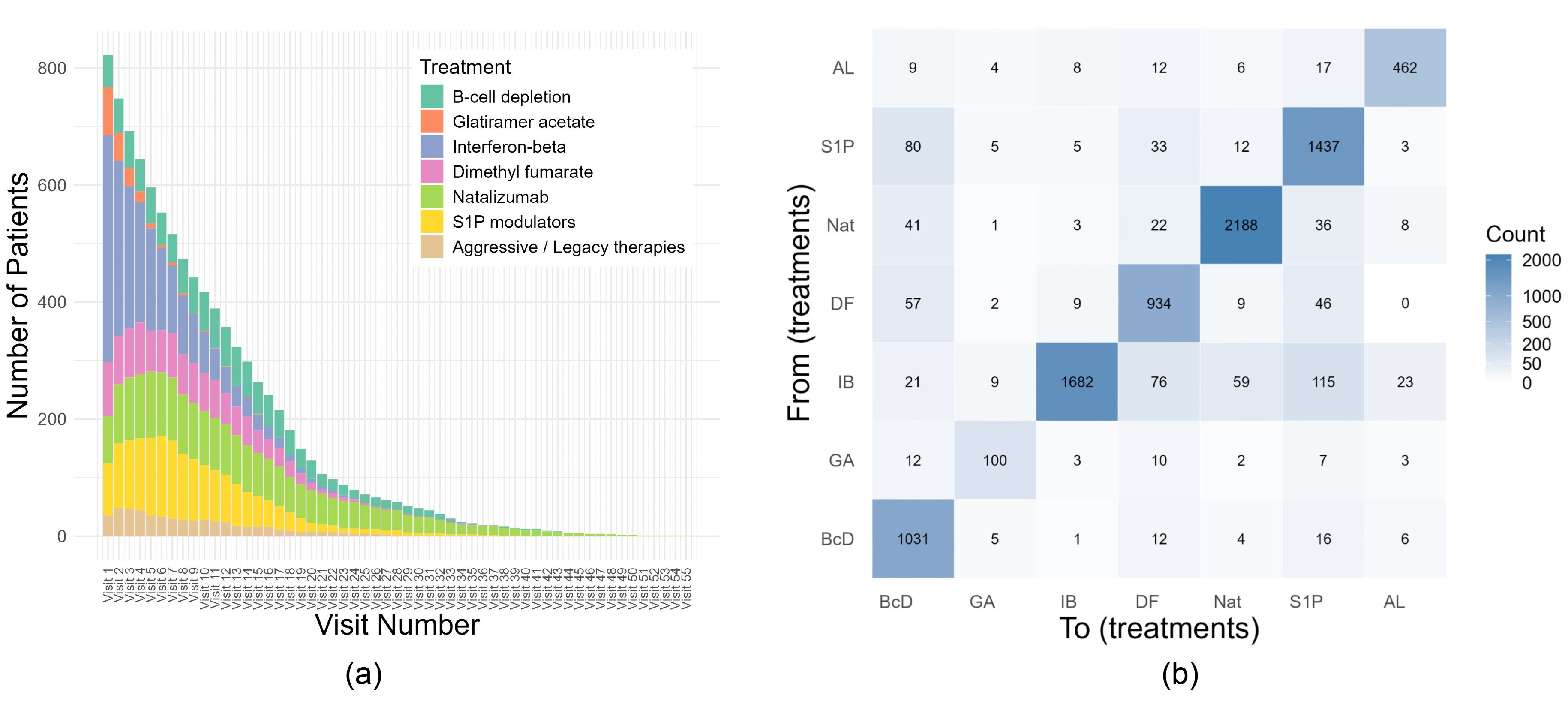}
	\caption{(a) Stacked alluvial-style plot displaying the longitudinal treatment sequences across visits.        The vertical bars represent treatment steps, and the stacked colors represent the distribution of              therapies at each step.
        (b) Empirical transition matrix heatmap showing the observed counts of transitions between therapies. Cells indicate the number of observed transitions from each treatment (rows) to the next treatment (columns), with darker shading indicating higher counts. 
        }
	\label{fig:exploratory}
\end{figure}

Exploratory analyses of treatment transitions are summarized in Figure~\ref{fig:exploratory}. In panel (a), the relative frequencies of DMT use across visits are displayed, where IB appears as the most common initial therapy and Nat emerges as the most frequent long-term maintenance option. Panel (b) shows the empirical transition matrix, revealing the sparse nature of transitions across treatment pairs. Among maintenance therapies (i.e., remaining on the same treatment), Nat, IB, and S1P were most frequently sustained. The most common across-treatment transitions were from IB to S1P (115 transitions), S1P to BcD (80 transitions), IB to DF (76 transitions), IB to Nat (59 transitions), and DF to BcD (57 transitions). These findings underscore the dynamic and heterogeneous nature of treatment sequences in MS, motivating a transition-based modeling framework.

To investigate the influence of patient-specific covariates on treatment transitions in the EHR cohort of MS patients, we employ a Markov chain framework in which transition probabilities are modeled as functions of patient-level covariates. While conceptually intuitive, this approach poses challenges related to model identifiability and the optimization of a potentially multi-modal likelihood function, as discussed later. To address these issues, we propose SMART-MC, a novel statistical modeling framework, complemented by MSCOR global optimization tool to maximize model likelihood. Subsequently, a detailed analysis is conducted to examine how patient covariates influence the DMT trajectory of MS patients throughout their treatment course. 
\vspace{-0.6cm}
\section{SMART-MC}
\label{smart_mc}
\vspace{-.6cm}
\subsection{SMART-MC Model Framework}
\vspace{-.4cm}
Consider a dataset of treatment sequences from $K$ patients, each prescribed one of $N$ treatments at various doctor visits. The treatment sequence for patient $k$ is denoted as $\bm{Y}_k = \{(Y_{k,1}, \dots, Y_{k,t_k})\}$, where $Y_{k,t} \in \{1, \dots, N\}$ is the treatment at time $t$, and $t_k$ is the sequence length. Each patient is also characterized by $p$ covariates, $\bm{X}_k = (X_{k1}, \dots, X_{kp})$. Assuming a Markovian framework, the treatment sequence is modeled using an initial state vector (ISV) $\bm{M}_{s}^{(k)}$ and a transition matrix (TM) $\bm{M}_{T}^{(k)}$, as follows:
\vspace{- 0.4\baselineskip}
\begin{align*}
    &\bm{M}_{s}^{(k)}=(s_{1}^{(k)}, s_{2}^{(k)},\dots,s_{N}^{(k)}), \; \; 0\leq s_{v}^{(k)}\leq1, \; \;  \sum_{v=1}^{N}{s_{v}^{(k)}}=1.\\
    &\bm{M}_{T}^{(k)} = \begin{bmatrix}
m_{1,1}^{(k)} & m_{1,2}^{(k)} & \dots & m_{1,N}^{(k)} \\ 
\vdots & \vdots & \ddots & \vdots \\ 
m_{N,1}^{(k)} & m_{N,2}^{(k)} & \dots & m_{N,N}^{(k)}
\end{bmatrix}, \;\; 0\leq m_{u,v}^{(k)}\leq1, \; \; \sum_{v=1}^Nm_{u,v}^{(k)} = 1,\; u,v = 1,\ldots,N.
\end{align*}
Before proceeding further, we briefly overview the contextual interpretation of the model parameters introduced herein. Let $s_{v}^{(k)}$ represent the initial state probability of treatment $v$ for the $k$-th patient, and let $m_{u,v}^{(k)}$ denote the probability of transitioning from treatment $u$ to treatment $v$ for the $k$-th patient. For convenience, we define $m_{0,v}^{(k)} \equiv s_{v}^{(k)}$ for the remainder of the article. By appending the ISV and TM for the $k$-th patient, we obtain $\bm{M}^{(k)} = \begin{bmatrix} \bm{M}_s^{(k)} \\ \bm{M}_T^{(k)} \end{bmatrix}$, a $(N+1)\times N$ matrix, such that $0 \leq m_{u,v}^{(k)} \leq 1$, and $\sum_{v=1}^N m_{u,v}^{(k)} = 1$, for $u = 0, 1, \dots, N$, $v = 1, \dots, N$. We aim to model each $m^{(k)}_{u,v}$ as a function of patient-specific phenotypes $\bm{X}_k$. To facilitate this, we introduce a matrix of coefficient vectors $\bm{B} = \big(\bm{\beta}_{u,v}\big)_{(N+1) \times N}$ for $u = 0, 1, \dots, N$, $v = 1, \dots, N$, where each $\bm{\beta}_{u,v} \in \mathds{R}^{p+1}$ is a coefficient vector of length $p+1$, representing the transition-specific coefficients, including the intercept. Following the multinomial logistic regression framework \citep{Theil1969}, we model $m^{(k)}_{u,v}$ as a function of the covariates $\bm{X}_k$ as:
\vspace{-0.4cm}
\begin{align}
m^{(k)}_{u,v} = \frac{\exp(\bm{X}_k^{\prime}\bm{\beta}_{u,v})}{\sum_{v = 1}^N \exp(\bm{X}_k^{\prime}\bm{\beta}_{u,v})}, \;\; u = 0, 1, \dots, N, \; v = 1, \dots, N,
\label{multinomial}
\end{align}
where $\bm{X}_k^{\prime} = [1 \;\; \bm{X}_k]_{1 \times (p+1)}$ is the covariate vector with an appended 1 to incorporate the intercept. This formulation allows the estimation of patient-specific initial state and transition probabilities as functions of covariates, while adhering to the constraints on $m^{(k)}_{u,v}$. While this framework assigns individualized transition matrices (including the initial state vector) to each patient, several caveats remain with regard to ensuring the model's identifiability, as discussed in the following subsection.
\vspace{-0.5cm}
\subsection{Imposed Constraints to Ensure Identifiability}
It is straightforward to verify that the model remains non-identifiable without additional constraints. Given its similarity to multinomial logistic regression, a natural solution is to set one state as the reference by fixing $\bm{\beta}_{u,v} = \bm{0}$ for some $v \in \{1, \dots, N\}$ within each row $u$. While this resolves identifiability, it forces at least one non-zero component per row, limiting our goal of inducing data-driven sparsity in the transition matrix (see Section \ref{intro}). In MS DMT sequence modeling, many transitions are rare, often with empirical counts near zero, and cannot be anticipated a priori. Imposing non-zero constraints on such transitions may yield non-zero estimates even when empirical counts are zero, reducing model flexibility. This motivates the need for an alternative identifiability strategy.

Instead of using the non-identifiability resolution technique from the previous approach, we propose constraining each $\bm{\beta}_{u,v}$ to have an $l_2$ norm of 1, defined as $||\bm{x}||_2 = \sqrt{x_1^2 + \cdots + x_n^2}$. This constraint, well-studied in single-index modeling \citep{Carroll1997, Das2017}, parsimoniuosly resolves the identifiability issue and improves control over $m^{(k)}_{u,v}$ uniformly for all $v \in \{1, \dots, N\}$. If a transition has zero empirical count, we bypass estimating the corresponding coefficient vector, making appropriate adjustments. 
In the following subsection, we explore how our model framework incorporates sparsity in the transition matrix in a fully data-driven manner.
\vspace{-0.6cm}
\subsection{Adjustments to the Model for Rare Transition Estimation}
\label{smart}
\vspace{-.3cm}
In this subsection, we outline adjustments that enable data-driven estimation of transition probabilities for rare events. As seen in \eqref{multinomial}, each transition probability depends on $p+1$ coefficients. When empirical transition counts fall below this threshold, estimating the corresponding coefficient vector becomes ill-posed. In such cases, we treat transition probabilities as constants derived from observed data. While this precludes inference on covariate effects for rare transitions, it is appropriate when sample sizes are insufficient to support model estimation. This strategy avoids attempting to estimate more parameters than available data points. We now describe the SMART mechanism for handling such rare transitions.

First, we find the empirical counts corresponding to each initial state and across-state transition. Let $\widehat{\bm{C}}_s$ and $\widehat{\bm{C}}_T$ denote the empirical initial state count vector and the empirical transition count matrix, respectively, given by $\widehat{\bm{C}}_s = \big(\widehat{c}_{0,v} \big)_{1 \times N}$ and $\widehat{\bm{C}}_T = \big(\widehat{c}_{u,v} \big)_{N \times N}$ for $u,v = 1,\ldots,N$. Next, by appending $\widehat{\bm{C}}_s$ and $\widehat{\bm{C}}_T$, we obtain the empirical count matrix $\widehat{\bm{C}} = \begin{bmatrix} \widehat{\bm{C}}_s \\ \widehat{\bm{C}}_T \end{bmatrix}$, a $(N+1) \times N$ dimensional matrix. Furthermore, by dividing each row of the empirical count matrix $\widehat{\bm{C}}$ by the corresponding row sums, we obtain the empirical probability matrix $\widehat{\bm{M}} = \big(\widehat{m}_{u,v} \big)_{(N+1) \times N}$, where $\widehat{m}_{u,v} = \frac{\widehat{c}_{u,v}}{\sum_{v=1}^N\widehat{c}_{u,v}}$, which accounts for both the initial state and the across-state transition probabilities.

The selection of which elements of $\bm{M}^{(k)}$ to model as functions of covariates is guided by the empirical count matrix $\widehat{\bm{C}}$. A transition (or initial state) is included in the covariate-dependent component if its empirical count is at least $p+1$, ensuring sufficient data to avoid over-parameterization. For greater estimation stability, however, a more conservative threshold, denoted as \textit{Tol} (e.g., \textit{Tol} = $2(p+1)$ or $5(p+1)$), may be used. Based on $\widehat{\bm{C}}$, we define an inclusion indicator matrix $\bm{I} = \big( \bm{I}(u,v) \big)_{(N+1) \times N}$, where $\bm{I}(u,v) = 1$ if $\widehat{c}_{u,v} \geq \textit{Tol}$, and 0 otherwise.

Empirically estimating rare transition (or initial state) probabilities calls for an adjustment to the corresponding rows of $\bm{M}^{(k)}$ to ensure that each row sums to 1. This adjustment involves additional scaling, especially when at least one element in the row is modeled as a function of covariates. In SMART-MC, we ensure that the transition probabilities for rare transitions (or initial states, i.e., locations where $\widehat{c}_{u,v} < \textit{Tol}$) remain equal to their empirical probabilities across all $\bm{M}^{(k)}$, for $k = 1, \ldots, K$. To implement this, we scale the probabilities of non-rare transitions (or initial states) in each row so that their sum equals one minus the sum of the empirical probabilities for the rare transitions. This is done by first computing the complementary indicator matrix $\bm{J}$, where $\bm{J}(u,v) = 1 - \bm{I}(u,v)$ for $u = 0, 1, \ldots, N$ and $v = 1, \ldots, N$. Next, we define the linear projection matrix of $\bm{X}^{\prime}$ with respect to $\bm{B}$ as $\bm{L}^{(k)} = \big(L_{uv}^{(k)}\big)_{(N+1)\times N}$, where, $L_{uv}^{(k)} = \exp\left( \bm{X}_k^{\prime} \bm{\beta}_{uv} \right)$. Taking Hadamard (element-wise) product of $\bm{L}^{(k)}$ and $\bm{I}$ we obtain $\bm{H}^{(k)} = \bm{L}^{(k)} \; \circ \; \bm{I}= \big(H_{u,v}^{(k)}\big)_{(N+1) \times N}$,
where $H_{u,v}^{(k)} = L^{(k)}_{u,v} \cdot I(\widehat{c}_{u,v} \geq \textit{Tol})$ for $u=0, 1,\ldots,N,\; v = 1,\ldots, N,\; k = 1,\ldots, K$. Now, taking Hadamard product of $\widehat{\bm{M}}$ and $\bm{J}$ we get $\bm{G} = \widehat{\bm{M}} \; \circ \; \bm{J}= \big(G_{u,v}\big)_{(N+1) \times N}$, where $G_{u,v} = \widehat{m}_{u,v} \cdot I(\widehat{c}_{u,v} < \textit{Tol})$ for $u=0, 1,\ldots,N,\; v = 1,\ldots, N$. 
Finally, the adjusted $m_{u,v}^{(k)}$ is given by
\vspace{-.9cm}
\begin{align}
    m_{u,v}^{(k)} = \; & G_{u,v} + \big(1 - \sum_{n=1}^N G_{u,n}\big)\frac{H_{u,v}^{(k)}}{\sum_{n = 1}^N H_{u,n}^{(k)}} \nonumber \\
    = \; & \widehat{m}_{u,v} \cdot I(\widehat{c}_{u,v} < \textit{Tol}) \; + \nonumber \\
    \; & \bigg(1 - \sum_{n=1}^N \widehat{m}_{u,n} \cdot I(\widehat{c}_{u,n} < \textit{Tol})\bigg)  \cdot   \frac{exp\big(\bm{X}_k\bm{\beta}_{u,v}\big) \cdot I(\widehat{c}_{u,v} \geq\textit{Tol})}{\sum_{n = 1}^N exp\big(\bm{X}_k\bm{\beta}_{u,n}\big) \cdot I(\widehat{c}_{u,n} \geq\textit{Tol})},
    \label{final_trans}
\end{align}
for $u=0, 1,\ldots,N,\; v = 1,\ldots, N$. This adjustment ensures the constraints $\sum_{v = 1}^Nm_{u,v}^{(k)} = 1$ and $m_{u,v}^{(k)} \geq 0$ for $u =0, \ldots, N$ are satisfied.
\vspace{-.6cm}

\subsection{Likelihood}
\vspace{-.4cm}
Suppose the treatment sequence for the $k$-th patient is denoted as $\bm{Y}_k = \{(Y_{k,1}, \dots, Y_{k,t_k})\}$, where $Y_{k,t} \in \{1, \dots, N\}$, additionally characterized by patient-specific covariates $\bm{X}_k$. Under the Markov assumption, where transitions depend only on the current treatment state and covariates, the full likelihood for the entire patient cohort is given by:
\vspace{-0.5cm}
\begin{align}
P(\bm{Y}_1, \dots ,\bm{Y}_K | \bm{B}, \bm{X}_1, \dots ,\bm{X}_K) = \prod_{k=1}^{K}m^{(k)}_{0,Y_{k,1}}m^{(k)}_{Y_{k,1},Y_{k,2}} \cdots m^{(k)}_{Y_{k,t_k - 1},Y_{k,t_k}}.
\label{likelihood}
\end{align}
The Markov formulation reflects clinical practice where treatment decisions at each visit are primarily driven by the patient’s current disease status, treatment response, side-effect profile, and updated clinical information, rather than the entire historical treatment sequence. This assumption has been widely used in modeling treatment switching patterns in chronic disease management, including MS \citep{Wolfson1985, Das2023MsiCOR}, where longitudinal treatment dynamics often exhibit memoryless or partially memoryless properties once covariates are properly incorporated. Moreover, the multiplicative likelihood structure reflects the independence across patients and the conditional independence of transitions across time under the Markov framework, allowing for efficient likelihood-based estimation.

\begin{figure}[h]
	\centering
	\includegraphics[width=.8\textwidth]{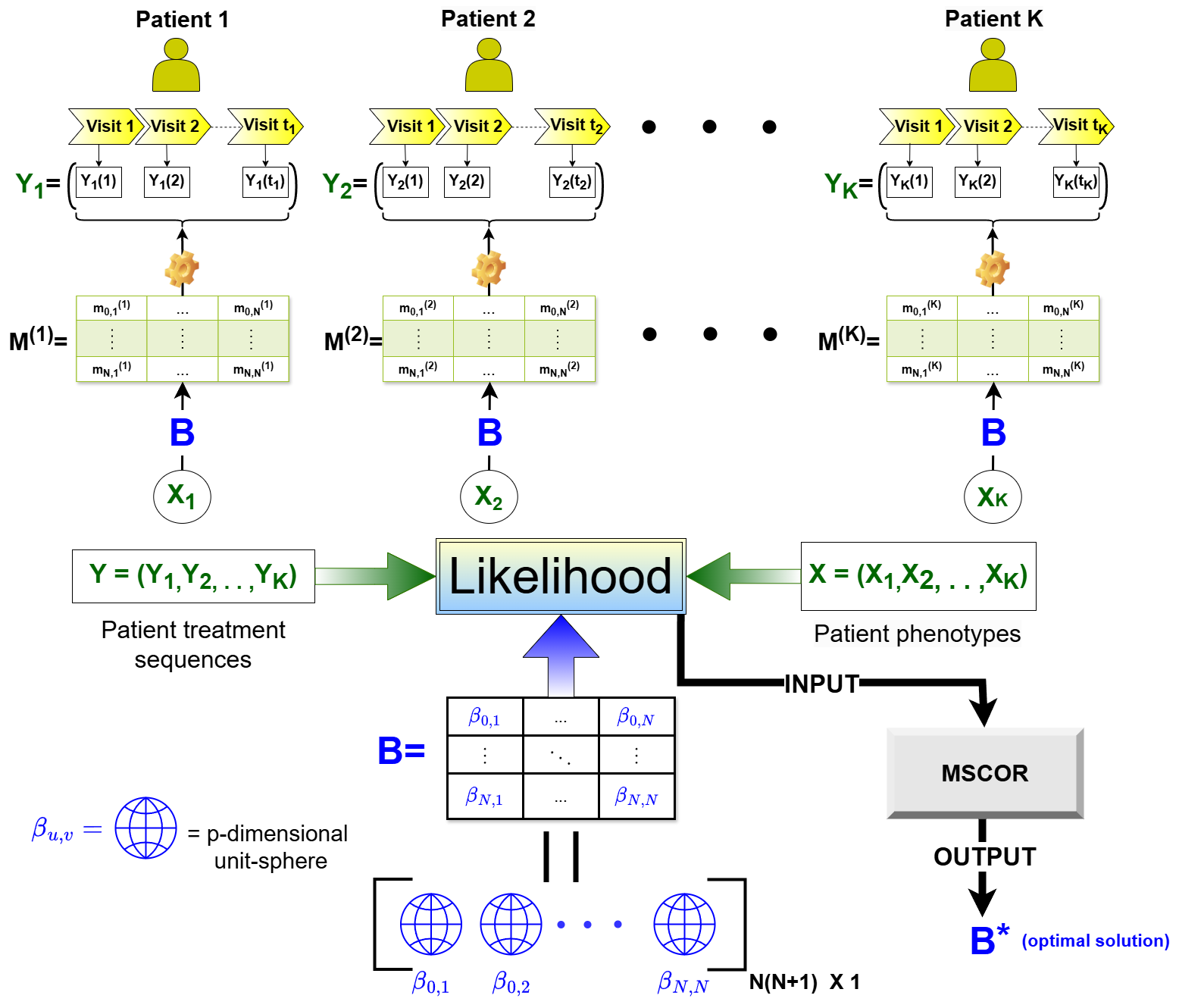}
	\caption{Concept diagram of SMART-MC visually depicting the data structure, likelihood, parameter-space and estimation steps of the analysis.}
	\label{SMART_concept}
\end{figure}

Due to our constraint of fixing the $\ell_2$ norm of each $\bm{\beta}_{u,v}$ to 1, and the non-concavity of \eqref{likelihood} (as discussed later), maximizing it requires a global optimization algorithm capable of maximizing a multi-modal function defined over a collection of unit spheres, which we develop in the following subsection. A visual illustration of SMART-MC is shown in Figure \ref{SMART_concept}. Due to space limitations, the simulation study assessing the performance of SMART-MC is presented in Section C of the supplementary material.
\vspace{-0.4cm}

\vspace{-0.4cm}
\subsection{Theoretical Properties}
The likelihood formulation in \eqref{likelihood} enables maximum likelihood estimation of the SMART-MC parameters under the proposed hybrid model structure, where rare transitions are estimated empirically and non-rare transitions are modeled via covariate-dependent multinomial logits. For each origin state $u \in \{0, 1, \dots, N\}$, define:
\vspace{-0.6cm}
\begin{align*}
\mathcal{V}_u := \left\{ v \in \{1, \dots, N\} \ : \ c_{u,v} \geq \textit{Tol} \right\}, \;
\mathcal{V}_u^c := \left\{ v \in \{1, \dots, N\} \ : \ c_{u,v} < \textit{Tol} \right\},
\end{align*}
where $\textit{Tol}$ denotes the user-specified threshold for sparse cell detection. To establish the large-sample behavior of the resulting estimators, we next provide a set of regularity conditions, which are revisited and justified in Section A of the supplementary material.
\vspace{-0.4cm}
\begin{enumerate}
    \item[(A1)] The data $\{(Y_k, X_k)\}_{k=1}^K$ are independent and identically distributed (i.i.d.) draws from the underlying population.
    \vspace{-0.4cm}
    \item[(A2)] The true transition probabilities satisfy
    \[
    m_{u,v}^{(k)} =
    \begin{cases}
    m_{u,v}^*, & \text{if } v \in \mathcal{V}_u^c, \\
    \left( 1 - \sum_{v' \in \mathcal{V}_u^c} m_{u,v'}^* \right) \cdot \dfrac{ \exp( X_k^\top \beta_{u,v}^* ) }{ \sum_{v' \in \mathcal{V}_u} \exp( X_k^\top \beta_{u,v'}^* ) }, & \text{if } v \in \mathcal{V}_u,
    \end{cases}
    \]
    where $\beta_{u,v}^* \in \mathbb{R}^{p+1}$ satisfy $\|\beta_{u,v}^*\|_2 = 1$, and $m_{u,v}^* \in (0,1)$.
    \item[(A3)] There exists a constant $C > 0$ such that $\| X_k \| \leq C$ for all $k$.
    \item[(A4)] For each $(u,v) \in \mathcal{V}_u$, the Fisher information matrix is full rank and covariates are not perfectly collinear.
\end{enumerate}
\spacingset{1.2}
Under these assumptions, we show the consistency and asymptotic normality of SMART-MC Estimator in Theorem~\ref{thm:consistency} and \ref{thm:asymp_normality}, respectively.
\begin{theorem}\label{thm:consistency}
Under assumptions \textnormal{(A1)}–\textnormal{(A4)}, as $K \to \infty$, the maximum likelihood estimator of SMART-MC satisfies:
\begin{itemize}
    \item [(a)] For all rare transitions $v \in \mathcal{V}_u^c$, $\hat{m}_{u,v} \xrightarrow{P} m_{u,v}^*$.
    \item [(b)] For all non-rare transitions $v \in \mathcal{V}_u$, $\hat{\beta}_{u,v} \xrightarrow{P} \beta_{u,v}^*$.
\end{itemize}
\end{theorem}
\begin{theorem}\label{thm:asymp_normality}
Under assumptions \textnormal{(A1)}–\textnormal{(A4)}, for each fixed origin state $u$ and each non-rare destination state $v \in \mathcal{V}_u$, let $\hat{\beta}_{u,v}$ denote the maximum pseudo-likelihood estimator under the unit-norm constraint:
\[
\hat{\beta}_{u,v} := \arg\max_{\beta \in \mathbb{R}^{p+1}, \| \beta \|_2 = 1} \ell_u(\beta),
\]
where $\ell_u(\beta)$ is the partial log-pseudo-likelihood defined over transitions from state $u$. Then,
\[
\sqrt{n_u} \, P_{u,v}^\top \left( \hat{\beta}_{u,v} - \beta_{u,v}^* \right) \xrightarrow{d} \mathcal{N}(0, \Sigma_{u,v}),
\]
where $P_{u,v} \in \mathbb{R}^{(p+1) \times p}$ is an orthonormal basis matrix for the tangent space $\mathcal{T}_{\beta_{u,v}^*} := \{ h \in \mathbb{R}^{p+1} : \beta_{u,v}^{*\top} h = 0 \}$; $\mathcal{I}_{u,v}$ is the Fisher information matrix evaluated at $\beta_{u,v}^*$ and $\Sigma_{u,v} := \left( P_{u,v}^\top \mathcal{I}_{u,v} P_{u,v} \right)^{-1}$.
\end{theorem}

\spacingset{1.8}
\noindent The detailed proofs of the theorems are provided in Section A of the supplementary material. To facilitate principled statistical inference based on asymptotic normality results, we further derive a Wald-type test, also detailed in Section A of the supplementary material. Although Theorem~\ref{thm:asymp_normality} provides a closed-form expression for asymptotic standard errors in the tangent space, enabling the construction of Wald-type statistical hypothesis tests -- in practice, we recommend using the bootstrap in finite samples (see Section A of the supplementary material) to construct confidence intervals and perform hypothesis tests, especially for derived quantities or when the sample size is moderate.

\section{MSCOR}\label{MSCOR}
\vspace{-0.3cm}
To estimate the matrix of coefficient vectors $\bm{B}$, we maximize the likelihood in \eqref{likelihood}. Each coefficient vector, corresponding to `non-rare' cases, lies on the surface of a $p$-dimensional unit sphere (the space of spherically constrained vectors in $\mathds{R}^{p+1}$). The optimization problem is formulated as:
\vspace{-1.2cm}
\begin{align}
    & \text{maximize:} \; f:\bm{S} \to \mathds{R}, \text{ where } \bm{S} = S^{n_1 - 1} \times \cdots \times S^{n_B - 1},
    \label{opt_problem}
\end{align}
where $S^{w-1} = \{(x_1, \ldots, x_w) \in \mathds{R}^w : \sum_{i=1}^w x_i^2 = 1\}$. Since the likelihood is not concave, a global optimization algorithm is required. In order to optimize the SMART-MC likelihood defined over high-dimensional, non-convex parameter spaces constrained to collections of unit spheres, we employ the Recursive Modified Pattern Search (RMPS) algorithm.
PS provides a derivative-free framework by generating candidate solutions around the current iterate and moving toward improvement. While PS offers some exploration, it may still converge prematurely. RMPS \citep{Das2023RMPSH} extends this approach via a recursive mechanism that adaptively adjusts exploration and search direction, achieving better balance between global and local search. Extensions of RMPS have demonstrated strong performance across constrained domains, including unit spheres \citep{Das2022}, simplexes \citep{Das2021}, and multi-simplex structures \citep{Das2023MsiCOR}. In this work, we further adapt RMPS for non-convex optimization over collections of unit spheres, integrating parallel threading to improve scalability.

\vspace{-0.4cm}
\subsection{MSCOR}
\label{MSCOR_sub}
\vspace{-0.4cm}
\subsubsection{Fermi's Principle}
\begin{figure}[h]
	\centering
	\includegraphics[width=0.7\textwidth]{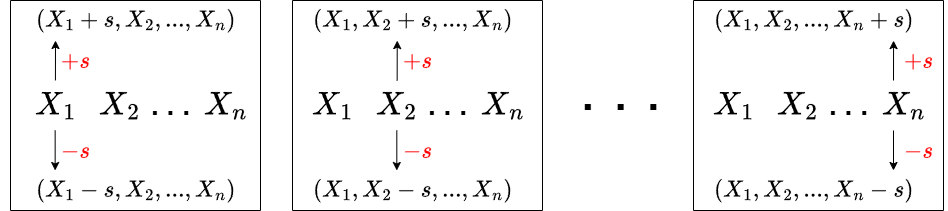}
	\caption{Fermi's principle : Possible $2n$ movements starting from initial point $(x_1,\ldots,x_n)$ inside an iteration with fixed step-size $s$, while optimizing any $n$-dimensional objective function over unconstrained parameter space.}
	\label{fermi}
\end{figure}
\vspace{-0.3cm}
\noindent The RMPS foundation, underlying MSCOR, is based on Fermi's principle \citep{Fermi1952}, which offers a strategy for optimizing an objective function over an unconstrained domain, even if non-differentiable or discontinuous. At each iteration, the function is evaluated at $2n$ neighboring points, corresponding to coordinate-wise movements in both positive and negative directions with a step-size, denoted by $s > 0$. The best-performing point is chosen as the updated solution. By adjusting $s$, candidate points can be sampled from nearby (small $s$) or distant (large $s$) neighborhoods, enabling adaptive exploration. Convergence occurs when no improvement is found as $s \to 0$ \citep{Torczon1997, Das2023RMPSH}. Figure \ref{fermi} shows the candidate points generated under this principle for a given $s$.
\vspace{-0.6cm}
\subsubsection{Movements Across Multiple Spherically Constrained Space}
In the case of a spherically constrained parameter space, starting from a solution on the unit sphere, moving one coordinate by step-size $s$ renders the updated point infeasible since it no longer resides on the unit sphere. To address this, we propose adjustments to the remaining coordinates to maintain the $l_2$-norm of the updated vector as 1. This adjustment, termed the \textit{adjustment step-size}, is computed to ensure feasibility under such constraints, a step unnecessary in unconstrained optimization. At the $j$-th iteration, let the current solution be $\boldsymbol{x}^{(j)} = (x_1, \ldots, x_n)$, where $||\boldsymbol{x}^{(j)}||_2 = 1$. We generate $2n$ candidate points around $\boldsymbol{x}^{(j)}$ using Fermi's principle. Denote the candidate solution after moving the $i$-th coordinate by $s$ in the positive direction as $\boldsymbol{x}^{(i+)} = (x_1^*, \ldots, x_n^*)$, where
\begin{align*}
x_q^* = 
\begin{cases} 
x_q + s & \text{if } q = i, \\
x_q + t_i & \text{if } q \in \{1, \ldots, n\} \setminus \{i\}.
\end{cases}
\end{align*}
To ensure $||\boldsymbol{x}^{(i+)}||_2 = 1$, $t_i$ is chosen such that $\sum_{q=1}^n (x_q^*)^2 = \sum_{q=1,q\neq i}^n (x_q + t_i)^2 + (x_i + s)^2 = 1$. Solving the resulting quadratic equation for $t_i$, we obtain two solutions:
\vspace{-0.3cm}
\begin{align}
t_i^{(1)} &= \frac{-2\sum_{q=1,q\neq i}^n x_q + \sqrt{D_i(s)}}{2(n-1)}, \quad
t_i^{(2)} = \frac{-2\sum_{q=1,q\neq i}^n x_q - \sqrt{D_i(s)}}{2(n-1)}, \nonumber 
\end{align}
where, $D_i(s) = \left(2\sum_{q=1,q\neq i}^n x_q\right)^2 - 4(n-1)(2s x_i + s^2)$. To ensure $t_i \to 0$ as $s \to 0$, a requirement for establishing convergence properties, the adjustment $t_i$ is set to $t_i^{(1)}$. However, scenarios where $D_i(s) < 0$ may arise, making $t_i$ nonexistent for certain step-sizes. In practice, these cases are rare; when encountered, $s$ is reduced iteratively until $D_i(s) > 0$. If this fails, the update is skipped, and subsequent steps are attempted. After generating the candidate points (up to $2n$), function values are evaluated, and the best candidate is chosen. If no candidate improves the objective, the current solution is retained, and $s$ is reduced further (detailed as follows). 

Using the updated Fermi's principle for spherically constrained space, as outlined above, starting from an initial solution, for a given step-size $s$, we can generate up to $2n$ candidate points. Now, consider $B$ unit spheres, each with a length $n_b$ for $b = 1, \dots, B$. Applying the same principle, we generate $2 \sum_{b=1}^{B} n_b$ candidate solutions. The current objective function value is then compared with those evaluated at the candidate points, and the best value is selected as the updated solution.
\vspace{-0.4cm}
\subsubsection{MSCOR Overview}
MSCOR proceeds through multiple \emph{runs}, each consisting of iterations until a convergence criterion is met, as detailed later. Each \emph{run} begins with the solution from the previous one, except the first, which starts from a user-provided initialization. It starts with a large step-size (inspired by Fermi’s principle), promoting exploration, which gradually decreases over iterations to near zero, shifting the focus toward local refinement. This mirrors the `cooling down' mechanism in simulated annealing (SA). At the start of each new \emph{run}, the step-size is reset to encourage renewed exploration. This alternating strategy helps MSCOR escape local minima. The algorithm terminates when solutions from two successive \emph{runs} are sufficiently close, indicating that further exploration is unlikely to yield improvement.
\noindent \textbf{Tuning Parameters:}  
Each \emph{run} is governed by the following tuning parameters: initial global step-size $s_{\text{initial}} > 0$, step decay rate $\rho > 1$, step-size threshold $\phi > 0$, and sparsity threshold $\lambda \geq 0$. These parameters are set by the user and remain constant across runs. Two additional parameters, $\tau_1$ and $\tau_2$, control the convergence criteria. Additionally, the maximum number of iterations per \emph{run} and the maximum number of runs are denoted as \textit{MaxIter} and \textit{MaxRun}, respectively.

\noindent \textbf{Global and Local Step-Sizes:}   
Let the objective function $f$ needs to be minimized over the parameter space consisting of multiple unit spheres, denoted by $\bm{O} = \bm{O}_1 \times \cdots \times \bm{O}_B \in \mathbf{S}^{n_1-1} \times \cdots \times \mathbf{S}^{n_b-1}$, with the $b$-th block being a $(n_b - 1)$-dimensional unit sphere, and denoted by $\bm{O}_b = (o_{b,1}, \dots, o_{b,n_b}) \in S^{n_b-1}$, for $b = 1, \dots, B$. The total number of parameters is $M = \sum_{j=1}^B n_j$. Within each \emph{run}, we use a global step-size, denoted by $s^{(j-1)}$ at the beginning of $j$-th \emph{run},  and $2M$ local step-sizes $\{(s_{b,i}^+, s_{b,i}^-)\}_{i=1}^{n_b}\}_{b=1}^B$ (denoted by $s_h$ in Algorithm \ref{euclid}; different index used in Algorithm \ref{euclid} to highlight parallelization), which adapt based on the tuning parameters and improvements in the objective function. 

In the first iteration, the global step-size is initialized to $s_{\text{initial}}$. This global step-size,  remains constant throughout the iteration (but periodically updated across iterations throughout a \emph{run}). At the end of each iteration, its value either remains the same or is divided by $\rho$ ($\rho > 1$), depending on whether a `sufficiently' better solution was discovered during that iteration (as detailed later). At the start of each iteration, the local step-sizes $s_{b,i}^+$ and $s_{b,i}^-$ are initialized to the current global step-size.

\noindent \textbf{Exploratory movements:} 
At the beginning of the $h$-th iteration, the current value of the parameters is denoted by $\bm{O}^{(h)} = (\bm{O}_1^{(h)}, \dots, \bm{O}_B^{(h)})$, where each $\bm{O}_b^{(h)} = (o_{b,1}^{(h)}, \dots, o_{b,n_b}^{(h)}) \in S^{n_b-1}$ for $b = 1, \dots, B$. During the iteration, the objective function is evaluated at up to $2M$ feasible points in the neighborhood of $\bm{O}^{(h)}$. These points are derived by exploring candidate points around $\bm{O}^{(h)}$, modulated by the local step-sizes $\{(s_{b,i}^+, s_{b,i}^-)\}_{i=1}^{n_b}\}_{b=1}^B$. The feasible exploration directions are classified into $M$ `positive' movements $(b,i,+)$ and $M$ `negative' movements $(b,i,-)$. A coordinate of the unit-sphere is termed `significant' if its value exceeds a sparsity threshold $\lambda$ (detailed later), where $\lambda$ can be set to zero to avoid thresholding. For each $b$, the $b$-th unit-sphere $\bm{O}_b^{(h)}$ has $m_b$ significant locations, excluding the $i$-th location $o_{b,i}^{(h)}$. Except for these $m_b + 1$ locations (including $i$-th), all others are replaced with zeros. The movement $(b,i,+)$ involves updating $o_{b,i}^{(h)}$ by adding $s_{b,i}^{+}$ to it, and adjusting the `significant' locations with an `adjustment step-size', ensuring the updated point maintains a zero $l_2$ norm. If the updated value exceeds the unit-sphere boundary, or the adjustment step-size is invalid, the local step-size is reduced by a factor of $\rho$ (ensuring $s_{b,i}^{+} > \phi$) and the update is attempted again until the point remains within the unit-sphere. In rare cases where no feasible candidate is found, $\bm{O}_b^{(h)}(i,+)$, proposal candidate point corresponding to movement $(b,i,+)$, remains unchanged, same as $\bm{O}_b^{(h)}$. The $(b,i,-)$ movement follows a similar process by subtracting $s_{b,i}^{-}$ followed by `adjustment' of the significant locations accordingly. Finally, the best candidate point is chosen from $2M + 1$ candidate points, including $\bm{O}^{(h)}$.

\begin{algorithm}
\caption{MSCOR}\label{euclid}
\begin{algorithmic}[1]
{\fontsize{8}{1}
\Statex{\textbf{Input}: {\scriptsize Initial guess; \big($B$ blocks of $(n_b-1)$-dimensional unit spheres; $b = 1,\ldots, B$\big)}}
\Statex{\textbf{Output}: {\scriptsize $\widehat{\boldsymbol{U}}$; MSCOR optimized final solution \big($B$ blocks of $(n_b-1)$-dimensional unit spheres; $b = 1,\ldots, B$\big)}}
\State \textbf{Initialization:} $R \gets 1$ {\tiny \bigg($R$ = \textit{run} index\bigg)}
\BState \emph{top}:
\If {$R = 1$}
\State $\boldsymbol{U}^{(0)} \gets \text{Initial guess},\;j \gets 1$ {\tiny \bigg($\bm{U}^{(j)}$ denotes the value of $\bm{U}$ at the end of $j$-th iteration\bigg)}
\Else
\State $\boldsymbol{U}^{(0)} \gets \widehat{\bm{U}}^{(R-1)},\;j \gets 1$ {\tiny \bigg($\widehat{\bm{U}}^{(r)}$ denotes the value of $\bm{U}$ at the end of $r$-th run\bigg)}
\EndIf
\State $s^{(0)} \gets s_{initial}$ {\tiny \bigg(we take $s_{initial} = 1$; $s^{(j-1)}$ denotes the value of \textit{global step-size} at the beginning of $j$-th iteration\bigg)}
\While {($j \leq max\_iter$ and $s^{(j)} > \phi$)} 
\State $F_1 \gets f(\bm{U}^{(j-1)})$, $s \gets s^{(j-1)}$ {\tiny \bigg(note that, $\bm{U}^{(j-1)} = (\bm{u}_1^{(j-1)}, \ldots, \bm{u}_B^{(j-1)})$\bigg)}
\For {$b = 1:B$}
\For {$h = 1:2n_b$}
\State $i \gets [\frac{(h+1)}{2} ]$ {\tiny \bigg($[\cdot]$ denotes largest smaller integer function\bigg)}
\State $\boldsymbol{u}_{b,h} \gets \boldsymbol{u}_b^{(j-1)}${\tiny\bigg(note that, 
$\bm{u}_{b,h} = \big(\bm{u}_{b,h}(1), \ldots, \bm{u}_{b,h}(n_b)\big)$\bigg)} 
\State $ s_h \gets (-1)^hs$
\State $\Lambda \gets \text{which}(|\bm{u}_{b,h}(k)| < \lambda), k \in \{1,\ldots,n_b\} 
\setminus \{i\}$ {\tiny \bigg(i.e., indexes, except $i$, whose absolute values are $< \lambda$\bigg)}
\State $\Gamma \gets \text{which}(|\bm{u}_{b,h}(k)| \geq \lambda), k \in \{1,\ldots,n_b\}
\setminus \{i\}$ {\tiny \bigg(i.e., indexes, except $i$, whose absolute values are $\geq \lambda$\bigg)}
\State $D \gets \big(2*sum(\boldsymbol{u}_{b,h}(\Gamma)))^2 - 4*length(\Gamma)*(2s_h\boldsymbol{u}_{b,h}(i) + s_h^2 - sumsquare(\boldsymbol{u}_{b,h}(\Lambda)))$.
\While {($D < 0$ and $|s_h|>\phi$)}
\State $s_h \gets \frac{s_h}{\rho}$
\State $D \gets \big(2*sum(\boldsymbol{u}_{b,h}(\Gamma)))^2 - 4*length(\Gamma)*(2s_h\boldsymbol{u}_{b,h}(i) + s_h^2 - sumsquare(\boldsymbol{u}_{b,h}(\Lambda)))$.
\EndWhile
\If {($D\geq 0$)}
\State $t \gets \frac{-2*sum(\boldsymbol{u}_{b,h}(\Gamma)) + \sqrt{D}}{2*length(\Gamma)}$
\State $\boldsymbol{u}_{b,h}(i) \gets \boldsymbol{u}_{b,h}(i) + s_h$
\State $\boldsymbol{u}_{b,h}(\Gamma) \gets \boldsymbol{u}_{b,h}(\Gamma) + t$
\State $\boldsymbol{u}_{b,h}(\Lambda) \gets 0$
\State $f_{b,h} \gets f(\bm{u}_1^{(j-1)}, \ldots, \bm{u}_{b-1}^{(j-1)}, \bm{u}_{b,h}, \bm{u}_{b+1}^{(j-1)},\ldots, \bm{u}_{B}^{(j-1)})$
\Else 
\State $f_{b,h} \gets F_1$ {\tiny \bigg($\bm{u}_{b,h}$ remains unchanged, no update is made\bigg)}
\EndIf
\EndFor
\EndFor
\State $(b_{best},h_{best}) \gets \argmin_{b,h} f_{b,h}$ {\tiny over} $b = 1,\ldots, B$, $h = 1,\ldots,2n_b$
\State $\bm{u}_{temp} \gets \bm{u}_{b_{best},h_{best}}$
\State $F_2 \gets f_{b_{best},h_{best}}$
\State $\boldsymbol{U}^{(j)} \gets \boldsymbol{U}^{(j-1)}$
\If {($F_2 < F_1$)} $\bm{u}^{(j)}_{b_{best}} \gets \bm{u}_{temp}$ {\tiny \bigg(hence $\boldsymbol{U}^{(j)}$ becomes $\big(\bm{u}_1^{(j-1)}, \ldots, \bm{u}_{b_{best}-1}^{(j-1)}, \bm{u}_{temp}, \bm{u}_{b_{best}+1}^{(j-1)},\ldots, \bm{u}_{B}^{(j-1)}\big)$ \bigg)}
\EndIf
\If {($j > 1$)}
\If {($|F_1-min(F_1,F_2)| < \tau_1$ and $s>\phi$)} $s \gets \frac{s}{\rho}$
\EndIf
\EndIf
\State $s^{(j)} \gets s$, $j \gets j+1$
\EndWhile
\State $\widehat{\boldsymbol{U}}^{(R)} \gets \boldsymbol{U}^{(j)}$ 
\algstore{myalg}}
\end{algorithmic}
\end{algorithm}

\begin{algorithm}                     
\begin{algorithmic} [1]                   
{\algrestore{myalg}
\fontsize{8}{1}
\If {$||\widehat{\boldsymbol{U}}^{(R)} - \widehat{\boldsymbol{U}}^{(R-1)}|| < \tau_2$ }
\State \Return $\widehat{\boldsymbol{U}} = \widehat{\boldsymbol{U}}^{(R)}$ {\tiny \big(returning MSCOR optimized final solution $\widehat{\boldsymbol{U}}$\big)}
\State \textbf{break} {\tiny \big(exiting MSCOR\big)}
\Else
\State $R \gets R+1$
\State \textbf{go to} \emph{top}
\EndIf 
}
\end{algorithmic}
\end{algorithm}
\begin{figure}[]
	\centering
	\includegraphics[width=.9\textwidth]{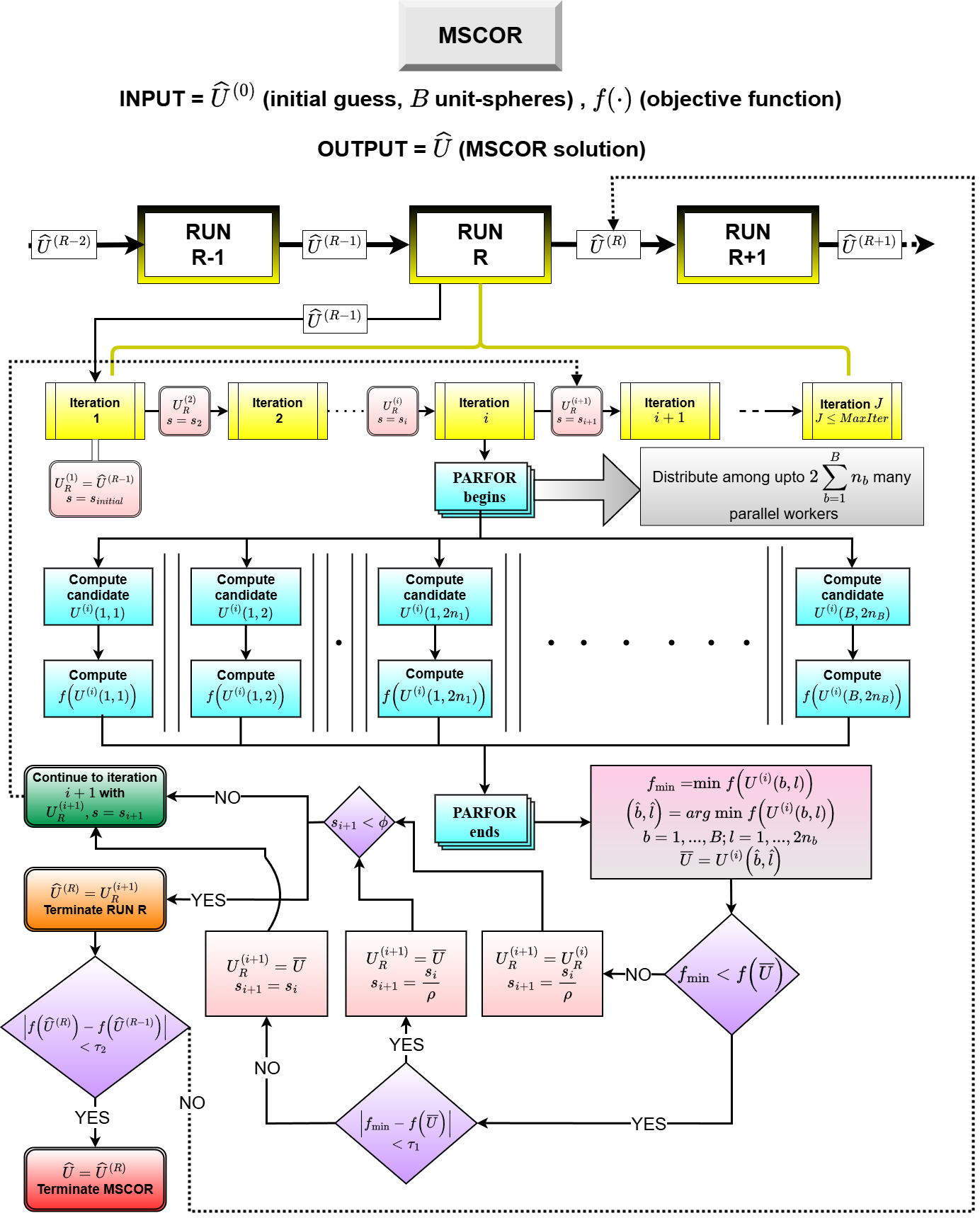}
	\caption{MSCOR flowchart.}
	\label{MSCOR_concept}
\end{figure}

\noindent \textbf{Sparsity control:}
We introduce a sparsity control step to promote sparse solutions. For each modified unit-sphere $\{\bm{O}_j^{(h)}(i,+)\}_{i=1}^{n_b}$ and $\{\bm{O}_j^{(h)}(i,-)\}_{i=1}^{n_b}$ for $b = 1,\dots,B$, we zero out the values of coordinates deemed ``insignificant'' (those less than $\lambda$). To preserve the $l_2$ constraint to be 1, the ``significant'' coordinates are updated by corresponding calculated ``adjustment step-size''. Note that the term `sparsity' is used here solely in the context of the nature of the final solution to a given objective function, and does not refer to the specific statistical modeling framework of SMART-MC.
\vspace{-0.3cm}
\begin{remark}
The parameter $\lambda$ should be set relatively large if prior knowledge suggests that the final solution is sparse; otherwise, it can be chosen to be smaller or set to zero.
\end{remark}
\noindent \textbf{Loop termination criteria:} 
At each iteration, the value of the global step-size either remains unchanged or is divided by $\rho$. If $|f(\bm{O}^{(h+1)}) - f(\bm{O}^{(h)})| < \tau_1$ at the end of iteration $(h+1)$, the global step-size get divided by $\rho$; otherwise, it remains unchanged. Once the global step-size becomes less than $\phi$, the \emph{run} terminates, forwarding the last obtained solution (denote it by $\mathbf{\hat{O}}^{(L)}$ for the $L$-th \emph{run}) to the next \emph{run} to serve as the starting point for that \emph{run}. MSCOR terminates when the solutions obtained by two consecutive \emph{runs}, say $R-1$ and $R$, satisfy $|f(\mathbf{\hat{O}}^{(R)}) - f(\mathbf{\hat{O}}^{(R-1)})| < \tau_2$. A flowchart of the MSCOR algorithm is shown in Figure \ref{MSCOR_concept}, and pseudo-code is provided in Algorithm \ref{euclid}.

\noindent \textbf{Parallelized MSCOR:} Close inspection of the MSCOR exploration strategy reveals that for any given step-size, the exploration and evaluation of the objective functions at the corresponding up to $2M$ candidate points are independent of each other, allowing these updates to be performed simultaneously within each iteration across $2M$ parallel threads, further alleviating the computational burden (as illustrated in Figure \ref{MSCOR_concept}). A comprehensive comparative analysis assessing the enhancement in computational speed achieved by parallelized MSCOR relative to MSCOR in maximizing the SMART-MC likelihood is presented in Table S5 of the supplementary material. 

Further details on MSCOR’s behavior in detecting non-convexity are provided in Section B of the supplementary material.
\vspace{-0.8cm}
\subsection{Theoretical property}
Here we establish the convergence property of MSCOR. Specifically, we show that the stopping criteria across all \emph{runs} ensure each solution is optimal under certain regularity conditions in Theorem \ref{MSCOR_theorem}. The proof of the theorem is detailed in Section B of the supplementary material. While this result does not strongly demonstrate MSCOR's global optimization capability, we validate it empirically through an extensive benchmark study in the following subsection.
\begin{definition}
\spacingset{1.3}
The `\textit{shadow}' of a point $\bm{W}$ (denoted by $\bm{W}^C$) belonging to the closure of $\bm{S}$ (i.e., $\bar{\bm{S}}$) is the point of intersection of the straight line connecting the origin to $\bm{W}$ with $\bm{S}$, where $\bm{S} = S^{n_1 - 1} \times \dots \times S^{n_B - 1}$.
\end{definition}
\vspace{-0.3cm}
\begin{theorem}
\spacingset{1.3}
\label{MSCOR_theorem}
Suppose $f:\bm{S} \mapsto \mathds{R}$ is convex, continuous and differentiable with extended definition on $\bar{\bm{S}}$, such that, $f(\bm{W})=f(\bm{W}^C)$ when $\bm{W} \in interior(\bm{S})$. Consider a sequence $\delta_{j,k} = \frac{s_j}{\rho^k}$ for $k\in \mathds{N}$ and $s_j>0, \rho>1$. Suppose $\mathbf{U} \in \bm{S}$ given by
\vspace{-0.4cm}
  \begin{align*}
      \mathbf{U} = (\mathbf{u}_1, \ldots, \mathbf{u}_B) \text{  where  } \mathbf{u}_b = (u_{j,1},\ldots, u_{j, n_b})\in S^{n_j - 1}, \; j = 1,\ldots, B.
  \end{align*}
Define, $\mathbf{u}_{j,k}^{(i+)} = (u_{j,1}+t_i(\delta_{j,k}),\ldots, u_{j,i-1}+t_i(\delta_{j,k}), u_{j,i}+\delta_{j,k},u_{j,i+1}+t_i(\delta_{j,k}), \ldots, u_{j,n_j} + t_i(\delta_{j,k}))$, $\mathbf{u}_{j,k}^{(i-)} = (u_{j,1}+t_i(-\delta_{j,k}),\ldots, u_{j,i-1}+t_i(-\delta_{j,k}), u_{j,i}-\delta_{j,k},u_{j,i+1}+t_i(-\delta_{j,k}), \ldots, u_{j,n_j} + t_i(-\delta_{j,k}))$ for $j = 1,\ldots,B$, $i=1,\cdots,n_j$, where $t_i(s)$ denotes the adjustment step-size corresponding to step-size $s$. Define $b_{j,i} = -\frac{u_{j,i}}{\big|\sum_{k=1,k\neq i}^{n_j} u_{j,k}\big|}$. If the following conditions hold true
\vspace{-0.4cm}
 \begin{enumerate}
 \item for all sufficiently large $k \in \mathrm{N}$, 
    $f(\mathbf{U}) \leq f(\mathbf{u}_1, \ldots, \mathbf{u}_{j-1}, \mathbf{u}_{j,k}^{(i+)}, \mathbf{u}_{j+1}, \ldots, \mathbf{u}_{B})$ and
    $f(\mathbf{U}) \leq f(\mathbf{u}_1, \ldots, \mathbf{u}_{j-1}, \mathbf{u}_{j,k}^{(i-)}, \mathbf{u}_{j+1}, \ldots, \mathbf{u}_{B})$
    \item $1-b_{j,i} \neq 0$
    \item $\big[(n_j-2) + \sum_{i=1}^{n_j-1}\frac{1}{1-b_{j,i}}\big] \neq 0$
\end{enumerate}
for $j = 1,\ldots,B$, $i=1,\ldots,n_j-1$, then a global minimum of $f$ over $\mathbf{S}$ occurs at $\mathbf{U}$. 
\end{theorem}
\spacingset{1.8}
\begin{table}[h]
\centering
\resizebox{0.99\columnwidth}{!}{%
\begin{tabular}{c|c|ccc|ccc|ccc}
\hline
\multirow{2}{*}{Functions} & \multirow{2}{*}{Algorithms} & \multicolumn{3}{c|}{$B = 5, n_b = 5$} & \multicolumn{3}{c|}{$B = 10, n_b = 20$} & \multicolumn{3}{c}{$B = 100, n_b = 5$} \\ \cline{3-11} 
 &  & min. value & se of solution & mean time (se) & min. value & se of solution & mean time (se) & min. value & se of solution & mean time (se) \\ \hline
\multirow{6}{*}{\begin{tabular}[c]{@{}c@{}}Ackley's\\ (modified)\end{tabular}} & MSCOR & \textbf{2.22e - 14} & 0.029 & 1.64 (0.008) & \textbf{3.61e - 13} & 0.000 & 312.58 (0.385) & \textbf{1.65e - 09} & 0.288 & 3600.04$^*$(0.006) \\
 & GA & 1.51e + 01 & 0.169 & 16.34 (0.769) & 2.59e + 01 & 0.080 & 78.81 (0.315) & 3.88e + 02 & 2.214 & 357.20 (3.682) \\
 & SA & 4.70e + 00 & 0.081 & 1.84 (0.092) & 2.16e + 01 & 0.039 & 53.48 (2.490) & 2.70e + 02 & 1.175 & 371.77 (23.565) \\
 & IP & \textbf{7.51e - 12} & 0.347 & 0.06 (0.003) & \textbf{3.67e - 03} & 0.023 & 0.09 (0.002) & 2.17e + 02 & 6.179 & 0.42 (0.026) \\
 & SQP & 9.50e - 04 & 0.414 & 0.03 (0.001) & 1.28e - 02 & 0.000 & 0.41 (0.001) & \textbf{8.33e + 01} & 9.219 & 5.15 (0.028) \\
 & AS & 2.35e + 00 & 0.328 & 0.03 (0.001) & 1.53e + 00 & 0.401 & 0.47 (0.003) & 1.56e + 02 & 4.564 & 5.60 (0.010) \\ \hline
\multirow{6}{*}{\begin{tabular}[c]{@{}c@{}}Griewank\\ (modified)\end{tabular}} & MSCOR & \textbf{\textless 1e - 16} & 0.000 & 1.54 (0.007) & \textbf{1.78e - 15} & 0.000 & 204.51 (0.444) & \textbf{1.46e - 09} & 0.000 & 3600.07$^*$(0.010) \\
 & GA & 8.04e - 01 & 0.040 & 19.59 (0.962) & 1.12e + 00 & 0.021 & 88.70 (0.287) & 3.60e + 01 & 0.400 & 461.57 (4.188) \\
 & SA & 1.06e - 01 & 0.008 & 2.03 (0.101) & 7.99e - 01 & 0.004 & 54.12 (2.392) & 2.72e + 01 & 0.166 & 372.25 (11.450) \\
 & IP & 2.47e - 13 & 0.000 & 0.02 (0.002) & 6.53e - 04 & 0.000 & 0.10 (0.002) & 2.03e + 00 & 0.175 & 0.50 (0.025) \\
 & SQP & \textbf{1.98e - 13} & 0.000 & 0.01 (0.000) & \textbf{5.96e - 12} & 0.000 & 0.24 (0.001) & \textbf{3.80e - 12} & 0.000 & 1.69 (0.015) \\
 & AS & 3.50e - 08 & 0.022 & 0.03 (0.002) & 2.77e - 07 & 0.005 & 0.43 (0.015) & 4.54e - 07 & 0.464 & 5.79 (0.722) \\ \hline
\multirow{6}{*}{\begin{tabular}[c]{@{}c@{}}Neg. sum\\ of squares\\ (modified)\end{tabular}} & MSCOR & \textbf{\textless 1e - 16} & 0.000 & 0.45 (0.005) & \textbf{\textless 1e - 16} & 0.000 & 43.81 (0.413) & \textbf{1.51e - 14} & 0.000 & 1602.09 (15.515) \\
 & GA & 5.17e + 00 & 0.198 & 16.47 (0.805) & 8.27e + 01 & 0.648 & 74.74 (0.258) & 1.89e + 02 & 2.398 & 325.61 (2.558) \\
 & SA & 2.19e + 00 & 0.044 & 1.85 (0.087) & 7.10e + 01 & 0.126 & 50.59 (2.549) & 1.65e + 02 & 0.435 & 358.06 (16.27) \\
 & IP & \textbf{7.99e - 15} & 0.000 & 0.02 (0.000) & 1.26e + 00 & 0.100 & 0.09 (0.002) & 3.83e + 00 & 1.520 & 0.40 (0.023) \\
 & SQP & 1.07e - 14 & 0.000 & 0.02 (0.000) & \textbf{4.26e - 07} & 0.000 & 0.41 (0.002) & \textbf{9.09e - 12} & 0.000 & 3.78 (0.102) \\
 & AS & 1.92e - 09 & 0.093 & 0.02 (0.001) & 1.60e + 01 & 0.714 & 0.45 (0.003) & 2.42e + 01 & 3.595 & 5.53 (0.093) \\ \hline
\multirow{6}{*}{\begin{tabular}[c]{@{}c@{}}Rastrigin\\ (modified)\end{tabular}} & MSCOR & \textbf{\textless 1e - 16} & 0.762 & 2.08 (0.417) & \textbf{8.53e - 13} & 0.000 & 135.99 (0.255) & \textbf{1.02e + 02} & 5.544 & 3600.04$^*$(0.011) \\
 & GA & 9.90e + 01 & 5.792 & 18.21 (0.835) & 1.59e + 03 & 9.215 & 79.37 (0.262) & 4.98e + 03 & 73.999 & 412.85 (51.696) \\
 & SA & 8.64e + 00 & 0.302 & 1.76 (0.082) & 3.47e + 01 & 2.006 & 93.74 (3.322) & 4.72e + 02 & 10.532 & 935.30 (60.66) \\
 & IP & 6.72e + 00 & 0.725 & 0.04 (0.001) & \textbf{1.68e - 04} & 5.922 & 0.10 (0.001) & 5.14e + 02 & 111.633 & 0.41 (0.010) \\
 & SQP & 8.18e + 00 & 0.637 & 0.03 (0.000) & 7.04e + 00 & 3.435 & 0.42 (0.002) & \textbf{4.71e + 02} & 10.107 & 5.32 (0.075) \\
 & AS & \textbf{1.20e + 00} & 0.969 & 0.03 (0.000) & 2.19e + 02 & 21.095 & 0.46 (0.001) & 8.49e + 02 & 104.721 & 5.78 (0.058) \\ \hline
\end{tabular}}
\caption{A comparative study of MSCOR, GA, SA, IP, SQP and AS methods for optimizing modified Ackley, Griewank, negative sum of squares, and Rastrigin functions is presented for cases with parameter settings $(B,n_b) = (5,5), (10,20), (100,5)$. Each experiment is repeated 100 times. S.e. denotes the standard error. Time is measured in seconds. For the scenarios where MSCOR's average computation time exceeds upper bound 3600 seconds, are labeled with $^*$. See Table S1 in the supplementary material for the median and maximum execution time summaries.}
\label{MSCOR_benchmark}
\end{table}
\vspace{-0.7cm}
\subsection{Benchmark Study of Global Optimization}
To evaluate the performance of MSCOR, we consider the minimization of four benchmark functions: Rastrigin, Ackley, Sphere, and Griewank \citep{Jamil2013}, with parameter spaces modified as collections of unit spheres (see Section B of the supplementary material). MSCOR is implemented in MATLAB and executed on a Windows 10 Enterprise system with 32 GB RAM and a 12th Gen Intel(R) Core(TM) i7-12700 processor (12 cores, 20 logical processors, 2100 MHz). MSCOR is compared with GA, SA, SQP, IP, and Active-set (AS); where GA and SA are global optimizers, and SQP, IP, and AS are convex optimizers. MATLAB's built-in functions \texttt{ga}, \texttt{simulannealbnd}, and \texttt{fmincon} are used for implementation. We consider scenarios $(B, n_b) = (5, 5), (10, 20), (100, 5)$, with the last scenario reflecting the dimensionality of the later case study. Each experiment is repeated 100 times with random initializations. Results are summarized in Table \ref{MSCOR_benchmark}. MSCOR consistently outperforms all competitors, yielding superior solutions within reasonable time frames. For Ackley's and Griewank functions, MSCOR terminated at the 1-hour upper bound but still produced better solutions than most competitors. While parallel MSCOR could further reduce computation time, it was not used to ensure fair comparison since not all competitor algorithms are parallelizable.

\vspace{-0.6cm}
\section{SMART-MC Analysis of Dynamic MS DMTs}
\label{MS_analysis}
\vspace{-0.3cm}
We applied SMART-MC to investigate how clinical and demographic factors shape MS treatment transitions across real-world DMT pathways. As outlined in Section~\ref{dataset}, we sought to evaluate the influence of disease duration, age, sex, and race on treatment sequencing, while identifying the most common transitions and those most sensitive to patient characteristics. Age and disease duration are re-centered and re-scaled, as detailed in Section D of the supplementary material. Race is encoded using two indicator variables for the White and Black populations, with individuals categorized as Other serving as the reference group. In order to ensure stability of the estimates and to restrict rare treatment transitions from unduly influencing the overall transition dynamics, we consider a conservative threshold of \textit{Tol} equal to $5(p+1)$, where the number of covariates (excluding the intercept) is $p = 5$ in our case. Standard errors and $p$-values for covariate effects were estimated via bootstrap using 1000 replicates (see Table S6 and S7 of the supplementary material). The full exploratory analysis answering all research questions in detail is presented in Section D of the supplementary material; here, we summarize the key highlights.

\begin{figure}[!ht]
 \centering
     \includegraphics[width=.95\textwidth]{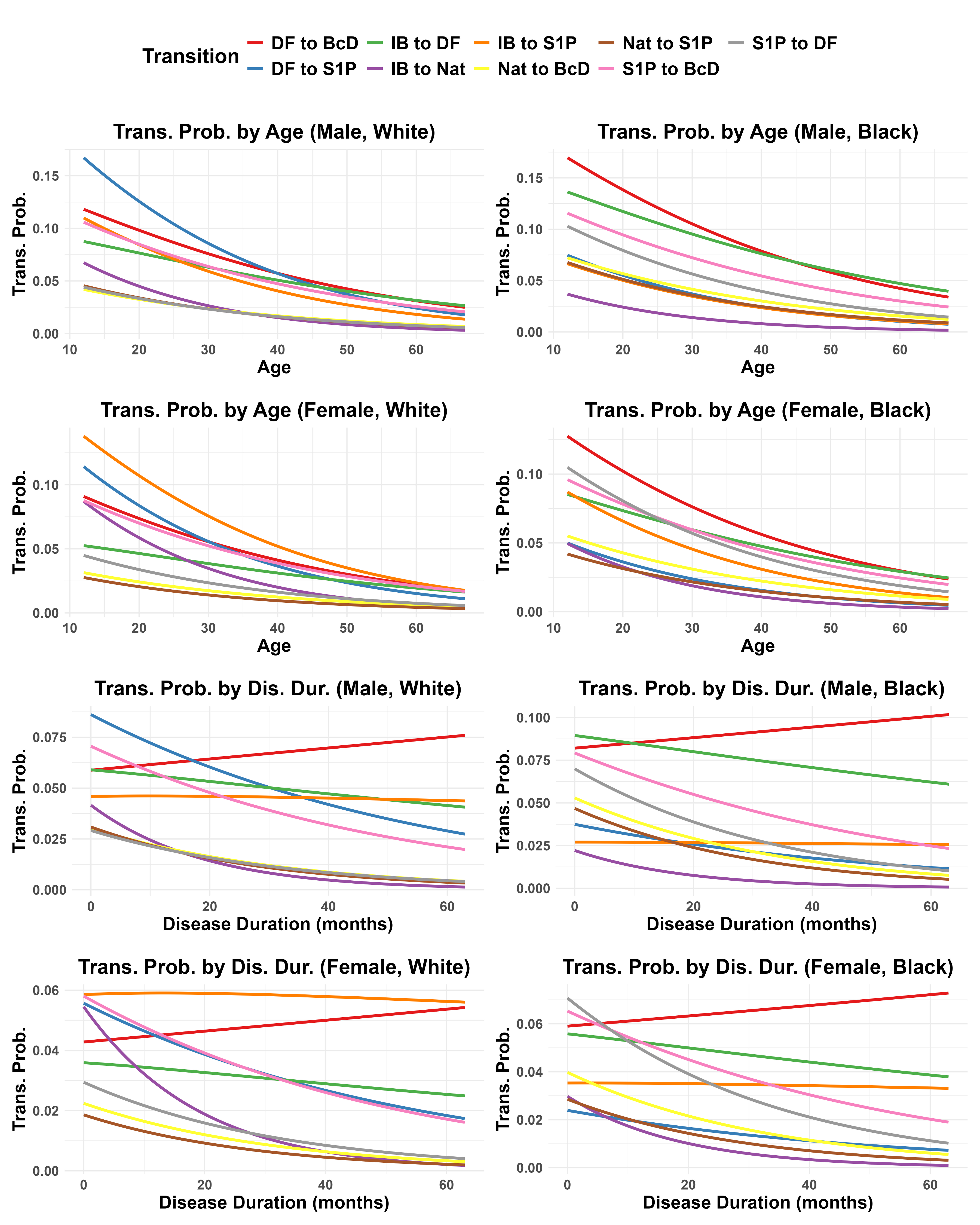}
      \caption{Estimated transition probabilities for non-rare across-DMT transitions across age and disease duration, stratified by key patient subgroups, as derived from the SMART-MC model.} 
      \label{fig:patient_transitions} 
 \end{figure}

\begin{figure}[!ht]
 \centering
     \includegraphics[width=.9\textwidth]{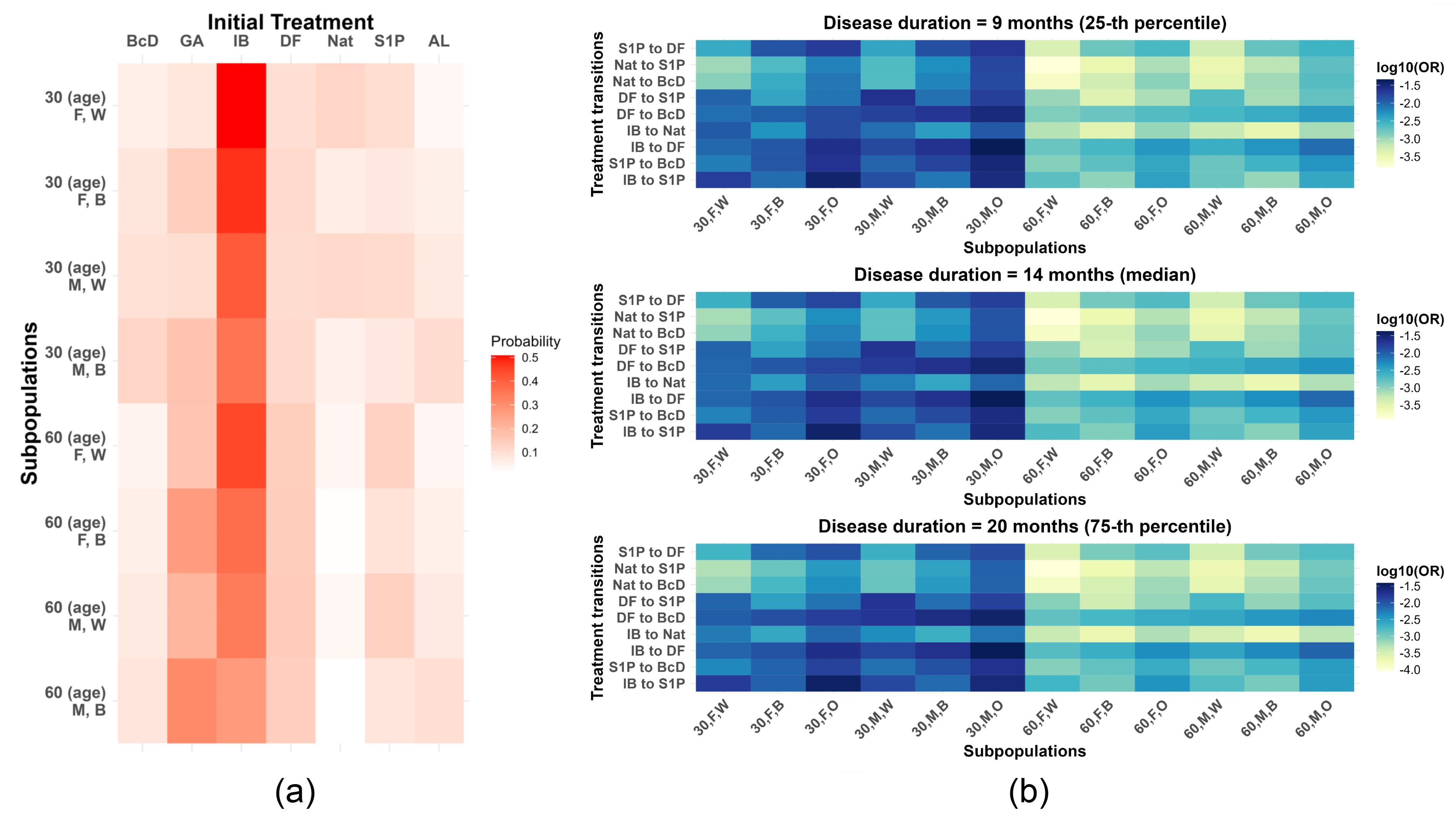}
     \vspace{0.0cm}
      \caption{(a) Estimated initial treatment probabilities across subpopulations defined by age (30 or 60 years), sex (M = Male, F = Female), and race (W = White, B = Black). (b) Odds ratios (OR) of transitioning to a different treatment versus continuing the same treatment for the top 7 most frequent MS DMT transitions (to a different one). The x-axis represents patients' phenotypes: age, sex, and race, where `O' denotes race category `Others'. Plots are shown for all three quartiles (0.25, 0.5, 0.75) of disease duration.} 
      \label{fig:InitialProb_odds_ratio} 
 \end{figure}

Longer disease duration was significantly associated with persistence on injectable therapies such as IB ($p = 0.001$) and with transitions from IB to fumarates ($p = 0.007$), but negatively associated with escalation from DF to S1P ($p = 0.016$), consistent with the clinical intuition that patients with longer disease history may stabilize on platform therapies or be less frequently escalated. Figure~\ref{fig:patient_transitions} illustrates how transition probabilities vary across patient subgroups defined by age, sex, and race, using SMART-MC fitted estimates. Each panel displays transition probabilities for a specific sex--race group as a function of either age (top two rows) or disease duration in months (bottom two rows). The aforementioned trend is visible in Figure~\ref{fig:patient_transitions}, where transition probabilities for DF to S1P and S1P to DF decline with disease duration, while IB to S1P exhibits a more stable or increasing trend. Furthermore, to demonstrate the odds ratios of non-rare across-DMT transitions relative to remaining on the same treatment, the trained model estimates such odds ratios across representative ages (30 and 60), sex (M/F), and race (W/B), calculated at all three quartiles of disease duration, as shown in Figure~\ref{fig:InitialProb_odds_ratio}(b). It is observed that, as disease duration increases from 9 to 20 months, the odds of transitioning from IB to S1P and from DF to BcD remain comparatively elevated across most subgroups, in contrast to other transitions.

Age at diagnosis also influenced transition dynamics: older patients were more likely to persist on existing treatments such as DF ($p < 0.001$) and Nat ($p < 0.001$), but less likely to escalate from IB to Nat ($p < 0.001$) or S1P ($p = 0.009$). These effects reflect a clinical preference for minimizing aggressive treatment changes in older individuals, which coincides with findings in \cite{Balusha2024}. Sex and race also played substantial roles in treatment persistence and escalation patterns. Female patients were more likely to stay on DF ($p = 0.001$) and Nat ($p = 0.048$), but less likely to transition from IB to DF ($p < 0.001$), Nat to S1P ($p = 0.035$), or S1P to BcD ($p = 0.028$), suggesting sex-related differences in treatment tolerance or access. Racial disparities emerged most notably in persistence on and transitions from high-efficacy agents. Black patients were more likely to remain on IB ($p = 0.009$) and Nat ($p = 0.019$), but less likely to transition to S1P from IB ($p = 0.013$), DF ($p = 0.002$), or Nat ($p = 0.004$), highlighting potential differences in care patterns or drug response. Among White patients, we observed higher persistence on S1P ($p < 0.001$), IB ($p = 0.008$), and Nat ($p < 0.001$), but reduced transitions from DF to BcD ($p = 0.018$), Nat to S1P ($p < 0.001$), and Nat to BcD ($p < 0.001$). A visual depiction of these trends is apparent in Figure~\ref{fig:patient_transitions}. Figure~\ref{fig:InitialProb_odds_ratio}(b) further underscores that younger patients exhibit a greater tendency to transition to a different treatment, a pattern more pronounced in non-Black and non-White populations.

We further identified the most common transition pathways (see Table S6 of the supplementary material for details). The top five transitions: Nat to Nat (24.8\%), IB to IB (19.5\%), S1P to S1P (16.6\%), BcD to BcD (11.9\%), and DF to DF (10.8\%) reflect a strong tendency toward treatment persistence. Among these, Nat to Nat was influenced by age ($p < 0.001$), sex ($p = 0.048$), and race (White: $p < 0.001$; Black: $p = 0.019$), while DF to DF was significantly associated with age ($p < 0.001$) and sex ($p = 0.001$), indicating greater persistence among older and female patients. IB to IB persistence was linked to disease duration ($p = 0.001$) and race (White: $p = 0.008$, Black: $p = 0.009$), and S1P to S1P was associated with White race ($p < 0.001$), suggesting strong demographic effects even among those remaining on the same DMT. Across-treatment transitions such as IB to S1P (1.33\%), IB to DF (0.88\%), DF to S1P (0.53\%), DF to BcD (0.66\%), and Nat to S1P (0.41\%) were relatively infrequent but covariate-sensitive. For instance, IB to S1P was more common among younger, non-Black patients with longer disease duration (age: $p = 0.009$; duration: $p = 0.012$; Black: $p = 0.013$), while DF to BcD was less likely among White patients ($p = 0.018$). The odds ratios in Figure~\ref{fig:InitialProb_odds_ratio}(b) reinforce these trends, showing how patient subgroups differ in their likelihood of switching versus persisting on therapies. Lastly, Figure~\ref{fig:InitialProb_odds_ratio}(a) shows the estimated initial treatment probabilities across subpopulations defined by age (30 and 60), sex (M/F), and race (White/Black). IB emerges as the most common initial therapy across all groups, particularly among younger female patients, with decreasing probability of IB initiation as age increases. GA also shows moderate initial uptake, while high-efficacy therapies such as S1P and AL are rarely used as first-line options, regardless of subgroup.

Together, these findings illustrate how SMART-MC enables granular, covariate-informed inference on MS treatment dynamics. By uncovering significant patterns in treatment persistence and transitions across demographic and clinical subgroups, the model directly addresses the research questions posed in Section~\ref{dataset}. These insights support the broader goals of individualized care and precision treatment strategies in real-world MS management.

\vspace{-\baselineskip}
\section{Conclusion}
\label{conclusion}
\vspace{-0.3cm}
In this article, we propose SMART-MC, a novel Markov model to study how patient covariates influence the likelihood of treatment transitions among MS patients. This approach enables us revealing the nature of the association between covariates and transition probabilities, in terms of both direction and magnitude. SMART-MC also promptly addresses the issue with rare transitions, ultimately proposing a framework that not only avoids the extra computational burden of imposing sparsity but also uses such occurrences to its advantage by alleviating the burden to some extent, through avoiding estimating them as a function of covariates. 
In order to handle the multi-modal likelihood arising in SMART-MC, we propose a Pattern Search-based global optimization technique, named MSCOR. Some of the attractive key features of MSCOR are noted as follows: (1) ability to escape local solutions, (2) parallelization using a number of threads linearly increasing with the dimension of the parameter space, (3) sparsity control, (4) automatic early termination capability while optimizing convex functions without prior knowledge, (5) non-convex detection. Further, MSCOR does not require the objective function to be differentiable; or even continuous,
which makes MSCOR very powerful and versatile Black-box optimization tool on multiple spherically constrained spaces, being extensively relevant across all domains, far beyond its limiting role in this considered case-study. Performing SMART-MC analysis of MS DMT sequence data from an EHR cohort at the Massachusetts General and Brigham Hospital system, we discovered key insights regarding how patient phenotypes, such as age at diagnosis, disease duration, sex, and race, inform the likelihood of persistence with certain DMTs across diverse patient cohorts.

To address the sticky behavior of observed DMT sequences, which typically exhibit infrequent changes, we aggregate treatments into 3-month intervals. This aggregation helps reveal long-term treatment patterns by smoothing short-term fluctuations. Nonetheless, future work could explore incorporating the full temporal resolution of the data to better capture rapid transitions. Another promising direction is to improve the robustness of the estimation procedure. While the current framework mitigates the influence of rare transitions by applying a conservative threshold \textit{Tol}, it may also be valuable to develop methods for handling outlier patients who experience an unusually high number of DMT transitions. In addition, the model can be extended to accommodate sparse covariate effects, enabling the integration of high-dimensional data such as biomarkers or neuroimaging, and ultimately advancing understanding of personalized treatment strategies.
\vspace{-0.3\baselineskip}
\begin{center}
{\large\bf SUPPLEMENTARY MATERIAL}
\end{center}
\vspace{-0.2\baselineskip}
\begin{description}
\vspace{-0.7\baselineskip}
\item[Supplementary text:] Supplementary material is provided as a separate pdf document.
\vspace{-0.5\baselineskip}
\if1\blind
{
\item[Code and data:] Code for SMART-MC and MSCOR, including demos to fit them to any similarly structured dataset, are made available on GitHub at\\ \href{https://github.com/priyamdas2/SMART-MC-MSCOR}{https://github.com/priyamdas2/SMART-MC-MSCOR}.
}
\fi
\if0\blind
{
\item[Code and data:] Code for SMART-MC and MSCOR, including demos to fit them to any similarly structured dataset, will be made available on GitHub.
}
\fi

\end{description}
\vspace{-1.5\baselineskip}
\if1\blind
{\section*{Disclosure statement}
\vspace{-0.4\baselineskip}
The authors report there are no competing interests to declare.
}
\fi
\bibliographystyle{agsm}
\bibliography{Bibliography-MM-MC}
\end{document}


\def\spacingset#1{\renewcommand{\baselinestretch}%
{#1}\small\normalsize} \spacingset{1}


\if1\blind
{
  \title{\bf Supporting Information for ``SMART-MC: Characterizing the Dynamics of Multiple Sclerosis Therapy Transitions Using a Covariate-Based Markov Model''}
  \author{Beomchang Kim 
  \hspace{.2cm}\\
    {\normalsize \medskip Department of Biostatistics, Virginia Commonwealth University}\\
    Zongqi Xia 
    \hspace{.2cm}\\
    {\normalsize \medskip Department of Neurology, Department of Biomedical Informatics,}\\
    {\normalsize \medskip University of Pittsburgh}\\
    and \\
    Priyam Das
    \hspace{.2cm}\\
 {\normalsize \medskip Department of Biostatistics, Virginia Commonwealth University}\\
 {\normalsize \medskip Department of Biomedical Informatics, Harvard Medical School}}
  \maketitle
} \fi

\if0\blind
{
  \bigskip
  \bigskip
  \bigskip
  \begin{center}
    {\large\bf Supporting Information for ``SMART-MC: Characterizing the Dynamics of Multiple Sclerosis Therapy Transitions Using a Covariate-Based Markov Model''}
\end{center}
  \medskip
} \fi

\bigskip
\tableofcontents

\newpage
\spacingset{1.45} 

\section{SMART-MC}

\subsection{Justifications of Assumptions for Theorem 1 \& 2 (main draft)}

We begin by restating and justifying the assumptions underlying Theorem 1 in the main draft, which establishes consistency of the SMART-MC estimator. These assumptions are standard in likelihood-based inference and are tailored to the structure of our model.

\vspace{0.2cm}
\noindent \textbf{(A1) Independent and Identically Distributed Data:}  
\begin{quote}
The data $\{(Y_k, X_k)\}_{k=1}^K$ are independent and identically distributed (i.i.d.).
\end{quote}

\noindent \textit{Justification:}  Each $(Y_k, X_k)$ pair represents an independent patient, where $Y_k$ is the observed treatment sequence and $X_k$ is the associated covariate vector. In longitudinal observational studies using EHR data, it is common and reasonable to assume that different patients are sampled independently from the population. Although the treatment sequence lengths $t_k$ may vary across patients, this does not violate the i.i.d. assumption because the treatment transitions and covariates are drawn from a common distribution. Identical distribution refers to the underlying generative process, not to fixed sequence lengths or the number of transitions per subject. This is consistent with standard practice in longitudinal and categorical data modeling.

\vspace{0.3cm}
\noindent \textbf{(A2) Correct Model Specification:}  
\begin{quote}
The true transition probabilities $m^{(k)}_{u,v}$ satisfy the hybrid model form described in Equation~(2) of the main manuscript:
\[
m^{(k)}_{u,v} =
\begin{cases}
m^*_{u,v}, & \text{if } v \in \mathcal{V}_u^c, \\
\left( 1 - \sum_{v' \in \mathcal{V}_u^c} m^*_{u,v'} \right)
\cdot \dfrac{ \exp(X_k^\top \beta^*_{u,v}) }{ \sum_{v' \in \mathcal{V}_u} \exp(X_k^\top \beta^*_{u,v'}) },
& \text{if } v \in \mathcal{V}_u,
\end{cases}
\]
where $\|\beta_{u,v}^*\|_2 = 1$ and $m^*_{u,v} \in (0,1)$.
\end{quote}

\noindent \textit{Justification:}  
The model assumes that rare transitions are captured by empirical probabilities, while more frequent transitions follow a multinomial logistic form conditioned on covariates. The constraint $\|\beta_{u,v}^*\|_2 = 1$ resolves the identifiability issue without enforcing arbitrary reference categories, which could be problematic in sparse data settings. This approach is standard in single-index and semiparametric modeling (e.g., \citealp{Carroll1997, Das2017}), ensuring parsimonious and identifiable parameterization.

\vspace{0.3cm}
\noindent \textbf{(A3) Bounded Covariates:}  
\begin{quote}
There exists a constant $C > 0$ such that $\|X_k\| \leq C$ for all $k$.
\end{quote}

\noindent \textit{Justification:}  
In real-world clinical datasets, patient-level covariates such as age, sex, race, and other biometrics are bounded by nature. Even continuous variables like age or disease duration are measured within physiologically plausible ranges. This assumption is essential for technical reasons—it ensures the continuity and compactness of the likelihood function over a bounded domain, which is critical in consistency and asymptotic analysis.

\vspace{0.3cm}
\noindent \textbf{(A4) Full Rank Fisher Information and Non-Collinearity of Covariates:}  
\begin{quote}
For each $(u,v) \in \mathcal{V}_u$, the Fisher information matrix is full rank, and the covariates are not perfectly collinear.
\end{quote}

\noindent \textit{Justification:}  To justify this assumption under our hybrid estimation setup, we explicitly consider the Fisher information for $\beta_{u,v}$ under the truncated multinomial likelihood restricted to $\mathcal{V}_u$. Let $Z_k$ be the indicator of the observed transition destination in $\mathcal{V}_u$ for a patient with covariates $X_k$. To characterize local identifiability and curvature of the log-likelihood, we consider the Fisher information matrix under the unconstrained multinomial logistic parameterization. Although our model constrains $\|\beta_{u,v}\|_2 = 1$, which restricts the parameter space to a product of spheres, identifiability and consistency are preserved under compactness and smoothness of the likelihood (e.g., \citealp{Carroll1997}). For each origin state $u$, let $n_u$ denote the total number of observed transitions from treatment $u$ to any destination $v \in \mathcal{V}_u$, pooled across all patients. Let $\{(X_i, Z_i)\}_{i=1}^{n_u}$ denote the covariate vector and destination state for each such transition, where $Z_i \in \mathcal{V}_u$ and $X_i$ is the corresponding patient-level covariate vector (which may be repeated across transitions for the same patient). Then the conditional log-likelihood for transitions from $u$ is given by:
\[
\ell_u(\beta_u) = \sum_{i=1}^{n_u} \log \left( \frac{\exp(X_i^\top \beta_{u,Z_i})}{\sum_{v \in \mathcal{V}_u} \exp(X_i^\top \beta_{u,v})} \right),
\]
where $\beta_u = \{\beta_{u,v}: v \in \mathcal{V}_u\}$.
The Fisher information matrix in the unconstrained setting is then:
\[
\mathcal{I}(\beta_u) = \mathbb{E}_X \left[ \sum_{v \in \mathcal{V}_u} \pi_v(X) \left( X X^\top - \sum_{v' \in \mathcal{V}_u} \pi_{v'}(X) X X^\top \right) \right],
\]
where $\pi_v(X) = \mathbb{P}(Z = v \mid X; \beta_u)$. This matrix can be simplified as:
\[
\mathcal{I}(\beta_u) = \mathbb{E}_X \left[ \sum_{v \in \mathcal{V}_u} \pi_v(X)(1 - \pi_v(X)) X X^\top \right],
\]
which is a positive semi-definite matrix reflecting the weighted covariance of the covariates. To ensure that this matrix is full rank (positive definite), it suffices that:

\begin{itemize}
    \item[(i)] The covariate vectors $X_k \in \mathbb{R}^p$ (rows of the design matrix) span $\mathbb{R}^p$, implying the matrix has full column rank;
    \item[(ii)] For each $X_k$, the transition probabilities $\pi_v(X_k)$ are strictly between 0 and 1 for at least two $v \in \mathcal{V}_u$, ensuring outcome variability.
\end{itemize}

These two conditions are met in our modeling framework, as detailed below.

\begin{itemize}
    \item[(i)] \textbf{Covariate design matrix has full column rank:}  
    Assumption (A3) ensures that each patient-level covariate vector $X_k \in \mathbb{R}^p$ is bounded. In practice, covariates are centered or standardized, and checked for multicollinearity using diagnostics such as variance inflation factors (VIFs), ensuring that the empirical design matrix constructed from $\{X_k\}_{k=1}^K$ has full column rank. In the SMART-MC model, a single patient may contribute multiple transitions from the same origin state $u$, resulting in repeated covariate vectors in the design matrix. While this introduces intra-patient dependence and violates strict row-wise independence, it does not affect identifiability or consistency. Patients are sampled independently, and as $K \to \infty$, the number of unique covariate vectors increases. Thus, the transition-level log-likelihood may be viewed as a pseudo-likelihood whose curvature is still informative as long as the collection of patient covariates spans $\mathbb{R}^p$. The rare transition threshold $\text{Tol} > p+1$ further ensures that covariate-dependent modeling for any $\mathcal{V}_u$ involves a sufficient number of distinct patients to avoid degeneracy. Consequently, the design matrix remains asymptotically full rank.

    \item[(ii)] \textbf{Outcome variability across destination states:}  
    For each origin state $u$, the SMART-MC model restricts covariate-dependent modeling to transitions in $\mathcal{V}_u$—those observed at least $\text{Tol}$ times across the dataset. We set $\text{Tol} > p + 1$ to ensure that for every such $u$, the number of transitions in each destination $v \in \mathcal{V}_u$ exceeds the number of covariates, and typically involve multiple patients. This guards against degenerate response vectors and ensures that for any given $X_k$, the multinomial transition probabilities $\pi_v(X_k)$ are non-extreme (i.e., strictly between 0 and 1 for at least two $v$). Consequently, the log-likelihood surface has meaningful curvature in the direction of $\beta_u$.
\end{itemize}
 
Thus the above arguments jointly imply that the Fisher information matrix $\mathcal{I}(\beta_u)$ is positive definite. For any nonzero $a \in \mathbb{R}^p$,
\[
a^\top \mathcal{I}(\beta_u) a = \mathbb{E}_X \left[ \sum_{v \in \mathcal{V}_u} \pi_v(X)(1 - \pi_v(X)) (a^\top X)^2 \right] > 0,
\]
unless $a^\top X = 0$ almost surely, which contradicts the full support of $X$. This establishes that under the two stated conditions, the Fisher information matrix is full rank, and hence Assumption (A4) is justified.

\clearpage

\subsection{Proof of Theorem~1 (Consistency of SMART-MC Estimator)}
\begin{theorem}\label{thm:consistency}
Under assumptions \textnormal{(A1)}–\textnormal{(A4)}, as $K \to \infty$, the maximum likelihood estimator of SMART-MC satisfies:
\begin{itemize}
    \item [(a)] For all rare transitions $v \in \mathcal{V}_u^c$, $\hat{m}_{u,v} \xrightarrow{P} m_{u,v}^*$.
    \item [(b)] For all non-rare transitions $v \in \mathcal{V}_u$, $\hat{\beta}_{u,v} \xrightarrow{P} \beta_{u,v}^*$.
\end{itemize}
\end{theorem}

\begin{proof}[Proof of Theorem~1]
We divide the proof into two parts corresponding to the parameter types: empirical transition probabilities for rare transitions, and regression coefficients for non-rare transitions.

\noindent \underline{Rare transitions scenario}: 
For $v \in \mathcal{V}_u^c$, SMART-MC estimates $m_{u,v}$ using the empirical frequency:
\[
\hat{m}_{u,v} = \frac{c_{u,v}}{\sum_{v'=1}^N c_{u,v'}},
\]
where $c_{u,v}$ denotes the total number of transitions from $u$ to $v$ across all patients. Let $\pi_{u,v}$ denote the true marginal transition probability from $u$ to $v$ in the population.

By the i.i.d. assumption (A1) and the law of large numbers (LLN), we have:
\[
\frac{1}{K} c_{u,v} \xrightarrow{a.s.} \pi_{u,v}, \quad \text{and} \quad
\frac{1}{K} \sum_{v'=1}^N c_{u,v'} \xrightarrow{a.s.} \sum_{v'=1}^N \pi_{u,v'}.
\]
Hence, by the continuous mapping theorem,
\[
\hat{m}_{u,v} \xrightarrow{P} \frac{\pi_{u,v}}{\sum_{v'=1}^N \pi_{u,v'}} = m_{u,v}^*,
\]
where the final equality follows from the structure of the SMART-MC model (Assumption A2). This establishes consistency for rare transition probabilities.\\

\noindent \underline{Non-rare transitions scenario}: For each origin state $u$, we consider transitions into $\mathcal{V}_u$, the subset of destination states with sufficient sample size (at least $\text{Tol} > p + 1$ transitions). For $v \in \mathcal{V}_u$, we estimate $\beta_{u,v}$ by maximizing the partial (pseudo-)likelihood over non-rare transitions.

Let $\{(X_i, Z_i)\}_{i=1}^{n_u}$ denote the sequence of transitions from $u$ into $\mathcal{V}_u$, where $X_i$ is the covariate vector associated with the transition and $Z_i \in \mathcal{V}_u$ is the destination state. Then the log-pseudo-likelihood for origin state $u$ is:
\[
\ell_u(\beta_u) = \sum_{i=1}^{n_u} \log \left( \frac{\exp(X_i^\top \beta_{u,Z_i})}{\sum_{v \in \mathcal{V}_u} \exp(X_i^\top \beta_{u,v})} \right),
\]
where $\beta_u = \{\beta_{u,v} : v \in \mathcal{V}_u\}$ is subject to the identifiability constraint $\| \beta_{u,v} \|_2 = 1$. Although the design matrix may include repeated covariates due to multiple transitions (from same treatment $u$) per patient, the pseudo-likelihood remains consistent as the number of patients $K \to \infty$ and patient-level sampling is i.i.d. This ensures that the law of large numbers applies at the population level, ensuring pseudo-likelihood consistency despite intra-subject dependence (see justification for assumption (A4) earlier for details). We now verify that standard conditions for consistency of the maximum (pseudo-)likelihood estimator are satisfied:

\begin{itemize}
    \item \textit{Identifiability:}  
    Assumption (A4) guarantees that the Fisher information matrix $\mathcal{I}(\beta_u)$ is positive definite. Under the multinomial logit structure, this implies identifiability of $\beta_u$ up to the norm constraint. Identifiability is preserved under the constraint $\| \beta_{u,v} \|_2 = 1$ by standard results in single-index models (see \citealp{Carroll1997}).
    
    \item \textit{Correct model specification:}  
    The log-likelihood is correctly specified under assumption (A2), which matches the data-generating model for transitions in $\mathcal{V}_u$.
    
    \item \textit{Compact parameter space:}  
    The constraint $\| \beta_{u,v} \|_2 = 1$ restricts each $\beta_{u,v}$ to a compact subset of the sphere in $\mathbb{R}^p$.
    
    \item \textit{Continuity and boundedness of log-likelihood:}  
    Assumption (A3) ensures $\| X_i \| \leq C$, hence the log-likelihood terms are continuous and uniformly bounded in $\beta_u$ over the constrained parameter space.
    
    \item \textit{Uniform convergence:}  
    By the Glivenko-Cantelli theorem and the boundedness of covariates, the average log-likelihood converges uniformly in $\beta_u$ to its expectation:
    \[
    \frac{1}{n_u} \ell_u(\beta_u) \xrightarrow{a.s.} \mathbb{E} \left[ \log \left( \frac{\exp(X^\top \beta_{u,Z})}{\sum_{v \in \mathcal{V}_u} \exp(X^\top \beta_{u,v})} \right) \right],
    \]
    where the expectation is over $(X, Z)$ distributed as the true transition distribution from $u$.
    
   \item \textit{Uniqueness of maximizer:}  
    Each parameter vector $\beta_{u,v}$ is constrained to lie on the unit sphere $\mathbb{S}^{p} := \{ b \in \mathbb{R}^{p+1} : \|b\|_2 = 1 \}$. The expected log-likelihood function over this domain remains strictly concave in directions orthogonal to the null space of the Fisher information matrix. Under Assumption (A4), which ensures that $\mathcal{I}(\beta_u)$ is positive definite, the population risk function admits a unique maximizer on the constrained domain up to model-identifiability constraints. The unit-norm constraint eliminates the non-identifiability due to scaling inherent in multinomial logistic models, yielding well-defined and isolated maximizers (see \citealp{Carroll1997, white1982maximum} for uniqueness in constrained M-estimation under smooth manifolds).

\end{itemize}

Therefore, all conditions for consistency of constrained maximum likelihood estimators are satisfied. Moreover, our setting falls under the general theory of M-estimators with constrained parameter spaces. In particular, the unit-norm constraint $\| \beta_{u,v} \|_2 = 1$ defines a compact and smooth submanifold of $\mathbb{R}^{p+1}$, and under identifiability and continuity of the objective function, consistency follows from Example 5.19 of \cite{vandervaart1998}. Hence by Theorem 5.7 of \cite{vandervaart1998}, the maximum pseudo-likelihood estimator satisfies:
\[
\hat{\beta}_{u,v} \xrightarrow{P} \beta_{u,v}^*, \quad \text{for all } v \in \mathcal{V}_u.
\]
This completes the proof.
\end{proof}


\subsection{Proof of Theorem~2 (Asymptotic Normality of SMART-MC Estimator)}

\begin{theorem}[Asymptotic Normality of SMART-MC Estimator]\label{thm:asymp_normality}
Under assumptions \textnormal{(A1)}–\textnormal{(A4)}, for each fixed origin state $u$ and each non-rare destination state $v \in \mathcal{V}_u$, let $\hat{\beta}_{u,v}$ denote the maximum pseudo-likelihood estimator under the unit-norm constraint:
\[
\hat{\beta}_{u,v} := \arg\max_{\beta \in \mathbb{R}^{p+1}, \| \beta \|_2 = 1} \ell_u(\beta),
\]
where $\ell_u(\beta)$ is the partial log-pseudo-likelihood defined over transitions from state $u$. Then,
\[
\sqrt{n_u} \, P_{u,v}^\top \left( \hat{\beta}_{u,v} - \beta_{u,v}^* \right) \xrightarrow{d} \mathcal{N}(0, \Sigma_{u,v}),
\]
where $P_{u,v} \in \mathbb{R}^{(p+1) \times p}$ is an orthonormal basis matrix for the tangent space $\mathcal{T}_{\beta_{u,v}^*} := \{ h \in \mathbb{R}^{p+1} : \beta_{u,v}^{*\top} h = 0 \}$; $\mathcal{I}_{u,v}$ is the Fisher information matrix evaluated at $\beta_{u,v}^*$ and $\Sigma_{u,v} := \left( P_{u,v}^\top \mathcal{I}_{u,v} P_{u,v} \right)^{-1}$.
\end{theorem}

\begin{proof}
We treat $\hat{\beta}_{u,v}$ as a constrained M-estimator that maximizes the pseudo-likelihood $\ell_u(\beta)$ over the unit sphere $\mathbb{S}^p := \{ \beta \in \mathbb{R}^{p+1} : \|\beta\|_2 = 1 \}$. This unit-norm constraint defines a compact and smooth Riemannian manifold embedded in $\mathbb{R}^{p+1}$, with tangent space at the true parameter $\beta_{u,v}^*$ given by
\[
\mathcal{T}_{\beta_{u,v}^*} := \left\{ h \in \mathbb{R}^{p+1} : \beta_{u,v}^{*\top} h = 0 \right\}.
\]

Let $P_{u,v} \in \mathbb{R}^{(p+1) \times p}$ be a matrix whose columns form an orthonormal basis of $\mathcal{T}_{\beta_{u,v}^*}$. Since $\hat{\beta}_{u,v}$ maximizes $\ell_u(\beta)$ subject to the unit-norm constraint, the first-order condition implies that the gradient of $\ell_u$ at $\hat{\beta}_{u,v}$ must lie in the normal space to the constraint manifold. Therefore, its projection onto the tangent space vanishes:
\[
P_{u,v}^\top \dot{\ell}_{u,v}(\hat{\beta}_{u,v}) = 0,
\]
where $\dot{\ell}_{u,v}(\beta) := \partial \ell_u(\beta)/\partial \beta$ is the gradient. Applying a Taylor expansion around $\beta_{u,v}^*$ and projecting onto the tangent space, we obtain
\[
0 = P_{u,v}^\top \dot{\ell}_{u,v}(\beta_{u,v}^*) 
+ P_{u,v}^\top \ddot{\ell}_{u,v}(\beta_{u,v}^*)(\hat{\beta}_{u,v} - \beta_{u,v}^*) 
+ o_p(\| \hat{\beta}_{u,v} - \beta_{u,v}^* \|),
\]
where $\ddot{\ell}_{u,v}(\beta) := \partial^2 \ell_u(\beta)/\partial \beta \partial \beta^\top$ is the Hessian.

Define the projected gradient and projected Hessian:
\[
\tilde{\ell}_{u,v} := P_{u,v}^\top \dot{\ell}_{u,v}(\beta_{u,v}^*), \quad 
\tilde{H}_{u,v} := P_{u,v}^\top \ddot{\ell}_{u,v}(\beta_{u,v}^*) P_{u,v}.
\]
Then the above expansion becomes:
\[
0 = \tilde{\ell}_{u,v} + \tilde{H}_{u,v} P_{u,v}^\top (\hat{\beta}_{u,v} - \beta_{u,v}^*) + o_p(\| \hat{\beta}_{u,v} - \beta_{u,v}^* \|).
\]
Solving for the projected difference yields:
\[
P_{u,v}^\top (\hat{\beta}_{u,v} - \beta_{u,v}^*) 
= - \tilde{H}_{u,v}^{-1} \tilde{\ell}_{u,v} + o_p\left( \| \hat{\beta}_{u,v} - \beta_{u,v}^* \| \right).
\]
Multiplying both sides by $\sqrt{n_u}$ gives the asymptotic linearization:
\[
\sqrt{n_u} P_{u,v}^\top (\hat{\beta}_{u,v} - \beta_{u,v}^*) 
= - \tilde{H}_{u,v}^{-1} \cdot \tilde{s}_{u,v} + o_p(1),
\]
where we define the normalized projected score as
\[
\tilde{s}_{u,v} := \frac{1}{\sqrt{n_u}} \tilde{\ell}_{u,v}.
\]
The remainder term becomes $o_p(1)$ since the Taylor expansion is valid uniformly in a neighborhood of $\beta_{u,v}^*$, and standard asymptotic theory for constrained M-estimators (e.g., \citealp{vandervaart1998}, Theorem~5.23), combined with consistency from Theorem 1 and compactness of the constraint set, ensures that $\| \hat{\beta}_{u,v} - \beta_{u,v}^* \| = O_p(n_u^{-1/2})$ under the conditions of Theorem~2.

Finally, since $\ell_u(\cdot)$ is a log-pseudo-likelihood and the second derivative is negative definite under Assumption~(A4), define the projected Fisher information matrix as $\tilde{\mathcal{I}}_{u,v} := -\tilde{H}_{u,v}$. Therefore,
\[
\sqrt{n_u} P_{u,v}^\top (\hat{\beta}_{u,v} - \beta_{u,v}^*) 
= \tilde{\mathcal{I}}_{u,v}^{-1} \cdot \tilde{s}_{u,v} + o_p(1),
\]
which establishes the asymptotic linear expansion.

By Assumptions (A1)–(A4), the pseudo-likelihood is correctly specified, and the standard regularity conditions for M-estimation apply. In particular, the covariates have bounded moments, the transition probabilities are smooth in $\beta$, and the log-pseudo-likelihood is twice continuously differentiable. Under these conditions, the projected score satisfies a Central Limit Theorem:
\[
\tilde{s}_{u,v} := \frac{1}{\sqrt{n_u}} \tilde{\ell}_{u,v} \xrightarrow{d} \mathcal{N}(0, \tilde{\mathcal{I}}_{u,v}),
\]
and the projected Hessian satisfies
\[
\tilde{H}_{u,v} = P_{u,v}^\top \ddot{\ell}_{u,v}(\beta_{u,v}^*) P_{u,v} = - P_{u,v}^\top \mathcal{I}_{u,v} P_{u,v} + o_p(1),
\]
where $\mathcal{I}_{u,v}$ is the Fisher information matrix of the pseudo-likelihood in the ambient space. Combining these, we obtain
\[
\sqrt{n_u} P_{u,v}^\top (\hat{\beta}_{u,v} - \beta_{u,v}^*) \xrightarrow{d} \mathcal{N}(0, \tilde{\mathcal{I}}_{u,v}^{-1}),
\]
where $\tilde{\mathcal{I}}_{u,v} := P_{u,v}^\top \mathcal{I}_{u,v} P_{u,v}$ is positive definite by Assumption~(A4).

This result follows from the general theory of M-estimators under smooth equality constraints, where the asymptotic distribution of the estimator is Gaussian in the tangent space of the constraint manifold (see \citealp{vandervaart1998}, Theorem~5.23 and Example~5.27; \citealp{amemiya1985advanced}, Section~4.5). Thus, the final form of the asymptotic distribution is
\[
\sqrt{n_u} P_{u,v}^\top (\hat{\beta}_{u,v} - \beta_{u,v}^*) \xrightarrow{d} 
\mathcal{N}\left( 0, \left( P_{u,v}^\top \mathcal{I}_{u,v} P_{u,v} \right)^{-1} \right).
\]

\end{proof}

\subsection{Inference on Covariate Effects via Wald-Type Test Statistics}

The asymptotic normality result in Theorem~\ref{thm:asymp_normality} enables principled statistical inference on covariate effects in the SMART-MC model. For each transition from origin state $u$ to destination state $v \in \mathcal{V}_u$, the pseudo-likelihood estimator $\hat{\beta}_{u,v}$ satisfies the constraint $\|\hat{\beta}_{u,v}\|_2 = 1$ and admits the asymptotic expansion:
\[
\sqrt{n_u} \, P_{u,v}^\top \left( \hat{\beta}_{u,v} - \beta_{u,v}^* \right) \xrightarrow{d} \mathcal{N}\left(0, \left(P_{u,v}^\top \mathcal{I}_{u,v} P_{u,v}\right)^{-1} \right),
\]
where $P_{u,v}$ is an orthonormal basis for the tangent space $\mathcal{T}_{\beta_{u,v}^*} := \{ h \in \mathbb{R}^{p+1} : \beta_{u,v}^{*\top} h = 0 \}$ and $\mathcal{I}_{u,v}$ is the Fisher information matrix for the pseudo-likelihood. In practice, $P_{u,v}$ can be computed via a Gram–Schmidt orthonormalization of any basis for the null space of $\hat{\beta}_{u,v}^\top$, or by performing a QR decomposition of a matrix whose columns span $\{ h \in \mathbb{R}^{p+1} : \hat{\beta}_{u,v}^\top h = 0 \}$. This construction ensures that $P_{u,v}$ spans the tangent space at the estimated parameter $\hat{\beta}_{u,v}$. To perform inference on individual covariate effects, we define the projected coefficient vector
\[
\theta_{u,v} := P_{u,v}^\top \hat{\beta}_{u,v} \in \mathbb{R}^p,
\]
which lies in an unconstrained Euclidean space. Under the asymptotic normality result, each component $(\theta_{u,v})_j$ admits an approximate standard error given by the square root of the $j$-th diagonal element of $\widehat{\Sigma}_{u,v} / n_u$, where
\[
\widehat{\Sigma}_{u,v} := \left( P_{u,v}^\top \widehat{\mathcal{I}}_{u,v} P_{u,v} \right)^{-1}
\]
is the plug-in estimator of the asymptotic covariance, with $\widehat{\mathcal{I}}_{u,v}$ denoting the observed Fisher information evaluated at $\hat{\beta}_{u,v}$.

A Wald-type test for the null hypothesis $H_0: (\beta_{u,v}^*)_j = 0$ proceeds via the test statistic
\[
z_j := \frac{(\theta_{u,v})_j}{\widehat{\mathrm{SE}}_j},
\]
where $\widehat{\mathrm{SE}}_j := \sqrt{[\widehat{\Sigma}_{u,v}]_{jj} / n_u}$ is the estimated standard error for covariate $j$. Under $H_0$, $z_j$ is approximately standard normal, and a two-sided $p$-value is given by $2\Phi(-|z_j|)$. This framework enables testing whether specific demographic or clinical factors (e.g., age, sex, disease duration) significantly influence transition probabilities between treatment classes. In principle, one may also construct confidence intervals for each $(\theta_{u,v})_j$ using the normal approximation:
\[
\text{CI}_{1-\alpha} = (\theta_{u,v})_j \pm z_{\alpha/2} \cdot \widehat{\mathrm{SE}}_j.
\]

For more complex hypotheses involving multiple covariates such as testing whether all demographic variables are jointly insignificant, Wald or score tests based on the projected log-pseudo-likelihood may be employed. The theory of constrained M-estimators supports such multivariate inference, though we do not pursue this direction further here.

\paragraph{Application in Practice:} In applied settings such as MS treatment modeling, this framework enables direct inference on whether specific covariates significantly influence the likelihood of transitioning between treatment categories. For example, one could test whether disease duration has a statistically significant effect on transitions from one DMT to another by evaluating the corresponding $z$-statistic in $\theta_{u,v}$. Similarly, age or race-related differences in transition probabilities across therapy classes (e.g., oral to monoclonal antibody treatments) can be assessed via their associated Wald $p$-values. Such hypothesis tests can inform clinical understanding of treatment personalization and equity.


\subsection{Bootstrap-Based Inference on Covariate Effects}

While asymptotic normality enables Wald-type tests in the projected tangent space, such inference can be sensitive to the accuracy of the estimated Fisher information matrix and may be unreliable when the number of patients transitioning from a given state ($n_u$) is limited. To improve robustness, we instead employ nonparametric bootstrap inference based on repeated estimation of the constrained pseudo-likelihood parameters.

For each transition pair $(u,v)$:
\begin{enumerate}
    \item We generate $B = 1000$ bootstrap resamples of the patient cohort by sampling individuals with replacement.
    \item For each bootstrap replicate \( b = 1, \dots, B \), we re-estimate the constrained pseudo-likelihood coefficient vector \( \hat{\beta}_{u,v}^{(b)} \in \mathbb{R}^{p+1} \), subject to the unit-norm constraint \( \| \hat{\beta}_{u,v}^{(b)} \|_2 = 1 \).
    \item The empirical standard deviation across the \( B \) bootstrap estimates is used to compute standard errors for each covariate coefficient, and approximate two-sided $p$-values are obtained via normal approximation.
\end{enumerate}

Note that we do not perform tangent space projection in each bootstrap sample. Instead, inference is carried out directly on the constrained estimators \( \hat{\beta}_{u,v}^{(b)} \), leveraging their stability and well-defined geometry on the unit sphere. Although this avoids the need to compute the local tangent basis \( P_{u,v} \), it still provides valid uncertainty quantification for inference on individual covariate effects.

In our MS treatment case study, all reported confidence intervals and $p$-values are derived from this bootstrap procedure. Empirically, bootstrap standard errors were more stable than those obtained from asymptotic covariance approximations, especially for transitions with moderate sample size. This approach allows us to assess the significance of demographic and clinical predictors in influencing treatment transitions, while accommodating potential deviations from large-sample theory.


\section{MSCOR}
\subsection{Proof of Theorem~3 (Convergence of MSCOR)}
Proof of `Theorem 3' in the main draft is detailed as the proof of Theorem \ref{MSCOR_theorem} below. To maintain notational consistency and identifiability, in the theorem statements and proofs, we denote the $w-1$ dimensional unit sphere as $O^{w-1}$ instead of $S^{w-1}$ (the notation used in the main paper).
\begin{theorem}
\label{appentheorem1}
 Suppose $\bm{S} = O^{n_1 - 1} \times \ldots \times O^{n_B - 1}$, and $O^{w-1} = \{(x_1,\cdots,x_w) \in \mathds{R}^w \; : \; \sum_{i=1}^w x_i^2 = 1, i=1,\cdots,w\}$. Consider a sequence of step sizes $\delta_{j,k} = \frac{s_j}{\rho^k}$ for $k\in \mathrm{N}$ and $s_j> 0, \rho>1$. Then there exists a $K$ such that for $k \geq K$, all adjustment step sizes $t_i(\delta_{j,k})$ are real for $j = 1,\ldots,B$, $i = 1,\ldots, n_j$. 
  \end{theorem}
\begin{proof}[Proof of Theorem \ref{appentheorem1}]
From Equation (4) of the main draft, the adjustment step size $t_i(\delta_{j,k})$ is given by ($\beta$ is replaced with $u$ to make the notation consistent across rest of the theorems)
 \begin{align*}
 & t_i(\delta_{j,k}) = \frac{-2\sum_{k=1,k\neq i}^{n_j}u_{j,k}^{(l)} + \sqrt{D_i(\delta_{j,k})}}{2(n_j-1)}, i=1,\ldots,n_j, \\
& D_i(\delta_{j,k}) = \bigg(2\sum_{k=1,k\neq i}^{n_j}u_{j,k}^{(l)}\bigg)^2 - 4(n_j-1)(2\delta_{j,k}u_{j,i}^{(l)} + \delta_{j,k}^2).
\end{align*}
Note that $\delta_{j,k} \to 0$ as $k \to \infty$. Hence, $$\lim_{k \to \infty} D_i(\delta_{j,k}) =  \bigg(2\sum_{k=1,k\neq i}^{n_j}u_{j,k}^{(l)}\bigg)^2$$ Since $D_i(\delta_{j,k})$ is a continuous function of $\delta_{j,k}$, if we take $k$ to be sufficiently large, we can make $D_i(\delta_{j,k}) \geq 0$. Suppose for $k\geq K_{j,i}$, $D_i(\delta_{j,k}) \geq 0$ holds true for $j = 1,\ldots,B$, $i = 1,\ldots,n_j$. Take $K = \max_{1\leq j \leq B,\; 1\leq i \leq n_j}{K_{j,i}}$. Hence, for all $k \geq K$, $t_i(\delta_{j,k})$ is real.
\end{proof}

\begin{proposition}
\label{appenprop1}
Consider a matrix $\mathbf{A} = (a_{ij})_{(n-1) \times (n-1)}$ such that $a_{ii} = 1$ for $i=1,\ldots,n-1$ and $a_{ij} = b_i$ for $i\neq j, i = 1,\ldots,n-1, j=1,\ldots, n-1$. Then A is full rank for $n \in \mathbb{N} \setminus \{1\}$ iff 
\begin{enumerate}
    \item $1-b_i \neq 0$ for $i=1,\ldots,n-1$.
    \item $\big[(n-2) + \sum_{i=1}^{n-1}\frac{1}{1-b_i}\big] \neq 0$.
\end{enumerate}
 \end{proposition} 
\begin{proof}
We have
\begin{align*}
\mathbf{A} = \begin{bmatrix}
    1 & b_1 & \cdots & b_1  \\
    b_2 & 1 &\cdots & a_2\\
    \vdots & \vdots & \ddots & \vdots \\
    b_{(n-1)} & b_{(n-1)} & \cdots & 1
  \end{bmatrix}.
\end{align*}
By performing a series of column operations $C_i : C_i - C_{n-1}$ for $i=1,\ldots, n-2$, we obtain $\mathbf{A}^\prime$ as follows: 
\begin{align*}
\mathbf{A}^\prime =
\begin{bmatrix}
    1-b_1 & 0 & \cdots & b_1  \\
   0 & 1 - b_2 &\cdots & b_2\\
    \vdots & \vdots & \ddots & \vdots \\
    b_{(n-1)}-1 & b_{(n-1)}-1 & \cdots & 1
  \end{bmatrix} 
 \end{align*}

 \noindent
 Now consider the following series of row and column operations 
  \begin{align*}
&\begin{bmatrix}
    1-b_1 & 0 & \cdots & b_1  \\
   0 & 1 - b_2 &\cdots & b_2\\
    \vdots & \vdots & \ddots & \vdots \\
    b_{(n-1)}-1 & b_{(n-1)}-1 & \cdots & 1
  \end{bmatrix}  \xrightarrow{C_{n-1} : C_{n-1}+\sum_{i=1}^{n-2} C_i} \\
&    \begin{bmatrix}
    1-b_1 & 0 & \cdots & 1  \\
   0 & 1 - b_2 &\cdots & 1\\
    \vdots & \vdots & \ddots & \vdots \\
    b_{(n-1)}-1 & b_{(n-1)}-1 & \cdots & (n-2)(b_{(n-1)}-1)+1
  \end{bmatrix} \xrightarrow{R_{n-1} : R_{n-1}/(b_{(n-1)}-1)}\\
&  \begin{bmatrix}
    1-b_1 & 0 & \cdots & 1  \\
   0 & 1 - b_2 &\cdots & 1\\
    \vdots & \vdots & \ddots & \vdots \\
    1 & 1 & \cdots & (n-2)+\frac{1}{b_{(n-1)}-1}
  \end{bmatrix}\xrightarrow{C_{n-1} : C_{n-1} + \sum_{i=1}^{n-2}\frac{1}{b_i-1}C_i}
    \end{align*}
  \begin{align*}
&   \begin{bmatrix}
    1-b_1 & 0 & \cdots & 0  \\
   0 & 1 - b_2 &\cdots & 0\\
    \vdots & \vdots & \ddots & \vdots \\
    1 & 1 & \cdots & (n-2)+\sum_{i=1}^{n-1}\frac{1}{b_i-1}
  \end{bmatrix}\xrightarrow{R_{n-1} : R_{n-1} + \sum_{i=1}^{n-2}\frac{1}{b_i-1}R_i}\\
&   \begin{bmatrix}
    1-b_1 & 0 & \cdots & 0  \\
   0 & 1 - b_2 &\cdots & 0\\
    \vdots & \vdots & \ddots & \vdots \\
    0 & 0 & \cdots & (n-2)+\sum_{i=1}^{n-1}\frac{1}{b_i-1}
  \end{bmatrix},  \text{ which we denote as } \mathbf{A}^{\prime \prime}.
 \end{align*} 
Since $\mathbf{A}^{\prime \prime}$ is diagonal matrix, the determinant of $\mathbf{A}^{\prime \prime}$ is given by $$det(\mathbf{A}^{\prime \prime}) = \big[(n-2)+\sum_{i=1}^{n-1}\frac{1}{b_i-1}\big]\prod_{i=1}^{n-2}(1-b_i)$$
 \noindent
 Clearly $r(\mathbf{A}) = r(\mathbf{A}^{\prime \prime})$ where $r(\mathbf{B})$ denotes the rank of any given matrix $\mathbf{B}$. Hence $\mathbf{A}$ is full rank iff 
 \begin{enumerate}
 \item $(1-b_i) \neq 0$ for $i=1,\ldots, n-1,$
 \item $\big[(n-2)+\sum_{i=1}^{n-1}\frac{1}{b_i-1}\big] \neq 0.$
 \end{enumerate}
\end{proof}

\begin{theorem}
\label{MSCOR_theorem}
Suppose $f:\bm{S} \mapsto \mathds{R}$ is convex, continuous and differentiable with extended definition on $\bar{\bm{S}}$, such that, $f(\bm{W})=f(\bm{W}^C)$ when $\bm{W} \in interior(\bm{S})$. Consider a sequence $\delta_{j,k} = \frac{s_j}{\rho^k}$ for $k\in \mathds{N}$ and $s_j>0, \rho>1$. Suppose $\mathbf{U} \in \bm{S}$ given by
  \begin{align*}
      \mathbf{U} = (\mathbf{u}_1, \ldots, \mathbf{u}_B) \text{  where  } \mathbf{u}_b = (u_{j,1},\ldots, u_{j, n_b})\in O^{n_j - 1}, \; j = 1,\ldots, B.
  \end{align*}
Define, $\mathbf{u}_{j,k}^{(i+)} = (u_{j,1}+t_i(\delta_{j,k}),\ldots, u_{j,i-1}+t_i(\delta_{j,k}), u_{j,i}+\delta_{j,k},u_{j,i+1}+t_i(\delta_{j,k}), \ldots, u_{j,n_j} + t_i(\delta_{j,k}))$, $\mathbf{u}_{j,k}^{(i-)} = (u_{j,1}+t_i(-\delta_{j,k}),\ldots, u_{j,i-1}+t_i(-\delta_{j,k}), u_{j,i}-\delta_{j,k},u_{j,i+1}+t_i(-\delta_{j,k}), \ldots, u_{j,n_j} + t_i(-\delta_{j,k}))$ for $j = 1,\ldots,B$, $i=1,\cdots,n_j$, where $t_i(s)$ denotes the adjustment step-size corresponding to step-size $s$. Define $b_{j,i} = -\frac{u_{j,i}}{\big|\sum_{k=1,k\neq i}^{n_j} u_{j,k}\big|}$. If the following conditions hold true
 \begin{enumerate}
 \item for all sufficiently large $k \in \mathrm{N}$, 
    $f(\mathbf{U}) \leq f(\mathbf{u}_1, \ldots, \mathbf{u}_{j-1}, \mathbf{u}_{j,k}^{(i+)}, \mathbf{u}_{j+1}, \ldots, \mathbf{u}_{B})$ and
    $f(\mathbf{U}) \leq f(\mathbf{u}_1, \ldots, \mathbf{u}_{j-1}, \mathbf{u}_{j,k}^{(i-)}, \mathbf{u}_{j+1}, \ldots, \mathbf{u}_{B})$
    \item $1-b_{j,i} \neq 0$
    \item $\big[(n_j-2) + \sum_{i=1}^{n_j-1}\frac{1}{1-b_{j,i}}\big] \neq 0$
\end{enumerate}
for $j = 1,\ldots,B$, $i=1,\ldots,n_j-1$, then a global minimum of $f$ over $\mathbf{S}$ occurs at $\mathbf{U}$. 
\end{theorem}

\begin{proof}[Proof of Theorem \ref{MSCOR_theorem}]
From Theorem \ref{appentheorem1}, there exists a $K \in \mathds{N}$ such that for all $k \geq K_1$, $t_i(\delta_{j,k})$ is real for $j = 1,\ldots, B,\; i = 1,\ldots, n_j$. Similarly it can be shown that there exists a $K_2 \in \mathds{N}$ such that for all $k \geq K_2$, $t_i(-\delta_{j,k})$ is real. Take $K = \max{(K_1,K_2)}$. Hence for all $k \geq K$, both $t_i(\delta_{j,k})$ and $t_i(-\delta_{j,k})$ are real for $j = 1,\ldots, B,\; i = 1,\ldots, n_j$; and therefore, $\mathbf{u}_{j,k}^{(i+)}$ and $\mathbf{u}_{j,k}^{(i-)} \in O^{n_j - 1}$ for all $j = 1,\ldots, B,\; i = 1,\ldots, n_j$. For the rest of the proof, we only consider the cases for $k \geq K$. Define
$$O^{w-1,+} = \{(x_1,\cdots,x_{w}) \in \mathds{R}^{w} \; : \; \sum_{i=1}^w x_i^2 = 1, x_w > 0, i=1,\cdots,w\}, $$ 
$$O^{w-1,-} = \{(x_1,\cdots,x_{w}) \in \mathds{R}^{w} \; : \; \sum_{i=1}^w x_i^2 = 1, x_w \leq 0, i=1,\cdots,w\}. $$ 
\noindent
Note that, $O^{w-1} = O^{w-1,+} \; \cup \; O^{w-1,-}$, and $O^{w-1,+} \; \cap \; O^{w-1,-} = \phi$ (null). Hence, if, $(\mathbf{u}_1, \ldots, \mathbf{u}_B) \in O^{n_1 - 1} \times \ldots \times O^{n_B - 1}$, then each $\mathbf{u}_j$ belongs to either $O^{n_j-1,+}$ or $O^{n_j-1,-}$.\\
To start with, we assume $\mathbf{u}_j \in O^{n_j - 1,+}$ for all $j = 1,\ldots,B$. We intend to prove the theorem for this particular sub-scenario, and eventually, following similar steps, the theorem can be established for all other $2^B - 1$ many sub-scenarios, depending on if individual $\mathbf{u}_j$ belongs to $O^{n_j-1,+}$ or $O^{n_j-1,-}$ for each $j = 1,\ldots, B$. For the rest of the proof (until very end), 
we assume $(\mathbf{u}_1, \ldots, \mathbf{u}_B) \in \hat{\bm{S}} \subset \bm{S}$ where $\hat{\bm{S}} = O^{n_1-1,+} \times \ldots \times O^{n_B-1,+}$.

Since $\bm{u}_j = (u_{j,1},\ldots, u_{j,n_j}) \in O^{n_j-1,+}$, $u_{j,n_j}$ can be derived as a unique function of first $n-1$ coordinates of $\bm{u}_j$ given by $u_{j,n_j} = \sqrt{1-\sum_{i=1}^{n_j-1}u_{j,i}^2}$.
Define $$O^{n_j-1}_* = \{(x_1,\cdots,x_{n_j-1}) \in \mathds{R}^{n-1} \; : \; \sum_{i=1}^{n_j-1} x_i^2 < 1, i=1,\cdots,n_j-1\}.$$ 

\noindent Define
\begin{itemize}
    \item $\mathbf{U}^*  = (\bm{u}^*_1,\ldots,\bm{u}^*_B),$ 
    \item $\mathbf{U}^{*(i+)}_{j,k}  = (\bm{u}^*_1,\ldots,\bm{u}^*_{j-1},\bm{u}_{j,k}^{*(i+)}, \bm{u}^*_{j+1}, \ldots \bm{u}^*_B),$
    \item $\mathbf{U}^{*(i-)}_{j,k}  = (\bm{u}^*_1,\ldots,\bm{u}^*_{j-1},\bm{u}_{j,k}^{*(i-)}, \bm{u}^*_{j+1}, \ldots \bm{u}^*_B),$ 
\end{itemize}
where 
\begin{align*}
   & \bm{u}^*_j = (u_{j,1},\ldots, u_{j,n_j-1}) \\
   & \mathbf{u}_{j,k}^{*(i+)} = (u_{j,1}+t_i(\delta_{j,k}),\ldots, u_{j,i-1}+t_i(\delta_{j,k}), u_{j,i}+\delta_{j,k},u_{j,i+1}+t_i(\delta_{j,k}), \ldots, u_{j,n_j - 1} + t_i(\delta_{j,k})),\\
   & \mathbf{u}_{j,k}^{*(i-)} = (u_{j,1}+t_i(-\delta_{j,k}),\ldots, u_{j,i-1}+t_i(-\delta_{j,k}), u_{j,i}-\delta_{j,k},u_{j,i+1}+t_i(-\delta_{j,k}), \ldots, u_{j,n_j-1} + t_i(-\delta_{j,k})),
\end{align*}
\noindent for $j = 1,\ldots, B,\; i = 1\ldots,n_j-1$.

Note that $\mathbf{u}^*, \mathbf{u}_k^{*(i+)},$ and $\mathbf{u}_k^{*(i-)}$ are the first $(n_j-1)$ coordinates of $\bm{u}_j, \bm{u}_{j,k}^{(i+)},$ and $\bm{u}_{j,k}^{(i-)}$, respectively. Define $f^* : \bm{S}^* \mapsto \mathds{R}$ such that 
\begin{align*}
 &f^*\bigg((x_{1,1},\ldots, x_{1,n_1-1}), \ldots,  (x_{B,1},\ldots, x_{B,n_B-1})\bigg) = \\
 &f\bigg((x_{1,1},\ldots, x_{1,n_1-1}, \sqrt{1 - \sum_{i=1}^{n_1 - 1} x_{1,i}^2}), \ldots, (x_{B,1},\ldots, x_{B,n_1-1}, \sqrt{1 - \sum_{i=1}^{n_B - 1} x_{B,i}^2})\bigg).
\end{align*}
where $\bm{S}^* = O_*^{n_1-1} \times \ldots \times O_*^{n_B-1}$. Hence we have $f^*(\mathbf{U}^*) = f(\mathbf{U})$ for any $\mathbf{U} \in \hat{\bm{S}}$. Thus for that $\mathbf{U}$, we also have $f^*(\mathbf{U^*}_{j,k}^{(i+)}) = f(\mathbf{U}_{j,k}^{(i+)})$ and $f^*(\mathbf{U^*}_{j,k}^{(i-)}) = f(\mathbf{U}_{j,k}^{(i-)})$ for sufficiently large $k$ (since $\mathbf{U}_{j,k}^{(i+)} \to \mathbf{U}$, $\mathbf{U}_{j,k}^{(i-)} \to \mathbf{U}$ and $\mathbf{U} \in \hat{\bm{S}}$, as $k \to \infty$; and $f$ is continuous).

Now \underline{we make a claim that $f^*$ is convex on $\bm{S}^*$}. $f$ is continuous and differentiable on $\hat{\bm{S}} \subset \mathbf{S}$. Since $f(\mathbf{U})= f^*(\mathbf{U}^*)$ for any $\mathbf{U} \in \hat{\bm{S}}$, therefore, $f^*$ is continuous and differentiable on $\mathbf{S}^*$. Since $f$ is convex on $\mathbf{S}$, with extension to $\bar{\mathbf{S}}$, $f$ is also convex on  $\hat{\mathbf{S}} \subset \mathbf{S}$, with extension to $\bar{\hat{\mathbf{S}}} \subset \bar{\mathbf{S}}$. . Note that, convexity of $f$ is only assumed over the surface of the multiple unit-spheres. Extensions are only performed to ensure $f$ still remains defined while satisfying the convexity property, which includes evaluation of the function on points lying on the line connecting two points on $\mathbf{S}$ (or $\hat{\mathbf{S}}$), which lies on the exterior surface. However, the points on the line connecting two surface points solely lie in the interior. Such extension helps $f$ in being defined for interior points as well. To this end, now suppose $\mathbf{X}_1, \mathbf{X}_2 \in \hat{\mathbf{S}}$. Consider $\mathbf{X}^*_1, \mathbf{X}^*_2 \in \mathbf{S}^*$. Hence,

\begin{align*}
    &\gamma f^*(\mathbf{X}^*_1) + (1-\gamma)f^*(\mathbf{X}^*_2) \\
   & = \gamma f(\mathbf{X}_1) + (1-\gamma)f(\mathbf{X}_2) \\
   & \geq f(\gamma \mathbf{X}_1 + (1-\gamma)\mathbf{X}_2) \; \text{(remains defined by extension)}\\
   & = f\bigg((\bar{x}_{1,1},\ldots, \bar{x}_{1,n_1}), \ldots, (\bar{x}_{B,1},\ldots, \bar{x}_{B,n_B}) \bigg) \\
   & = f^*\bigg((\bar{x}_{1,1},\ldots, \bar{x}_{1,n_1 - 1}), \ldots, (\bar{x}_{B,1},\ldots, \bar{x}_{B,n_B - 1}) \bigg)\\
   & = f^*(\gamma \mathbf{X}^*_1 + (1-\gamma)\mathbf{X}^*_2) 
\end{align*}
where $\mathbf{X}_v = \bigg((x^{(v)}_{1,1},\ldots, x^{(v)}_{1,n_1}), \ldots, (x^{(v)}_{B,1},\ldots, x^{(v)}_{B,n_B}) \bigg)$ for $v=1,2$; and $$\bar{x}_{j,i} = \gamma x^{(1)}_{j,i} + (1-\gamma)x^{(2)}_{j,i}$$ for $j=1,\ldots,B$, $i = 1,\ldots, n_j$.
\underline{Hence $f^*$ is also convex}.\\
\noindent
Define $h_{j,i} : U_{j,i} \mapsto \mathbf{S}^*$ such that 
\begin{align*}
  h_{j,i}(z) =\bigg(\bm{u}_1^*, \ldots, \bm{u}_{j-1}^*, \bm{u}_j^{*i}(z),\bm{u}_{j+1}^*, \ldots, \bm{u}_B^* \bigg)  
\end{align*}
where, $U_{j,i} = [-\delta_{j,K},\delta_{j,K}]$ and
$$\bm{u}_j^{*i}(z) = (u_{j,1}+t_{j,i}(z),\ldots,u_{j,i-1}+t_{j,i}(z), u_{j,i} + z,u_{i+1}+t_{j,i}(z), \ldots u_{j,n_j-1}+t_{j,i}(z)),$$ for $j = 1,\ldots, B$, $i=1,\ldots, n_j-1$. 

Note that $h_{j,i}(U_{j,i}) \subset \mathbf{S}^*$. Define $g_{j,i} :  U_{j,i} \mapsto \mathds{R}$ for $j = 1,\ldots, B, \; i=1,\ldots,n_j-1$ such that $g_{j,i} = f^* \circ h_{j,i}$. Hence, 
$$g_{j,i}(z) = f^*\bigg(\bm{u}_1^*, \ldots, \bm{u}^*_{j-1}, \bm{u}_j^{*i}(z),\bm{u}_{j+1}, \ldots, \bm{u}^*_B \bigg)$$
for $j = 1,\ldots, B, \; i=1,\ldots, n_j-1$.

Note that $h_{j,i}$ is continuous on $U_{j,i}=[-\delta_{j,K}, \delta_{j,K}]$ and differentiable on $(-\delta_K, \delta_K)$ for $i=1,\ldots,n_j-1$. Also $f^*$ is continuous and differentiable on $\mathbf{S}^*$. Since the composition of any two continuous functions is continuous, and the composition of two differentiable functions is differentiable; hence $g_{j,i}$ is continuous on $U_{j,i}=[-\delta_{j,K}, \delta_{j,K}]$ and differentiable on $(-\delta_{j,K}, \delta_{j,K})$. 

For any $i \in \{1,\ldots,n_j-1\}$, $g_{j,i}(\delta_{j,K}) = f^*(\mathbf{U}_{j,K}^{*(i+)})= f(\mathbf{U}_{j,K}^{(i+)}),\, g_{j,i}(-\delta_K) = f^*(\mathbf{U}_{j,K}^{*(i-)})= f(\mathbf{U}_{j,K}^{(i-)})$ where,
\begin{align*}
   &\mathbf{U}_{j,K}^{(i+)} =  \big(\mathbf{u}_1, \ldots, \mathbf{u}_{j-1}, \mathbf{u}_{j,K}^{(i+)}, \mathbf{u}_{j+1}, \ldots, \mathbf{u}_{B}\big),\\
   &\mathbf{U}_{j,K}^{(i-)} =  \big(\mathbf{u}_1, \ldots, \mathbf{u}_{j-1}, \mathbf{u}_{j,K}^{(i-)}, \mathbf{u}_{j+1}, \ldots, \mathbf{u}_{B}\big).
\end{align*}

From the conditions provided in the theorem, we have $g_{j,i}(0) \leq g_{j,i}(-\delta_{j,K})$ and $g_{j,i}(0) \leq g_{j,i}(\delta_{j,K})$. Without loss of generality, suppose $f(\mathbf{U}_{j,K}^{(i-)}) \leq f(\mathbf{U}_{j,K}^{(i+)})$ which implies $g_{j,i}(0) \leq g_{j,i}(-\delta_{j,K}) \leq g_{j,i}(\delta_{j,K})$. 

Since $g_{j,i}(0) \leq g_{j,i}(-\delta_{j,K}) \leq g_{j,i}(\delta_{j,K})$, from the continuity of $g_{j,i}$ it can be said that there exists a $w\in [0,\delta_{j,K}]$ such that $g_{j,i}(w) = g_{j,i}(-\delta_{j,K}) \geq g_{j,i}(0)$.  Since $g_{j,i}$ is continuous on $[-\delta_{j,K}, \delta_{j,K}]$ and differentiable on $(-\delta_{j,K}, \delta_{j,K})$, $g_{j,i}$ is also continuous on $[-\delta_{j,K}, w]$ and differentiable on $(-\delta_{j,K}, w)$. Using the mean value theorem, there exists a point $v \in [-\delta_{j,K}, w]$ such that $g_{j,i}^{\prime}(v) = 0$.

We claim that $g_{j,i}^{\prime}(v) = 0$ holds for $v = 0$. Suppose $g_{j,i}^{\prime}(0) \neq 0$ and $g_{j,i}^{\prime}(v^*) = 0$  for some $v^* \in (-\delta_{j,N},w) \setminus \{0\}$. Without loss of generality, take $v^* > 0$. Since $h_{j,i}$ and $f^*$ are convex on $U_{j,i}$ and $\mathbf{S}^*$ respectively, $g_{j,i}$ ($= f^* \circ h_{j,i}$) is also convex on $(-\delta_{j,K},w) \subset U_{j,i}$. Now $g_{j,i}^{\prime}(v^*) = 0$ implies $v^*$ is a local minimum. Also $g_{j,i}^{\prime}(0) \neq 0$, implies $0$ is not a local minimum (or critical point). Therefore, $g_{j,i}(0) > g_{j,i}(v^*)$. Take $M \in \mathds{N}$ such that it satisfies $0 <\delta_{j,M} < v^*$. Clearly, $K < M$ since $\delta_{j,M} < v^* \leq \delta_K$. Hence there exists a $\lambda \in (0,1)$ such that $\delta_{j,M} = (1-\lambda). v^* + \lambda. 0$. So,
\begin{align*}
g_{j,i}(\delta_{j,M}) &= g_{j,i}((1-\lambda).v^* + \lambda. 0) \\
& \leq (1-\lambda)g_{j,i}(v^*) + \lambda g_{j,i}(0)  \\
& = g_{j,i}(0) - (1-\lambda)(g_{j,i}(0) - g_{j,i}(v^*)) \\
& < g_{j,i}(0).
\end{align*}
But, for all $k \geq K$, $g_{j,i}(0) \leq g_{j,i}(\delta_{j,k})$ (since $f(\mathbf{U}) \leq f(\mathbf{U}_{j,k}^{(i+)})$), which implies $g_{j,i}(0) \leq g_{j,i}(\delta_{j,M})$ (since $K < M$). It is a contradiction. Thus, $g_{j,i}^{\prime}(0) = 0$. Now 
\begin{align*}
g_{j,i}^{\prime}(0) &= \bigg[\frac{\partial}{\partial \epsilon}g_{j,i}(\epsilon)\bigg]_{\epsilon = 0} \\
&= \bigg[\frac{\partial}{\partial \epsilon}f^*(h_{j,i}(\epsilon))\bigg]_{\epsilon = 0} \\ 
& = \bigg[\frac{\partial}{\partial h_{j,i}(\epsilon)}f^*(h_{j,i}(\epsilon))\bigg]_{\epsilon = 0} \bigg[\frac{\partial}{\partial \epsilon}h_{j,i}(\epsilon)\bigg]_{\epsilon = 0}.
\end{align*}

Now $h_{j,i}(0) = \mathbf{U}^*$. Hence 
\begin{align*}
&\bigg[\frac{\partial}{\partial h_{j,i}(\epsilon)}f^*(h_{j,i}(\epsilon))\bigg]_{\epsilon = 0} \\
=&\bigg[\frac{\partial}{\partial x_{1,1}}f^*(\mathbf{U}^*), \ldots, \frac{\partial}{\partial x_{j-1,n_{j-1}-1}}f^*(\mathbf{U}^*), \; \frac{\partial}{\partial x_{j,1}}f^*(\mathbf{U}^*), \ldots, \frac{\partial}{\partial x_{j,n_j-1}}f^*(\mathbf{U}^*), \\
& \frac{\partial}{\partial x_{j+1,1}}f^*(\mathbf{U}^*), \ldots, \frac{\partial}{\partial x_{B,n_B-1}}f^*(\mathbf{U}^*)\bigg]\\
=&\bigg[0,\ldots, 0, \; \frac{\partial}{\partial x_{j,1}}f^*(\mathbf{U}^*), \ldots, \frac{\partial}{\partial x_{j,n_j-1}}f^*(\mathbf{U}^*),\; 0,\ldots, 0\bigg]_{1 \times (\sum_{j=1}^B n_j - B)} \\
=&\bigg[0, \ldots, 0, \nabla_{j,1}, \ldots, \nabla_{j,n_j-1}, 0, \ldots, 0 \bigg]_{1 \times (\sum_{j=1}^B n_j - B)} 
\end{align*}

where $\nabla_{j,i} = \frac{\partial}{\partial x_{j,i}}f^*(\mathbf{U}^*)$ for $j = 1,\ldots, B, \; i=1,\ldots,n_j-1$.
$$
\big[\frac{\partial}{\partial \epsilon}h_{j,i}(\epsilon)\big]_{\epsilon=0} =
\begin{bmatrix}
    \bm{0}   \\
    (A_{j,i})_{(n_j-1) \times 1}  \\
    \bm{0} 
  \end{bmatrix}_{(\sum_{j=1}^B n_j - B) \times 1}.$$
  where
\begin{align*}
A_{j,i} = \begin{bmatrix}a^{(j)}_{i,1}\\
\vdots\\
a^{(j)}_{i,n_j-1}
\end{bmatrix}_{(n_j-1) \times 1}
\end{align*}
where $a^{(j)}_{i,i} = 1$ and $$a^{(j)}_{ik}  = \frac{\partial t_{j,i}(s)}{\partial s}\bigg |_{s=o} =  \frac{1}{2}\frac{-8(n_j-1)u_{j,i}}{2(n_j-1)\sqrt{(2\sum_{k=1,k\neq i}^{n_j}u_{j,k})^2}} = -\frac{u_{j,i}}{\big|\sum_{k=1,k\neq i}^{n_j}u_{j,k}\big|} = b_{j,i},$$
for $k \in \{1,\ldots,n_j-1\} \setminus \{i\}$.\\
\noindent
Hence 
\begin{align*}
&g_{j,i}^{\prime}(0) = \bigg[\frac{\partial}{\partial \epsilon}g_{j,i}(\epsilon)\bigg]_{\epsilon = 0}\\
&= \bigg[\mathbf{0}, \nabla_{j,1}, \ldots, \nabla_{j,n_j-1}, \mathbf{0} \bigg] \begin{bmatrix}\mathbf{0}\\
a^{(j)}_{i,1}\\
\vdots\\
a^{(j)}_{i,n_j-1}\\
\mathbf{0}
\end{bmatrix} \\
& = \bigg[a^{(j)}_{i,1},\ldots, a^{(j)}_{i,n_j-1}\bigg] \begin{bmatrix}
    \nabla_{j,1}  \\
    \vdots \\
    \nabla_{j, n_j-1} 
  \end{bmatrix} \\
  & = 0.
\end{align*}
Since this equation holds for all $i= 1,\cdots,n_j-1$, we have $\mathbf{A^{(j)}}\mathbf{x}^{(j)} = \mathbf{0}$ where
\begin{align*}
\mathbf{A}^{(j)}_{(n-1) \times (n-1)} = \begin{bmatrix}
    1 & b_{j,1} & \cdots & b_{j,1}  \\
    b_{j,2} & 1 &\cdots & b_{j,2}\\
    \vdots & \vdots & \ddots & \vdots \\
    b_{n_j-1,1} & b_{n_j-1,1} & \cdots & 1
  \end{bmatrix}, \; \mathbf{x}^{(j)}_{(n_j-1) \times 1} = \begin{bmatrix}
    \nabla_{j,1}  \\
    \vdots \\
    \nabla_{j,n_j-1} 
  \end{bmatrix}.
\end{align*}

 By Proposition \ref{appenprop1}, $\mathbf{A}^{(j)}_{(n_j-1) \times (n_j-1)}$ is full-rank, hence $\mathbf{A}^{(j)}\mathbf{x}^{(j)} = \mathbf{0}$ implies $\mathbf{x}^{(j)} = \mathbf{0}$. Hence $\frac{\partial}{\partial x_{j,i}}f^*(\mathbf{U}^*)=0$ for all $i = 1,\ldots, n_j-1$ for any given $j \in \{1,\ldots, B\}$. Hence, it follows that $\frac{\partial}{\partial x_{j,i}}f^*(\mathbf{U}^*)=0$ holds true  all $j=1,\ldots, B,\; i = 1,\ldots, n_j-1$. Therefore, $\mathbf{U}^*$ is a critical point. Since $f^*$ is convex on $\mathbf{S}^*$, a local minimum occurs at $\mathbf{U}^*$. But for a convex function, the global minimum occurs at any local minimum. Hence the global minimum of $f^*$ occurs at $\mathbf{U}^*$, which clearly implies the global minimum of $f$ on $\hat{\mathbf{S}} \subset \mathbf{S}$ occurs at $\mathbf{U}$.

 Thus the theorem is established for the scenario 
 when $\mathbf{u}_j \in O^{n_j-1,+}$ for all $j = 1,\ldots, B$. Following similar steps, the theorem can be further established for all other $2^B - 1$ many scenarios, depending on if $\mathbf{u}_j \in O^{n_j-1,+}$ or $\mathbf{u}_j \in O^{n_j-1,-}$ for each $j = 1,\ldots, B$. Hence the global minimum of $f$ occurs at $\mathbf{U}$ for $\mathbf{U} \in \mathbf{S}$.
\end{proof}

\subsection{MSCOR Tuning Parameters}
Here we provide the considered values of the tuning parameters in MSCOR. For benchmark comparisons, we take $MaxTime = 3600$ (seconds), $MaxRuns = 1000$, $MaxIter = 10000$, $s_{\text{initial}} = 1$, $\rho = 2$, $\phi = 10^{-20}$, $\tau_1 =10^{-6}$, $\tau_2 =10^{-20}$, $\lambda = 10^{-6}$. For SMART-MC likelihood maximization (both in the simulations study and real data analysis), we take, $MaxTime = 3600$, $MaxRuns = 10$, $MaxIter = 5000$, $s_{\text{initial}} = 1$, $\rho = 2$, $\phi = 10^{-20}$, $\tau_1 =10^{-1}$, $\tau_2 =10^{-1}$, $\lambda = 10^{-6}$. Such modification of $\tau_1$ and $\tau_2$ in the latter case allows faster convergence, spending less time for extensive refining of the solution to higher decimal places, while roaming in a small neighborhood, which seems unnecessary, given our negative log-likelihood is observed to lie somewhere between $10^4$ and $10^6$. 

\subsection{Convex Optimization and Non-convexity Detection}
If the objective function is known to be convex a priori, a single \emph{run} suffices, as the stopping criterion ensures local optimality (details provided in the following subsection). In the absence of prior information about convexity, MSCOR automatically terminates after the second \emph{run}, since each \emph{run} converges to an optimal solution. For convex functions, this solution is unique, resulting in identical outcomes in the first two consecutive \emph{runs}, thereby satisfying the stopping criterion. Extending this logic, if MSCOR converges after \emph{run} $R>2$, it indicates at least one successful escape from a local solution, confirming the presence of multiple optima and hence the non-convexity of the objective function. When optimizing the SMART-MC likelihood with MSCOR, the observed number of \emph{runs} required for convergence ranges from 5 to 10, corroborating the non-convexity of the likelihood.

\subsection{Modified Benchmark Functions}
\label{sec_appendix_compare}
For the benchmark study (see Section 4.3 of the main draft), we modify the actual benchmark functions \citep{Jamil2013} on the parameter space $\bm{S}$, where $$\bm{S} = O^{n - 1} \times \ldots \times O^{n - 1},$$ and we subsequently use them for the benchmark study considered in Section 4.3 of the main draft.
We modify the original form of the following objective functions in such a way that their global minimum lies in $\bm{S}$. For all of the following functions, we take, $$f(\bm{x}_1,\ldots,\bm{x}_B) = \sum_{b=1}^B f(\bm{x}_b).$$
Now we describe the structure of $f(\bm{x}_b)$ after modification. For ease of notation, we denote $\bm{x} \equiv \bm{x}_b$, where, $\bm{x} = (x_1,\ldots, x_n)$.
\paragraph{Modified Ackley's function:}
$n$-dimensional Ackley's function is given by
\begin{align*}
f(x_1,\ldots, x_n) =& -20 \exp\bigg[-0.2\sqrt{0.5\sum_{i=1}^n\bigg(x_i - \frac{1}{\sqrt{n}}\bigg)^2}\bigg] -\\
& \exp\bigg[\frac{1}{n}\sum_{i=1}^n\cos\bigg\{2\pi \bigg(x_i - \frac{1}{\sqrt{n}}\bigg)\bigg\}\bigg]+e+20.
\end{align*}
The global minimum value is $0$, which is attained at $(x_1,\ldots,x_n) = (\frac{1}{\sqrt{n}}, \ldots, \frac{1}{\sqrt{n}})$.

\paragraph{Modified Griewank function:}  
\begin{align*}
f(x_1,\ldots,x_n) = \frac{1}{4000}\sum_{i=1}^nn\bigg(x_i -\frac{1}{\sqrt{n}}\bigg) ^2 - \prod_{i=1}^n\cos\bigg[\frac{x_i-\frac{1}{\sqrt{n}}}{\sqrt{i}}\bigg] + 1.
\end{align*}
The global minimum value is $0$, which is attained at $(x_1,\ldots,x_n) = (\frac{1}{\sqrt{n}}, \ldots,  \frac{1}{\sqrt{n}})$.

\paragraph{Negative sum of squares function:}  
\begin{align*}
f(x_1,\ldots,x_n) = n -\sum_{i=1}^n ix_i^2.
\end{align*}
The global minimum value is $0$, which is attained at $(x_1,\ldots,x_n) = (0, \ldots,0,\pm 1).$

\paragraph{Modified Rastrigin function:} 
\begin{align*}
f(x_1,\ldots,x_n) = 10n + \sum_{i=1}^n \bigg[ \bigg(x_i - \frac{1}{\sqrt{n}}\bigg)^2 - 10 \cos\bigg\{2 \pi \bigg(x_i - \frac{1}{\sqrt{n}}\bigg)\bigg\} \bigg],
\end{align*}
The global minimum value is $0$, which is attained at $(x_1,\ldots,x_n) = (\frac{1}{\sqrt{n}}, \ldots,  \frac{1}{\sqrt{n}})$.

\subsection{Results from MSCOR Benchmark Study}
\begin{figure}[]
	\centering
	\includegraphics[width=.9\textwidth]{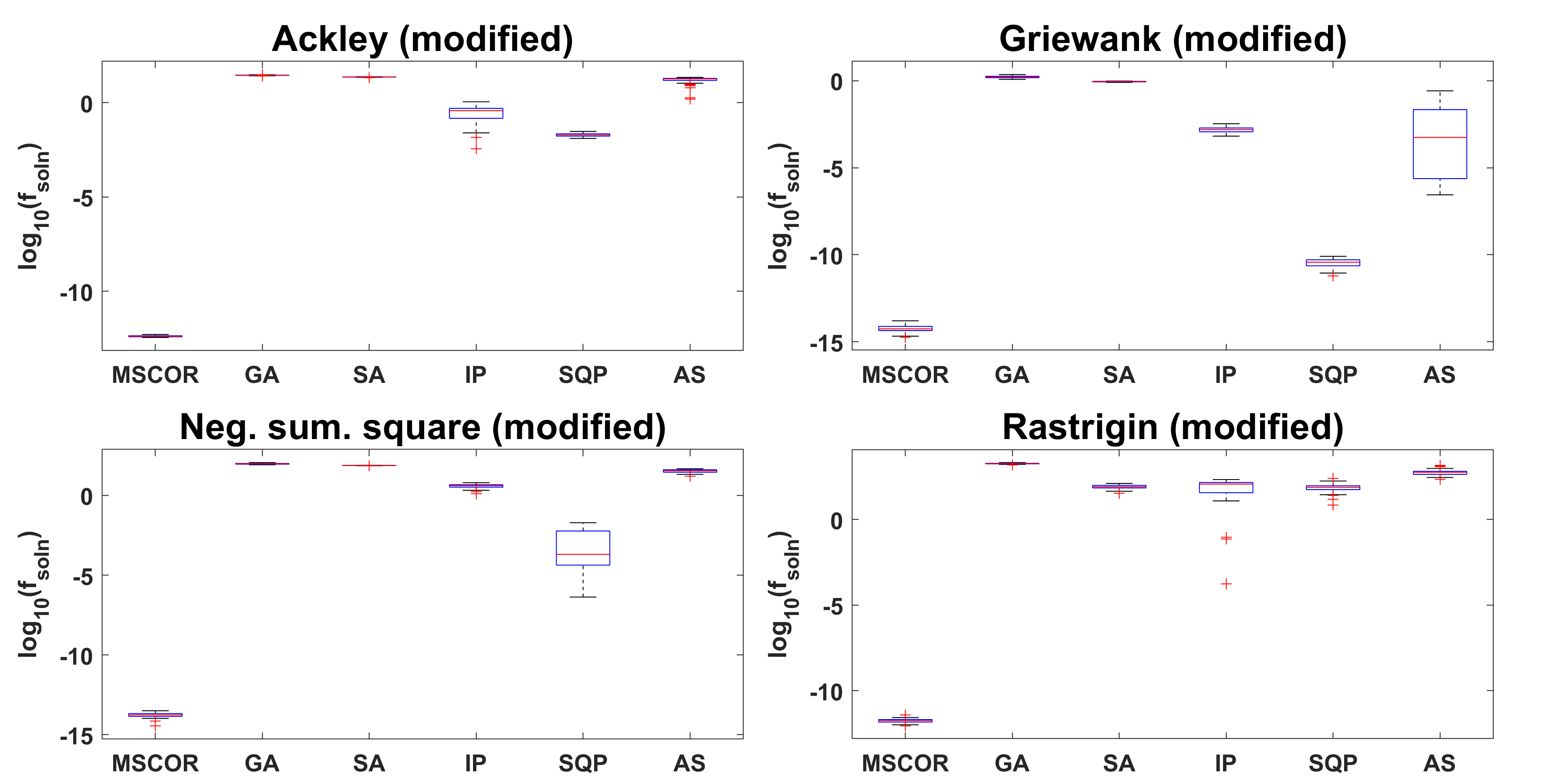}
	\caption{Distribution of logarithm (base 10) of the final objective function values at the solutions obtained minimizing modified benchmark functions (Ackley, Griewank, negative sum of squares, Rastrigin) over 100 experiments for $B = 10, n_b=20$ (for $b=1,\ldots,B)$ scenario using MSCOR, genetic algorithm (GA), simulated annealing (SA), interior-point (IP), sequential quadratic programming (SQP) and active-set(AS) are visually depicted.}
	\label{boxplot}
\end{figure}

In Figure \ref{boxplot}, we show box-plots of the values of the considered benchmark functions obtained at respective final solutions over 100 experiments (see Section 4.3 of the main draft) found by all considered algorithms. Objective function values at final MSCOR solutions are observed to be consistently smaller than those of its competitors. In Table \ref{full_comparison_table}, we provide the full table from the benchmark study conducted in Section 4.3 of the main draft, including summaries of the median and maximum execution times.

\begin{sidewaystable} 
\centering
\resizebox{0.99\columnwidth}{!}{%
\begin{tabular}{c|c|ccccc|ccccc|ccccc}
\hline
\multirow{2}{*}{Functions} & \multirow{2}{*}{Algorithms} & \multicolumn{5}{c|}{$B = 5, n_b = 5$} & \multicolumn{5}{c|}{$B = 10, n_b = 20$} & \multicolumn{5}{c}{$B = 100, n_b = 5$} \\ \cline{3-17} 
 &  & min. value & se of solution & mean time (se) & \begin{tabular}[c]{@{}c@{}}median \\ time\end{tabular} & \begin{tabular}[c]{@{}c@{}}max \\ time\end{tabular} & min. value & se of solution & mean time (se) & \begin{tabular}[c]{@{}c@{}}median \\ time\end{tabular} & \begin{tabular}[c]{@{}c@{}}max \\ time\end{tabular} & min. value & se of solution & mean time (se) & \begin{tabular}[c]{@{}c@{}}median \\ time\end{tabular} & \begin{tabular}[c]{@{}c@{}}max \\ time\end{tabular} \\ \hline
\multirow{6}{*}{\begin{tabular}[c]{@{}c@{}}Ackley's\\ (modified)\end{tabular}} & MSCOR & \textbf{2.22e - 14} & 0.029 & 1.64 (0.008) & 1.65 & 1.84 & \textbf{3.61e - 13} & 0.000 & 312.58 (0.385) & 312.23 & 326.17 & \textbf{1.65e - 09} & 0.288 & 3600.04$^*$(0.006) & 3600.04 & 3600.06 \\
 & GA & 1.51e + 01 & 0.169 & 16.34 (0.769) & 14.29 & 37.10 & 2.59e + 01 & 0.080 & 78.81 (0.315) & 79.03 & 85.23 & 3.88e + 02 & 2.214 & 357.20 (3.682) & 357.00 & 384.03 \\
 & SA & 4.70e + 00 & 0.081 & 1.84 (0.092) & 1.63 & 4.79 & 2.16e + 01 & 0.039 & 53.48 (2.490) & 51.90 & 94.11 & 2.70e + 02 & 1.175 & 371.77 (23.565) & 414.56 & 416.64 \\
 & IP & \textbf{7.51e - 12} & 0.347 & 0.06 (0.003) & 0.06 & 0.12 & \textbf{3.67e - 03} & 0.023 & 0.09 (0.002) & 0.09 & 0.23 & 2.17e + 02 & 6.179 & 0.42 (0.026) & 0.39 & 0.65 \\
 & SQP & 9.50e - 04 & 0.414 & 0.03 (0.001) & 0.04 & 0.05 & 1.28e - 02 & 0.000 & 0.41 (0.001) & 0.41 & 0.47 & \textbf{8.33e + 01} & 9.219 & 5.15 (0.028) & 5.14 & 5.29 \\
 & AS & 2.35e + 00 & 0.328 & 0.03 (0.001) & 0.04 & 0.06 & 1.53e + 00 & 0.401 & 0.47 (0.003) & 0.46 & 0.60 & 1.56e + 02 & 4.564 & 5.60 (0.010) & 5.60 & 5.68 \\ \hline
\multirow{6}{*}{\begin{tabular}[c]{@{}c@{}}Griewank\\ (modified)\end{tabular}} & MSCOR & \textbf{\textless 1e - 16} & 0.000 & 1.54 (0.007) & 1.54 & 1.76 & \textbf{1.78e - 15} & 0.000 & 204.51 (0.444) & 204.42 & 215.71 & \textbf{1.46e - 09} & 0.000 & 3600.07$^*$(0.010) & 3600.08 & 3600.11 \\
 & GA & 8.04e - 01 & 0.040 & 19.59 (0.962) & 18.03 & 40.03 & 1.12e + 00 & 0.021 & 88.70 (0.287) & 88.76 & 95.92 & 3.60e + 01 & 0.400 & 461.57 (4.188) & 461.23 & 482.58 \\
 & SA & 1.06e - 01 & 0.008 & 2.03 (0.101) & 1.76 & 4.58 & 7.99e - 01 & 0.004 & 54.12 (2.392) & 49.62 & 93.40 & 2.72e + 01 & 0.166 & 372.25 (11.450) & 391.98 & 398.10 \\
 & IP & 2.47e - 13 & 0.000 & 0.02 (0.002) & 0.02 & 0.19 & 6.53e - 04 & 0.000 & 0.10 (0.002) & 0.10 & 0.31 & 2.03e + 00 & 0.175 & 0.50 (0.025) & 0.47 & 0.72 \\
 & SQP & \textbf{1.98e - 13} & 0.000 & 0.01 (0.000) & 0.01 & 0.03 & \textbf{5.96e - 12} & 0.000 & 0.24 (0.001) & 0.23 & 0.31 & \textbf{3.80e - 12} & 0.000 & 1.69 (0.015) & 1.71 & 1.73 \\
 & AS & 3.50e - 08 & 0.022 & 0.03 (0.002) & 0.02 & 0.11 & 2.77e - 07 & 0.005 & 0.43 (0.015) & 0.52 & 0.70 & 4.54e - 07 & 0.464 & 5.79 (0.722) & 7.09 & 7.59 \\ \hline
\multirow{6}{*}{\begin{tabular}[c]{@{}c@{}}Neg. sum\\ of squares\\ (modified)\end{tabular}} & MSCOR & \textbf{\textless 1e - 16} & 0.000 & 0.45 (0.005) & 0.45 & 0.61 & \textbf{\textless 1e - 16} & 0.000 & 43.81 (0.413) & 43.48 & 62.79 & \textbf{1.51e - 14} & 0.000 & 1602.09 (15.515) & 1613.57 & 1683.67 \\
 & GA & 5.17e + 00 & 0.198 & 16.47 (0.805) & 14.40 & 37.30 & 8.27e + 01 & 0.648 & 74.74 (0.258) & 75.21 & 80.19 & 1.89e + 02 & 2.398 & 325.61 (2.558) & 327.37 & 336.93 \\
 & SA & 2.19e + 00 & 0.044 & 1.85 (0.087) & 1.70 & 5.57 & 7.10e + 01 & 0.126 & 50.59 (2.549) & 42.34 & 96.03 & 1.65e + 02 & 0.435 & 358.06 (16.27) & 393.59 & 395.41 \\
 & IP & \textbf{7.99e - 15} & 0.000 & 0.02 (0.000) & 0.02 & 0.05 & 1.26e + 00 & 0.100 & 0.09 (0.002) & 0.09 & 0.28 & 3.83e + 00 & 1.520 & 0.40 (0.023) & 0.38 & 0.61 \\
 & SQP & 1.07e - 14 & 0.000 & 0.02 (0.000) & 0.02 & 0.02 & \textbf{4.26e - 07} & 0.000 & 0.41 (0.002) & 0.40 & 0.54 & \textbf{9.09e - 12} & 0.000 & 3.78 (0.102) & 3.71 & 4.31 \\
 & AS & 1.92e - 09 & 0.093 & 0.02 (0.001) & 0.02 & 0.05 & 1.60e + 01 & 0.714 & 0.45 (0.003) & 0.44 & 0.56 & 2.42e + 01 & 3.595 & 5.53 (0.093) & 5.45 & 6.34 \\ \hline
\multirow{6}{*}{\begin{tabular}[c]{@{}c@{}}Rastrigin\\ (modified)\end{tabular}} & MSCOR & \textbf{\textless 1e - 16} & 0.762 & 2.08 (0.417) & 1.30 & 26.15 & \textbf{8.53e - 13} & 0.000 & 135.99 (0.255) & 135.81 & 141.42 & \textbf{1.02e + 02} & 5.544 & 3600.04$^*$(0.011) & 3600.04 & 3600.14 \\
 & GA & 9.90e + 01 & 5.792 & 18.21 (0.835) & 16.87 & 37.68 & 1.59e + 03 & 9.215 & 79.37 (0.262) & 79.52 & 85.07 & 4.98e + 03 & 73.999 & 412.85 (51.696) & 337.64 & 783.48 \\
 & SA & 8.64e + 00 & 0.302 & 1.76 (0.082) & 1.53 & 4.80 & 3.47e + 01 & 2.006 & 93.74 (3.322) & 90.75 & 151.49 & 4.72e + 02 & 10.532 & 935.30 (60.66) & 946.8 & 1212.52 \\
 & IP & 6.72e + 00 & 0.725 & 0.04 (0.001) & 0.03 & 0.09 & \textbf{1.68e - 04} & 5.922 & 0.10 (0.001) & 0.10 & 0.15 & 5.14e + 02 & 111.633 & 0.41 (0.010) & 0.39 & 0.47 \\
 & SQP & 8.18e + 00 & 0.637 & 0.03 (0.000) & 0.03 & 0.04 & 7.04e + 00 & 3.435 & 0.42 (0.002) & 0.42 & 0.53 & \textbf{4.71e + 02} & 10.107 & 5.32 (0.075) & 5.25 & 5.88 \\
 & AS & \textbf{1.20e + 00} & 0.969 & 0.03 (0.000) & 0.02 & 0.04 & 2.19e + 02 & 21.095 & 0.46 (0.001) & 0.45 & 0.53 & 8.49e + 02 & 104.721 & 5.78 (0.058) & 5.73 & 6.26 \\ \hline
\end{tabular}}
\caption{A comparative study of MSCOR, GA, SA, IP, SQP and AS methods for optimizing modified Ackley, Griewank, negative sum of squares, and Rastrigin functions is presented for cases with parameter settings $(B,n_b) = (5,5), (10,20), (100,5)$. The first two experiments are repeated 100 times, and the last one is repeated 10 times. S.e. denotes the standard error. Time is measured in seconds. For the scenarios where MSCOR's average computation time exceeds upper bound 3600 seconds, are labeled with $^*$.}
\label{full_comparison_table}
\end{sidewaystable}

\section{Simulation Study}
\label{sim_study}
\vspace{-0.3cm}
To evaluate the performance of SMART-MC, backed by MSCOR for optimizing the likelihood, we generate synthetic data with parameter dimensions similar to the real data used in the case study, detailed in the main draft. We consider $N = 10$ states, $K = 1000$ patients, and a sample state sequence length of $t_k = 20$ across all patients. We generate $p = 5$ patient-level covariates for each subject. The true transition matrix, including the initial state vector, is taken to be $67\%$ sparse, ensuring that each row contains at least two non-zero elements, including transitions within the same state. This is inspired by the fact that it is typical for an MS patient to mostly remain on the same treatment, occasionally moving to a different one. We generate patient-level covariate values and coefficient vectors corresponding to non-zero transitions under the following scenarios:
\begin{itemize}
    \item \textbf{Scenario 1:} Each patient-level covariate is drawn from $N(0,1)$, and the coefficients for non-zero transition locations are drawn from $N(0, 10^2)$.
    \item \textbf{Scenario 2:} Each patient-level covariate is drawn from $U(-10,10)$, and the coefficients for non-zero transition locations are drawn from $U(-1, 1)$.
\end{itemize}
After generating the coefficient vectors, they are scaled to have an $l_2$ norm of 1 in both scenarios. MSCOR is then fitted to estimate the coefficient vectors. We also consider another naive model in which, unlike SMART-MC, no sparsity-based adjustments are applied. Consequently, this naive model estimates all transition probabilities as functions of patient-level covariates.

We estimate the coefficient vectors corresponding to the top 10 most frequent transitions, as well as the most frequent initial state. Table \ref{simulation_table} and and \ref{simulation_table_2} show the estimated and true coefficient values corresponding to the top 10 most frequent transitions, along with the most frequent initial state for scenarios 1 and 2. We observe that MSCOR performs well in estimating these coefficients, with values close to the true ones, while the estimated coefficients from the naive model are far from the true values. For SMART-MC, in order to empirically assess whether the estimated coefficients converge to their true values, we calculate the mean absolute deviation (MAD) for the coefficients corresponding to the top 10 transitions, excluding less frequent transitions due to their potential unreliability arising from lower empirical transition counts. MAD is computed for $K = 1000, 2000, 3000$ and $t_k = 20, 40, 60$ for all patients, keeping $N = 10$ under scenario 1 setup. In Table \ref{MAD_table} we note that as the sample size and/or observed state sequence length increases, MAD decreases, reducing from $0.0296$ to $0.0096$ as we move from $(K, t_k) = (1000, 20)$ to $(K, t_k) = (3000, 60)$. It is observed that as we move to less frequent transitions (beyond those considered in the tables), the estimation performance decreases, yielding estimated values that are somewhat farther from the true values due to the smaller number of corresponding empirical transitions present in the dataset.

To assess the computational gain of parallelized MSCOR, we compare computation times across cases $(K, t_k) = (1000, 10), (1000, 20), (2000, 20)$ and $N = 6, 9, 12$, keeping $t_k = 20$ under scenario 1 setup. Computations are performed in MATLAB using 12 CPU cores. The results are presented in Table \ref{time_comparison}. Parallelized MSCOR achieves a 3--7 fold speedup over regular MSCOR, with greater gains observed as the parameter dimensions increase. This is expected since computational gains with parallel computing tend to increase as the objective function evaluation becomes more expensive \citep{MathWorks2024}.
\begin{table}[]
\centering
\resizebox{0.92\columnwidth}{!}{%
\begin{tabular}{|c|c|c|c|c|c|c|c|c|}
\hline
Transitions & \begin{tabular}[c]{@{}c@{}}Transition\\ counts\end{tabular} & Scenario & $\beta_0$ & $\beta_1$ & $\beta_2$ & $\beta_3$ & $\beta_4$ & $\beta_5$ \\ \hline
\multirow{3}{*}{4 (initial state)} & \multirow{3}{*}{203 (20.30\%)} & True & 0.23 & -0.87 & 0.00 & -0.35 & -0.04 & -0.27 \\
 &  & SMART-MC & 0.24 (0.019) & -0.80 (0.014) & 0.04 (0.020) & -0.45 (0.021) & -0.25 (0.014) & -0.20 (0.025) \\
 &  & Naive & 0.88 (0.019) & -0.29 (0.039) & -0.01 (0.037) & -0.07 (0.037) & -0.35 (0.035) & -0.11 (0.035) \\ \hline
\multirow{3}{*}{6 $\mapsto$ 6} & \multirow{3}{*}{2342 (12.33\%)} & True & 0.37 & -0.82 & -0.16 & -0.18 & -0.37 & 0.05 \\
 &  & SMART-MC & 0.40 (0.032) & -0.83 (0.017) & -0.13 (0.025) & -0.12 (0.042) & -0.34 (0.017) & 0.08 (0.032) \\
 &  & Naive & 0.82 (0.033) & -0.34 (0.034) & 0.23 (0.038) & -0.03 (0.035) & -0.34 (0.032) & -0.20 (0.033) \\ \hline
\multirow{3}{*}{4 $\mapsto$ 4} & \multirow{3}{*}{1796 (9.45\%)} & True & 0.30 & -0.57 & -0.45 & 0.32 & 0.27 & 0.44 \\
 &  & SMART-MC & 0.32 (0.011) & -0.57 (0.012) & -0.45 (0.018) & 0.31 (0.011) & 0.27 (0.023) & 0.44 (0.011) \\
 &  & Naive & 0.83 (0.030) & -0.41 (0.034) & -0.22 (0.036) & 0.27 (0.033) & 0.05 (0.037) & 0.15 (0.030) \\ \hline
\multirow{3}{*}{7 $\mapsto$ 7} & \multirow{3}{*}{1307 (6.88\%)} & True & 0.16 & 0.29 & -0.60 & 0.65 & 0.32 & -0.10 \\
 &  & SMART-MC & 0.11 (0.023) & 0.22 (0.029) & -0.64 (0.016) & 0.64 (0.013) & 0.27 (0.019) & -0.21 (0.023) \\
 &  & Naive & 0.73 (0.035) & 0.03 (0.035) & -0.32 (0.035) & 0.59 (0.035) & 0.02 (0.037) & -0.13 (0.030) \\ \hline
\multirow{3}{*}{8 $\mapsto$ 8} & \multirow{3}{*}{1261 (6.64\%)} & True & 0.02 & 0.58 & 0.42 & -0.06 & -0.69 & -0.01 \\
 &  & SMART-MC & 0.15 (0.029) & 0.58 (0.017) & 0.40 (0.013) & -0.08 (0.022) & -0.69 (0.023) & -0.07 (0.029) \\
 &  & Naive & 0.74 (0.034) & 0.33 (0.038) & 0.48 (0.033) & -0.29 (0.037) & -0.16 (0.039) & -0.02 (0.034) \\ \hline
\multirow{3}{*}{6 $\mapsto$ 7} & \multirow{3}{*}{1070 (5.63\%)} & True & 0.10 & -0.01 & -0.71 & 0.05 & 0.51 & 0.47 \\
 &  & SMART-MC & 0.07 (0.032) & 0.03 (0.017) & -0.67 (0.025) & 0.09 (0.041) & 0.54 (0.0017) & 0.50 (0.032) \\
 &  & Naive & 0.77 (0.038) & 0.37 (0.036) & -0.27 (0.033) & 0.11 (0.038) & 0.39 (0.040) & 0.18 (0.038) \\ \hline
\multirow{3}{*}{5 $\mapsto$ 4} & \multirow{3}{*}{900 (4.74\%)} & True & 0.50 & 0.05 & -0.24 & -0.71 & -0.23 & -0.36 \\
 &  & SMART-MC & 0.53 (0.032) & 0.16 (0.017) & -0.16 (0.028) & -0.73 (0.023) & -0.24 (0.035) & -0.27 (0.032) \\
 &  & Naive & 0.95 (0.034) & -0.08 (0.034) & -0.12 (0.033) & -0.15 (0.035) & 0.14 (0.032) & -0.18 (0.034) \\ \hline
\multirow{3}{*}{8 $\mapsto$ 2} & \multirow{3}{*}{872 (4.59\%)} & True & 0.33 & -0.19 & -0.81 & 0.42 & -0.14 & 0.08 \\
 &  & SMART-MC & 0.34 (0.030) & -0.18 (0.019) & -0.80 (0.014) & 0.43 (0.022) & -0.15 (0.022) & 0.09 (0.030) \\
 &  & Naive & 0.84 (0.033) & -0.21 (0.035) & -0.40 (0.036) & 0.08 (0.036) & 0.25 (0.040) & 0.12 (0.033) \\ \hline
\multirow{3}{*}{2 $\mapsto$ 5} & \multirow{3}{*}{754 (3.97\%)} & True & 0.20 & -0.59 & -0.25 & 0.05 & -0.10 & -0.73 \\
 &  & SMART-MC & 0.20 (0.025) & -0.54 (0.019) & -0.26 (0.016) & 0.04 (0.024) & -0.10 (0.030) & -0.77 (0.025) \\
 &  & Naive & 0.83 (0.033) & -0.29 (0.036) & -0.05 (0.033) & -0.07 (0.038) & 0.32 (0.036) & -0.33 (0.033) \\ \hline
\multirow{3}{*}{4 $\mapsto$ 2} & \multirow{3}{*}{684 (3.60\%)} & True & -0.43 & 0.45 & -0.09 & -0.61 & 0.06 & -0.49 \\
 &  & SMART-MC & -0.40 (0.010) & 0.50 (0.010) & -0.10 (0.018) & -0.58 (0.011) & 0.14 (0.022) & -0.48 (0.010) \\
 &  & Naive & 0.53 (0.039) & 0.48 (0.038) & 0.15 (0.039) & -0.43 (0.036) & -0.04 (0.036) & -0.53 (0.039) \\ \hline
\multirow{3}{*}{3 $\mapsto$ 5} & \multirow{3}{*}{678 (3.57\%)} & True & 0.26 & 0.44 & 0.57 & 0.39 & 0.37 & -0.36 \\
 &  & SMART-MC & 0.24 (0.031) & 0.42 (0.015) & 0.54 (0.026) & 0.37 (0.047) & 0.39 (0.026) & -0.43 (0.031) \\
 &  & Naive & 0.90 (0.034) & 0.29 (0.034) & 0.14 (0.032) & 0.06 (0.030) & 0.19 (0.034) & -0.21 (0.034) \\ \hline
\end{tabular}}
\caption{Scenario 1: The true and estimated coefficients of the subject covariates corresponding to the most frequent initial treatment and the top 10 most frequent treatment transitions (empirically) are reported (considering 1,000 subjects, 10 treatments, and treatment sequences of length 20 per subject). Transition counts in the simulated dataset are presented, with transition proportions provided in parentheses. Bootstrap standard errors are listed in parentheses next to the estimated coefficient values.} 
\label{simulation_table} 
\end{table}

\begin{table}[]
\centering
\resizebox{0.92\columnwidth}{!}{%
\begin{tabular}{|c|c|c|c|c|c|c|c|c|}
\hline
Transitions & \begin{tabular}[c]{@{}c@{}}Transition\\ counts\end{tabular} & Scenario & $\beta_0$ & $\beta_1$ & $\beta_2$ & $\beta_3$ & $\beta_4$ & $\beta_5$ \\ \hline
\multirow{3}{*}{10 (initial state)} & \multirow{3}{*}{230 (23.00\%)} & True & 0.28 & 0.02 & 0.17 & 0.05 & 0.68 & 0.66 \\
 &  & SMART-MC & 0.39 (0.050) & -0.02 (0.020) & 0.16 (0.029) & 0.04 (0.020) & 0.67 (0.034) & 0.61 (0.029) \\
 &  & Naive & 0.76 (0.018) & -0.25 (0.018) & 0.03 (0.016) & 0.25 (0.015) & 0.29 (0.016) & 0.47 (0.015) \\ \hline
\multirow{3}{*}{6 $\mapsto$ 6} & \multirow{3}{*}{2562 (13.48\%)} & True & 0.30 & 0.60 & 0.58 & 0.27 & 0.25 & 0.29 \\
 &  & SMART-MC & 0.39 (0.042) & 0.58 (0.024) & 0.55 (0.027) & 0.25 (0.038) & 0.26 (0.039) & 0.27 (0.042) \\
 &  & Naive & 0.46 (0.024) & 0.51 (0.019) & 0.39 (0.017) & -0.38 (0.016) & 0.43 (0.017) & -0.22 (0.024) \\ \hline
\multirow{3}{*}{4 $\mapsto$ 4} & \multirow{3}{*}{1882 (9.91\%)} & True & 0.18 & 0.40 & 0.59 & 0.08 & 0.66 & 0.09 \\
 &  & SMART-MC & 0.12 (0.025) & 0.41 (0.037) & 0.59 (0.042) & 0.09 (0.030) & 0.67 (0.021) & 0.10 (0.025) \\
 &  & Naive & 0.68 (0.022) & 0.16 (0.022) & 0.12 (0.018) & -0.31 (0.018) & 0.57 (0.017) & -0.27 (0.022) \\ \hline
\multirow{3}{*}{8 $\mapsto$ 8} & \multirow{3}{*}{1870 (9.84\%)} & True & 0.19 & 0.54 & 0.51 & 0.05 & 0.63 & 0.11 \\
 &  & SMART-MC & 0.22 (0.040) & 0.54 (0.041) & 0.50 (0.040) & 0.05 (0.021) & 0.63 (0.040) & 0.08 (0.040) \\
 &  & Naive & 0.87 (0.019) & 0.19 (0.016) & 0.10 (0.018) & -0.43 (0.017) & 0.14 (0.019) & 0.01 (0.019) \\ \hline
\multirow{3}{*}{3 $\mapsto$ 3} & \multirow{3}{*}{1326 (6.98\%)} & True & 0.48 & 0.21 & 0.15 & 0.28 & 0.67 & 0.41 \\
 &  & SMART-MC & 0.48 (0.037) & 0.20 (0.033) & 0.14 (0.028) & 0.29 (0.044) & 0.67 (0.027) & 0.41 (0.037) \\
 &  & Naive & 0.78 (0.018) & -0.01 (0.021) & 0.34 (0.017) & -0.15 (0.017) & 0.34 (0.019) & -0.38 (0.018) \\ \hline
\multirow{3}{*}{6 $\mapsto$ 7} & \multirow{3}{*}{1232 (6.48\%)} & True & 0.78 & 0.02 & 0.08 & 0.30 & 0.20 & 0.50 \\
 &  & SMART-MC & 0.79 (0.042) & 0.02 (0.024) & 0.09 (0.027) & 0.29 (0.038) & 0.20 (0.039) & 0.50 (0.042) \\
 &  & Naive & 0.60 (0.017) & -0.52 (0.016) & -0.45 (0.017) & -0.27 (0.017) & 0.28 (0.014) & 0.14 (0.017) \\ \hline
\multirow{3}{*}{5 $\mapsto$ 5} & \multirow{3}{*}{1094 (5.76\%)} & True & 0.56 & 0.01 & 0.40 & 0.40 & 0.27 & 0.54 \\
 &  & SMART-MC & 0.56 (0.034) & 0.01 (0.024) & 0.40 (0.029) & 0.40 (0.031) & 0.27 (0.035) & 0.54 (0.034) \\
 &  & Naive & 0.91 (0.017) & -0.22 (0.017) & 0.25 (0.018) & 0.14 (0.019) & 0.05 (0.015) & 0.18 (0.017) \\ \hline
\multirow{3}{*}{1 $\mapsto$ 1} & \multirow{3}{*}{801 (4.22\%)} & True & 0.57 & 0.43 & 0.54 & 0.12 & 0.21 & 0.37 \\
 &  & SMART-MC & 0.57 (0.029) & 0.43 (0.041) & 0.55 (0.033) & 0.06 (0.037) & 0.21 (0.026) & 0.38 (0.029) \\
 &  & Naive & 0.33 (0.023) & -0.22 (0.022) & 0.70 (0.023) & 0.18 (0.022) & -0.54 (0.022) & 0.19 (0.023) \\ \hline
\multirow{3}{*}{9 $\mapsto$ 9} & \multirow{3}{*}{794 (4.18\%)} & True & 0.26 & 0.26 & 0.19 & 0.55 & 0.46 & 0.56 \\
 &  & SMART-MC & 0.25 (0.023) & 0.27 (0.036) & 0.18 (0.021) & 0.54 (0.042) & 0.45 (0.043) & 0.57 (0.023) \\
 &  & Naive & 0.79 (0.017) & -0.28 (0.017) & -0.21 (0.021) & 0.10 (0.020) & -0.03 (0.020) & 0.50 (0.017) \\ \hline
\multirow{3}{*}{2 $\mapsto$ 2} & \multirow{3}{*}{744 (3.92\%)} & True & 0.01 & 0.81 & 0.38 & 0.24 & 0.12 & 0.36 \\
 &  & SMART-MC & 0.08 (0.031) & 0.81 (0.023) & 0.38 (0.032) & 0.21 (0.027) & 0.13 (0.031) & 0.37 (0.031) \\
 &  & Naive & 0.83 (0.015) & 0.52 (0.015) & -0.16 (0.014) & -0.10 (0.016) & -0.05 (0.018) & 0.01 (0.015) \\ \hline
\multirow{3}{*}{7 $\mapsto$ 2} & \multirow{3}{*}{742 (3.91\%)} & True & 0.42 & 0.49 & 0.06 & 0.55 & 0.32 & 0.42 \\
 &  & SMART-MC & 0.45 (0.036) & 0.50 (0.042) & 0.06 (0.029) & 0.54 (0.031) & 0.31 (0.044) & 0.41 (0.036) \\
 &  & Naive & 0.94 (0.016) & 0.00 (0.017) & -0.30 (0.015) & 0.11 (0.016) & -0.09 (0.016) & -0.09 (0.016) \\ \hline
\end{tabular}}
\caption{Scenario 2: The true and estimated coefficients of the subject covariates corresponding to the most frequent initial treatment and the top 10 most frequent treatment transitions (empirically) are reported (considering 1,000 subjects, 10 treatments, and treatment sequences of length 20 per subject). Transition counts in the simulated dataset are presented, with transition proportions provided in parentheses. Bootstrap standard errors are listed in parentheses next to the estimated coefficient values.} 
\label{simulation_table_2} 
\end{table}

\begin{table}[]
\centering
\resizebox{0.65\columnwidth}{!}{%
\begin{tabular}{c|ccc}
\hline
\multirow{2}{*}{\begin{tabular}[c]{@{}c@{}}Trt. seq.\\ length\end{tabular}} & \multicolumn{3}{c}{Number of patients} \\ \cline{2-4} 
 & 1000 & 2000 & 3000 \\ \hline
20 & 0.0296 (0.0017) & 0.0209 (0.0011) & 0.0171 (0.0009)\\
40 & 0.0209 (0.0010) & 0.0150 (0.0010) & 0.0120 (0.0007)\\
60 & 0.0169 (0.0011) & 0.0121 (0.0006) & 0.0096 (0.0007)\\ \hline
\end{tabular}}
\caption{The mean absolute deviation (MAD) between the true and estimated coefficients for the top 10 most frequent treatment transitions (determined empirically) is evaluated across all combinations of the number of patients $(1000,2000,3000)$ and treatment sequence lengths per subject $(20,40,60)$, over 10 experiments under scenario 1. Mean MAD values are reported along with standard errors in parentheses.} 
\label{MAD_table} 
\end{table}

\begin{table}[]
\centering
\resizebox{0.9\columnwidth}{!}{%
\begin{tabular}{c|c|c|ccc|ccc|ccc}
\hline
\multirow{2}{*}{\begin{tabular}[c]{@{}c@{}}Num.\\ covariates\end{tabular}} & \multirow{2}{*}{\begin{tabular}[c]{@{}c@{}}Num.\\ treatments\end{tabular}} & \multirow{2}{*}{\begin{tabular}[c]{@{}c@{}}Number of\\ parameters\end{tabular}} & \multicolumn{3}{c|}{$K = 1000, t_k=10$} & \multicolumn{3}{c|}{$K = 1000, t_k=20$} & \multicolumn{3}{c}{$K = 2000, t_k=20$} \\ \cline{4-12} 
 &  &  & \begin{tabular}[c]{@{}c@{}}MSCOR\\ time (sec)\end{tabular} & \begin{tabular}[c]{@{}c@{}}par-MSCOR\\ time (sec)\end{tabular} & \begin{tabular}[c]{@{}c@{}}Speed\\ improvement\end{tabular} & \begin{tabular}[c]{@{}c@{}}MSCOR\\ time (sec)\end{tabular} & \begin{tabular}[c]{@{}c@{}}par-MSCOR\\ time (sec)\end{tabular} & \begin{tabular}[c]{@{}c@{}}Speed\\ improvement\end{tabular} & \begin{tabular}[c]{@{}c@{}}MSCOR\\ time (sec)\end{tabular} & \begin{tabular}[c]{@{}c@{}}par-MSCOR\\ time (sec)\end{tabular} & \begin{tabular}[c]{@{}c@{}}Speed\\ improvement\end{tabular} \\ \hline
\multirow{3}{*}{$p=3$} & $N=6$ & 168 & 38 & 10 & 3.8x & 43 & 11 & 3.9x & 95 & 31 & 3.1x \\
 & $N=9$ & 360 & 204 & 32 & 6.4x & 252 & 40 & 6.3x & 587 & 96 & 6.1x \\
 & $N=12$ & 624 & 1198 & 178 & 6.7x & 1502 & 211 & 7.1x & 3253 & 501 & 6.5x \\ \hline
\multirow{3}{*}{$p=5$} & $N=6$ & 252 & 53 & 10 & 5.3x & 57 & 11 & 5.2x & 136 & 22 & 6.2x \\
 & $N=9$ & 540 & 328 & 52 & 6.3x & 443 & 71 & 6.2x & 1014 & 163 & 6.2x \\
 & $N=12$ & 936 & 2158 & 344 & 6.3x & 2825 & 455 & 6.2x & 7077 & 1082 & 6.5x \\ \hline
\multirow{3}{*}{$p=8$} & $N=6$ & 378 & 120 & 24 & 5.0x & 159 & 29 & 5.5x & 315 & 54 & 5.8x \\
 & $N=9$ & 810 & 744 & 119 & 6.3x & 923 & 143 & 6.5x & 2057 & 337 & 6.1x \\
 & $N=12$ & 1404 & 4127 & 634 & 6.5x & 4881 & 765 & 6.4x & 12697 & 1931 & 6.6x \\ \hline
\end{tabular}}
\caption{Time comparisons between MSCOR and parallel-MSCOR for different numbers of covariates ($p$) and distinct treatment options ($N$) are evaluated for three data sizes under scenario 1. Specifically, $(K, t_k) = (1000, 10), (1000, 20), (2000, 20)$, where $K$ denotes the number of patients, and $t_k$ denotes the length of the generated treatment sequence for each patient.} 
\label{time_comparison}
\end{table}
\newpage
\section{Additional Results from Real Data analysis}
\vspace{-0.0cm}
\subsection{Scaling Real Data}
We rescale the age at diagnosis by first subtracting 12 (the minimum age of diagnosis observed in the dataset), followed by division by 10. The disease duration is rescaled by diving it by 10. Sex is taken to be 1 for females, and 0 for males. Race variables corresponding to White and Black are represented by two indicator variables. Cases with both of those indicator values being 0 represents other races.
\vspace{-0.6cm}
\subsection{Extra Tables and Figures from MS Case Study}
\begin{table}[]
\centering
\resizebox{0.95\columnwidth}{!}{%
\begin{tabular}{|c|c|c|c|c|c|c|c|}
\hline
Transitions & \begin{tabular}[c]{@{}c@{}}Transition\\ counts\end{tabular} & Intercept & \begin{tabular}[c]{@{}c@{}}Age at\\ diagnosis\end{tabular} & \begin{tabular}[c]{@{}c@{}}Disease\\ duration\end{tabular} & \begin{tabular}[c]{@{}c@{}}Sex\\ (Female)\end{tabular} & \begin{tabular}[c]{@{}c@{}}Race:\\ White\end{tabular} & \begin{tabular}[c]{@{}c@{}}Race:\\ Black\end{tabular} \\ \hline
Nat $\mapsto$ Nat & 2188 (24.84\%) & {\color[HTML]{000000} 0.70 (0.088)} & {\color[HTML]{000000} 0.27 (0.075)} & {\color[HTML]{000000} 0.13 (0.098)} & {\color[HTML]{000000} 0.21 (0.106)} & {\color[HTML]{000000} 0.57 (0.108)} & {\color[HTML]{000000} 0.24 (0.102)} \\ \hline
IB $\mapsto$ IB & 1682 (19.45\%) & {\color[HTML]{000000} 0.76 (0.072)} & {\color[HTML]{000000} 0.12 (0.091)} & {\color[HTML]{000000} 0.35 (0.107)} & {\color[HTML]{FE0000} -0.20 (0.119)} & {\color[HTML]{000000} 0.33 (0.125)} & {\color[HTML]{000000} 0.36 (0.137)} \\ \hline
S1P $\mapsto$ S1P & 1437 (16.62\%) & {\color[HTML]{000000} 0.51 (0.099)} & {\color[HTML]{000000} 0.19 (0.131)} & {\color[HTML]{000000} 0.18 (0.142)} & {\color[HTML]{FE0000} -0.27 (0.160)} & {\color[HTML]{000000} 0.77 (0.170)} & {\color[HTML]{FE0000} -0.11 (0.231)} \\ \hline
BcD $\mapsto$ BcD & 1031 (11.92\%) & {\color[HTML]{000000} NA} & {\color[HTML]{000000} NA} & {\color[HTML]{000000} NA} & {\color[HTML]{000000} NA} & {\color[HTML]{000000} NA} & {\color[HTML]{000000} NA} \\ \hline
DF $\mapsto$ DF & 934 (10.80\%) & {\color[HTML]{000000} 0.61 (0.129)} & {\color[HTML]{000000} 0.49 (0.116)} & {\color[HTML]{FE0000} -0.18 (0.144)} & {\color[HTML]{000000} 0.55 (0.172)} & {\color[HTML]{FE0000} -0.12 (0.198)} & {\color[HTML]{000000} 0.17 (0.232)} \\ \hline
AL $\mapsto$ AL & 462 (5.34\%) & {\color[HTML]{000000} NA} & {\color[HTML]{000000} NA} & {\color[HTML]{000000} NA} & {\color[HTML]{000000} NA} & {\color[HTML]{000000} NA} & {\color[HTML]{000000} NA} \\ \hline
IB $\mapsto$ S1P & 115 (1.33\%) & {\color[HTML]{FE0000} -0.71 (0.193)} & {\color[HTML]{FE0000} -0.31 (0.119)} & {\color[HTML]{000000} 0.33 (0.131)} & {\color[HTML]{000000} 0.05 (0.181)} & {\color[HTML]{FE0000} -0.04 (0.212)} & {\color[HTML]{FE0000} -0.54 (0.218)} \\ \hline
GA $\mapsto$ GA & 100 (1.16\%) & {\color[HTML]{000000} NA} & {\color[HTML]{000000} NA} & {\color[HTML]{000000} NA} & {\color[HTML]{000000} NA} & {\color[HTML]{000000} NA} & {\color[HTML]{000000} NA} \\ \hline
S1P $\mapsto$ BcD & 80 (0.93\%) & {\color[HTML]{FE0000} -0.69 (0.152)} & {\color[HTML]{FE0000} -0.14 (0.151)} & {\color[HTML]{FE0000} -0.04 (0.168)} & {\color[HTML]{FE0000} -0.48 (0.219)} & {\color[HTML]{000000} 0.24 (0.239)} & {\color[HTML]{FE0000} -0.47 (0.356)} \\ \hline
IB $\mapsto$ DF & 76 (0.88\%) & {\color[HTML]{FE0000} -0.53 (0.072)} & {\color[HTML]{FE0000} -0.14 (0.091)} & {\color[HTML]{000000} 0.28 (0.103)} & {\color[HTML]{FE0000} -0.69 (0.125)} & {\color[HTML]{FE0000} -0.36 (0.113)} & {\color[HTML]{000000} 0.08 (0.127)} \\ \hline
IB $\mapsto$ Nat & 59 (0.68\%) & {\color[HTML]{FE0000} -0.73 (0.317)} & {\color[HTML]{FE0000} -0.48 (0.124)} & {\color[HTML]{FE0000} -0.20 (0.170)} & {\color[HTML]{000000} 0.08 (0.222)} & {\color[HTML]{000000} 0.32 (0.249)} & {\color[HTML]{FE0000} -0.28 (0.325)} \\ \hline
DF $\mapsto$ BcD & 57 (0.66\%) & {\color[HTML]{FE0000} -0.79 (0.298)} & {\color[HTML]{000000} 0.16 (0.145)} & {\color[HTML]{FE0000} -0.15 (0.195)} & 0.18 (0.275) & {\color[HTML]{FE0000} -0.55 (0.233)} & {\color[HTML]{FE0000} 0.05 (0.380)} \\ \hline
DF $\mapsto$ S1P & 46 (0.53\%) & {\color[HTML]{FE0000} -0.82 (0.158)} & {\color[HTML]{000000} 0.03 (0.124)} & {\color[HTML]{FE0000} -0.37 (0.154)} & {\color[HTML]{000000} 0.06 (0.189)} & {\color[HTML]{000000} 0.18 (0.198)} & {\color[HTML]{FE0000} -0.39 (0.128)} \\ \hline
Nat $\mapsto$ BcD & 41 (0.47\%) & {\color[HTML]{FE0000} -0.82 (0.064)} & {\color[HTML]{FE0000} -0.08 (0.088)} & {\color[HTML]{FE0000} -0.20 (0.119)} & {\color[HTML]{FE0000} -0.12 (0.126)} & {\color[HTML]{FE0000} -0.48 (0.089)} & {\color[HTML]{FE0000} -0.20 (0.119)} \\ \hline
Nat $\mapsto$ S1P & 36 (0.41\%) & {\color[HTML]{FE0000} -0.65 (0.088)} & {\color[HTML]{FE0000} -0.12 (0.112)} & {\color[HTML]{FE0000} -0.23 (0.149)} & {\color[HTML]{FE0000} -0.32 (0.152)} & {\color[HTML]{FE0000} -0.51 (0.112)} & {\color[HTML]{FE0000} -0.38 (0.131)} \\ \hline
S1P $\mapsto$ DF & 33 (0.38\%) & {\color[HTML]{FE0000} -0.80 (0.090)} & {\color[HTML]{FE0000} -0.21 (0.123)} & {\color[HTML]{FE0000} -0.15 (0.141)} & {\color[HTML]{FE0000} -0.27 (0.152)} & {\color[HTML]{FE0000} -0.36 (0.142)} & {\color[HTML]{FE0000} -0.30 (0.161)} \\ \hline
\end{tabular}}
\caption{SMART-MC estimated coefficient values corresponding to the most frequent treatment transitions (with at least 30 transitions) are reported, along with the corresponding transition counts (transition proportions). Bootstrap standard errors are listed in parentheses next to the estimated coefficient values. Estimated negative coefficient values are highlighted in red.} 
\label{MS_coeffs}
\end{table}

\begin{table}[]
\centering
\resizebox{0.65\columnwidth}{!}{%
\begin{tabular}{|c|c|c|c|c|c|}
\hline
{\color[HTML]{000000} Transitions} & {\color[HTML]{000000} \begin{tabular}[c]{@{}c@{}}Age at\\ diagnosis\end{tabular}} & {\color[HTML]{000000} \begin{tabular}[c]{@{}c@{}}Disease\\ duration\end{tabular}} & {\color[HTML]{000000} \begin{tabular}[c]{@{}c@{}}Sex\\ (Female)\end{tabular}} & {\color[HTML]{000000} \begin{tabular}[c]{@{}c@{}}Race:\\ White\end{tabular}} & {\color[HTML]{000000} \begin{tabular}[c]{@{}c@{}}Race:\\ Black\end{tabular}} \\ \hline
{\color[HTML]{000000} Nat $\mapsto$ Nat} & {\color[HTML]{000000} $<$0.001$^*$} & {\color[HTML]{000000} 0.185} & {\color[HTML]{000000} 0.048$^*$} & {\color[HTML]{000000} $<$0.001$^*$} & {\color[HTML]{000000} 0.019$^*$} \\ \hline
{\color[HTML]{000000} IB $\mapsto$ IB} & {\color[HTML]{000000} 0.187} & {\color[HTML]{000000} 0.001$^*$} & {\color[HTML]{FE0000} 0.093} & {\color[HTML]{000000} 0.008$^*$} & {\color[HTML]{000000} 0.009$^*$} \\ \hline
{\color[HTML]{000000} S1P $\mapsto$ S1P} & {\color[HTML]{000000} 0.147} & {\color[HTML]{000000} 0.205} & {\color[HTML]{FE0000} 0.092} & {\color[HTML]{000000} $<$0.001$^*$} & {\color[HTML]{FE0000} 0.634} \\ \hline
{\color[HTML]{000000} BcD $\mapsto$ BcD} & {\color[HTML]{000000} NA} & {\color[HTML]{000000} NA} & {\color[HTML]{000000} NA} & {\color[HTML]{000000} NA} & {\color[HTML]{000000} NA} \\ \hline
{\color[HTML]{000000} DF $\mapsto$ DF} & {\color[HTML]{000000} $<$0.001$^*$} & {\color[HTML]{FE0000} 0.211} & {\color[HTML]{000000} 0.001$^*$} & {\color[HTML]{FE0000} 0.544} & {\color[HTML]{000000} 0.464} \\ \hline
{\color[HTML]{000000} AL $\mapsto$ AL} & {\color[HTML]{000000} NA} & {\color[HTML]{000000} NA} & {\color[HTML]{000000} NA} & {\color[HTML]{000000} NA} & {\color[HTML]{000000} NA} \\ \hline
{\color[HTML]{000000} IB $\mapsto$ S1P} & {\color[HTML]{FE0000} 0.009$^*$} & {\color[HTML]{000000} 0.012$^*$} & {\color[HTML]{000000} 0.782} & {\color[HTML]{FE0000} 0.850} & {\color[HTML]{FE0000} 0.013$^*$} \\ \hline
{\color[HTML]{000000} GA $\mapsto$ GA} & {\color[HTML]{000000} NA} & {\color[HTML]{000000} NA} & {\color[HTML]{000000} NA} & {\color[HTML]{000000} NA} & {\color[HTML]{000000} NA} \\ \hline
{\color[HTML]{000000} S1P $\mapsto$ BcD} & {\color[HTML]{FE0000} 0.354} & {\color[HTML]{FE0000} 0.812} & {\color[HTML]{FE0000} 0.028$^*$} & {\color[HTML]{000000} 0.315} & {\color[HTML]{FE0000} 0.187} \\ \hline
{\color[HTML]{000000} IB $\mapsto$ DF} & {\color[HTML]{FE0000} 0.124} & {\color[HTML]{000000} 0.007$^*$} & {\color[HTML]{FE0000} $<$0.001$^*$} & {\color[HTML]{FE0000} $<$0.001$^*$} & {\color[HTML]{000000} 0.529} \\ \hline
{\color[HTML]{000000} IB $\mapsto$ Nat} & {\color[HTML]{FE0000} $<$0.001$^*$} & {\color[HTML]{FE0000} 0.239} & {\color[HTML]{000000} 0.719} & {\color[HTML]{000000} 0.199} & {\color[HTML]{FE0000} 0.389} \\ \hline
{\color[HTML]{000000} DF $\mapsto$ BcD} & {\color[HTML]{000000} 0.270} & {\color[HTML]{FE0000} 0.442} & {\color[HTML]{000000} 0.513} & {\color[HTML]{FE0000} 0.018$^*$} & {\color[HTML]{FE0000} 0.895} \\ \hline
{\color[HTML]{000000} DF $\mapsto$ S1P} & {\color[HTML]{000000} 0.809} & {\color[HTML]{FE0000} 0.016$^*$} & {\color[HTML]{000000} 0.751} & {\color[HTML]{000000} 0.363} & {\color[HTML]{FE0000} 0.002$^*$} \\ \hline
{\color[HTML]{000000} Nat $\mapsto$ BcD} & {\color[HTML]{FE0000} 0.363} & {\color[HTML]{FE0000} 0.093} & {\color[HTML]{FE0000} 0.341} & {\color[HTML]{FE0000} $<$0.001$^*$} & {\color[HTML]{FE0000} 0.093} \\ \hline
{\color[HTML]{000000} Nat $\mapsto$ S1P} & {\color[HTML]{FE0000} 0.284} & {\color[HTML]{FE0000} 0.123} & {\color[HTML]{FE0000} 0.035$^*$} & {\color[HTML]{FE0000} $<$0.001$^*$} & {\color[HTML]{FE0000} 0.004$^*$} \\ \hline
{\color[HTML]{000000} S1P $\mapsto$ DF} & {\color[HTML]{FE0000} 0.088} & {\color[HTML]{FE0000} 0.287} & {\color[HTML]{FE0000} 0.076} & {\color[HTML]{FE0000} 0.011$^*$} & {\color[HTML]{FE0000} 0.062} \\ \hline
\end{tabular}%
}
\caption{Bootstrap-based two-sided p-values for estimated SMART-MC coefficients corresponding to treatment transitions with at least 30 observed cases. Values are based on standard normal approximation using coefficient estimate and bootstrap standard error. Significant p-values are marked with asterisks; p-values corresponding to negative estimated coefficient values are highlighted in red.}
\label{MS_pvalues}
\end{table}

\begin{figure}[]
	\centering
	\includegraphics[width=0.95\textwidth]{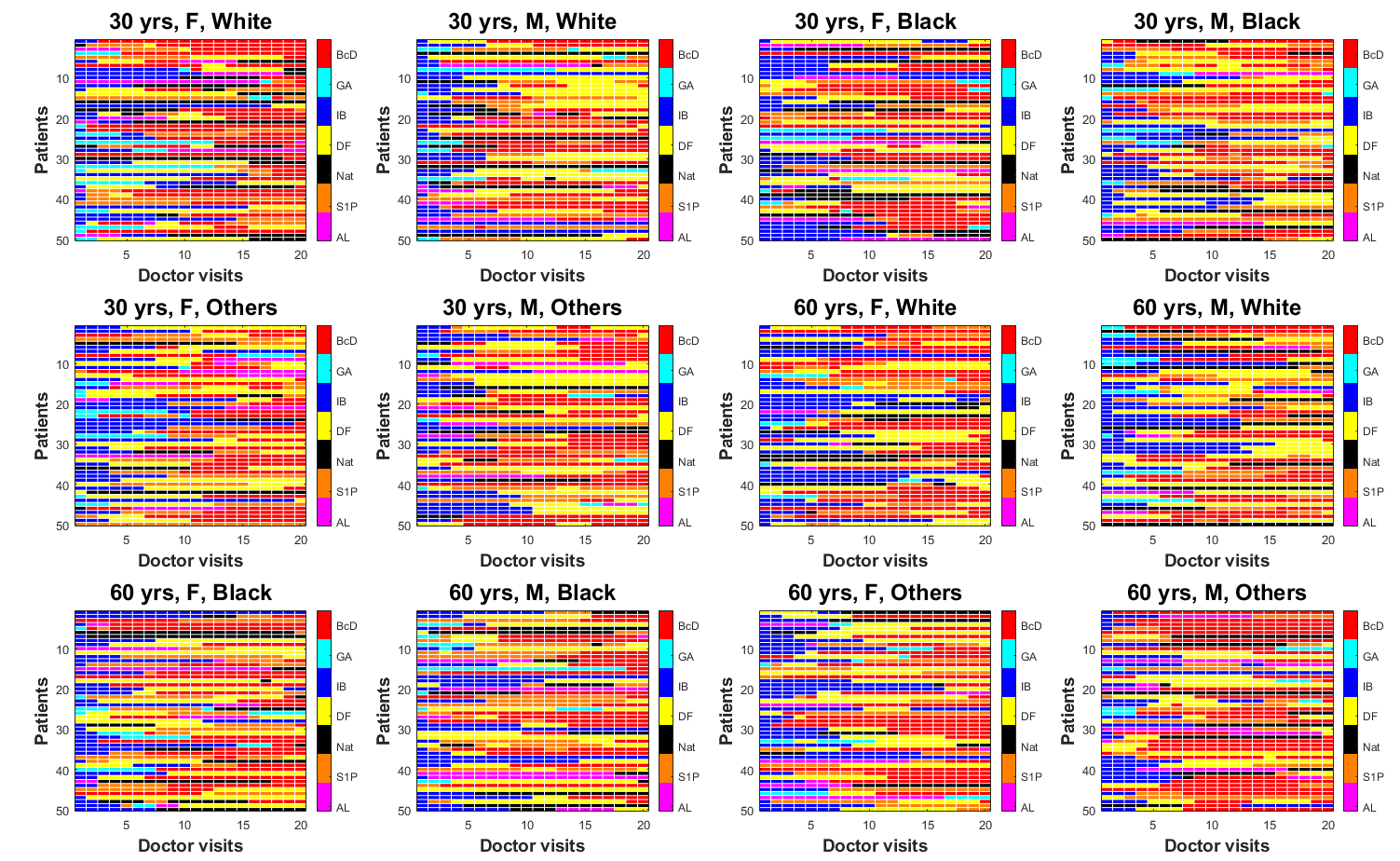}
	\caption{Multiple Sclerosis treatment sequences are generated for different types of patients based on age at diagnosis (30 years/60 years), sex (M/F), and race (White/Black/Others), using underlying Markov chain mechanism with optimal values of treatment transition-specific and covariate-specific coefficients estimated via SMART-MC. For each scenario, 50 generated realizations (y-axis) are depicted over the first 20 doctor visits (x-axis).}
	\label{dummy_seq}
\end{figure}

Training SMART-MC on the real data, we use the estimated model parameters as a generative model to obtain DMT sequences of hypothetical patients varying across age (30/60 years), sex (male/female) and race (White/Black/Others). Such dummy DMT sequences, each of length 20, are obtained 50 times for each combination of aforementioned covariate levels, and shown in Figure \ref{dummy_seq}. Across all cases, the disease duration for the dummy patients is set at the population mean level (15.4 months). SMART-MC estimated coefficient values (with bootstrap-based standard errors in parentheses) corresponding to the non-rare treatment transitions are reported in Table \ref{MS_coeffs}. Estimated coefficients corresponding to the transitions BcD to BcD, AL to AL, and GA to GA are not shown, as the corresponding rows in the transition matrix do not contain any other non-rare treatments, rendering these coefficients non-interpretable. Finally we report two-sided p-values for SMART-MC coefficient estimates using a standard normal approximation based on bootstrap standard errors in Table~\ref{MS_pvalues}.

\subsection{Interpretation of Covariate Effects on MS Treatment Transitions}

We summarize findings that correspond to the following reviewer-posed questions, using the bootstrap-based coefficient estimates and p-values reported in Tables~\ref{MS_coeffs} and~\ref{MS_pvalues} of the Supplement.

\begin{itemize}
    \item[(i)] \textbf{Effect of Clinical Factors (Disease Duration).} 

    Statistically significant effects of disease duration ($p < 0.05$) were identified in the following transitions:
     \begin{itemize}
        \item IB $\mapsto$ IB ($p = 0.001$; positive effect),
        \item IB $\mapsto$ DF ($p = 0.007$; positive effect),
        \item IB $\mapsto$ S1P ($p = 0.012$; positive effect),
        \item DF $\mapsto$ S1P ($p = 0.016$; negative effect).
    \end{itemize}
    These results indicate that longer disease duration is associated with increased persistence on first-line injectables (IB) and a greater likelihood of transitioning to fumarates (DF), but decreased likelihood of escalating from DF to S1P modulators. This pattern suggests disease progression and patient history may play a role in treatment de-escalation decisions.

    \item[(ii)] \textbf{Effect of Demographic Factors (Age, Sex, Race).} 

    \textit{Age at Diagnosis:} Significant transitions influenced by age include:
    \begin{itemize}
        \item Nat $\mapsto$ Nat ($p < 0.001$; positive effect),
        \item DF $\mapsto$ DF ($p < 0.001$; positive effect),
        \item IB $\mapsto$ S1P ($p = 0.009$; negative effect),
        \item IB $\mapsto$ Nat ($p < 0.001$; negative effect).
    \end{itemize}
    Older patients are more likely to persist on current treatments (e.g., Nat, DF) but less likely to escalate from injectables to higher efficacy agents like S1P or natalizumab.

    \textit{Sex (Female):} Statistically significant effects include:
    \begin{itemize}
        \item Nat $\mapsto$ Nat ($p = 0.048$; positive effect),
        \item DF $\mapsto$ DF ($p = 0.001$; positive effect),
        \item IB $\mapsto$ DF ($p < 0.001$; negative effect),
        \item Nat $\mapsto$ S1P ($p = 0.035$; negative effect),
        \item S1P $\mapsto$ BcD ($p = 0.028$; negative effect).
        \end{itemize}
    These results suggest that female patients are more likely to persist on certain therapies (e.g., DF, Nat) but less likely to escalate to B-cell depletion or S1P modulators, and may be less likely to switch from injectables to fumarates.

    \textit{Race (Black):} Significant covariate effects include:
    \begin{itemize}
    \item Nat $\mapsto$ Nat ($p = 0.019$; positive effect),
    \item IB $\mapsto$ IB ($p = 0.009$; positive effect),
    \item IB $\mapsto$ S1P ($p = 0.013$; negative effect),
    \item DF $\mapsto$ S1P ($p = 0.002$; negative effect),
    \item Nat $\mapsto$ S1P ($p = 0.004$; negative effect).
    \end{itemize}
    These results suggest that Black patients are more likely to persist on certain therapies such as interferons and natalizumab, but less likely to transition into S1P modulators from other classes. This pattern may reflect differences in care pathways, access, or patient-level tolerability considerations, and underscores the importance of accounting for racial differences when modeling treatment dynamics.

    \textit{Race (White):} Notable effects include:
    \begin{itemize}
    \item Nat $\mapsto$ Nat ($p < 0.001$; positive effect),
    \item IB $\mapsto$ IB ($p = 0.008$; positive effect),
    \item S1P $\mapsto$ S1P ($p < 0.001$; positive effect),
    \item IB $\mapsto$ DF ($p < 0.001$; negative effect),
    \item DF $\mapsto$ BcD ($p = 0.018$; negative effect),
    \item Nat $\mapsto$ BcD ($p < 0.001$; negative effect),
    \item Nat $\mapsto$ S1P ($p < 0.001$; negative effect),
    \item S1P $\mapsto$ DF ($p = 0.011$; negative effect)
    \end{itemize}
    These findings imply that White patients show higher persistence on several DMT classes (e.g., IB, S1P, Nat), but lower likelihood of transitioning between certain therapy classes such as IB to DF, or Nat to BcD/S1P. This may reflect clinical preference for maintaining stability or differential access/tolerability profiles across treatment types.

    \item[(iii)] \textbf{Most Frequent Transitions and Sensitivity to Covariates.} 
    Based on Tables~\ref{MS_coeffs} and \ref{MS_pvalues}, we summarize the most common treatment transitions in real-world MS care and identify those most sensitive to patient-level covariates. The five most frequent transitions involved continuation of the same DMT as follows:

    \begin{itemize}
    \item Nat $\mapsto$ Nat (24.84\%)
    \item IB $\mapsto$ IB (19.45\%)
    \item S1P $\mapsto$ S1P (16.62\%)
    \item BcD $\mapsto$ BcD (11.92\%)
    \item DF $\mapsto$ DF (10.80\%)
    \end{itemize}

    Among these, Nat $\mapsto$ Nat showed significant effects of age ($p < 0.001$), sex ($p = 0.048$), and race (White: $p < 0.001$; Black: $p = 0.019$), suggesting that both demographic and racial factors influence persistence on natalizumab. Similarly, DF $\mapsto$ DF was associated with age ($p < 0.001$) and sex ($p = 0.001$), with older patients and females more likely to continue on fumarates. IB $\mapsto$ IB persistence was significantly impacted by disease duration ($p = 0.001$) and race (White: $p = 0.008$, Black: $p = 0.009$), while S1P $\mapsto$ S1P showed strong association with race (White: $p < 0.001$). These findings indicate that both clinical history and demographic profiles shape continuation patterns across multiple DMTs.

    Among across-treatment transitions, the most frequent were:
    \begin{itemize}
    \item IB $\mapsto$ S1P (115 transitions; 1.33\%),
    \item IB $\mapsto$ DF (76 transitions; 0.88\%),
    \item DF $\mapsto$ S1P (46 transitions; 0.53\%),
    \item DF $\mapsto$ BcD (57 transitions; 0.66\%),
    \item Nat $\mapsto$ S1P (36 transitions; 0.41\%).
    \end{itemize}

    These transitions exhibited sensitivity to multiple covariates. For instance, IB $\mapsto$ DF was associated with disease duration ($p = 0.007$; positive), sex ($p < 0.001$; negative), and race (White: $p < 0.001$; negative), suggesting that longer disease history and male sex increase the likelihood of this injectable-to-oral switch, while White patients were less likely to undergo it. IB $\mapsto$ S1P showed significant effects of age ($p = 0.009$; negative), disease duration ($p = 0.012$; positive), and race (Black: $p = 0.013$; negative), indicating that younger non-Black patients with longer disease duration were more likely to escalate to S1P modulators. DF $\mapsto$ S1P was associated with disease duration ($p = 0.016$; negative) and Black race ($p = 0.002$; negative), suggesting that patients with longer disease course or Black patients were less likely to escalate from fumarate to S1P. DF $\mapsto$ BcD showed a significant negative association with White race ($p = 0.018$), indicating that White patients were less likely to escalate from dimethyl fumarate to B-cell therapies. Nat $\mapsto$ S1P was significantly associated with sex ($p = 0.035$; negative), and both White ($p < 0.001$; negative) and Black ($p = 0.004$; negative) race indicating that females and racial minorities were less likely to de-escalate from natalizumab to S1P modulators.
    Collectively, these patterns reflect nuanced, individualized care decisions that are sensitive to patient demographics and disease history, especially during escalation or de-escalation between DMT classes.
\end{itemize}

In summary, SMART-MC facilitates detailed covariate-level inference on MS treatment transitions. Bootstrap-based uncertainty quantification supports robust statistical conclusions, addressing key reviewer concerns regarding covariate significance and enabling hypothesis-driven analysis of real-world treatment sequences.

\newpage
\bibliographystyle{agsm}
\bibliography{Bibliography-MM-MC}